\documentclass[11pt,a4paper]{article}
\pdfoutput=1
\usepackage{jheppub}
\usepackage{lscape}
\usepackage{array}

\usepackage{amsmath}
\usepackage{amsfonts}
\usepackage{amssymb}
\usepackage{bbm} 
\usepackage{graphicx}
\usepackage{caption}
\usepackage{subcaption}

\usepackage[normalem]{ulem}   

\usepackage{diagbox}

\usepackage{mathtools}

\def\rme{{\rm e}}
\newcommand{\ii}{\mathrm{i}}

\newcommand\be{\begin{equation}}
\newcommand\ee{\end{equation}}
\newcommand\bea{\begin{eqnarray}}
\newcommand\eea{\end{eqnarray}}

\newcommand{\nn}{\nonumber}
\newcommand{\dd}{\mathrm{d}}

\renewcommand{\=}{\,= \,}

\renewcommand{\a}{\alpha}

\newcommand{\s}{\sigma}
\renewcommand{\t}{\tau}

\newcommand\G{\Gamma}

\renewcommand{\nn}{\nonumber}

\newcommand{\wt}{\widetilde}

\renewcommand{\Re}{\text{Re}}
\renewcommand{\Im}{\text{Im}}

\newcommand{\ve}{\varepsilon}
\newcommand{\CN}{\mathcal{N}}

\newcommand{\CH}{\mathcal{H}}

\newcommand{\p}{\partial}

\newcommand{\CI}{\mathcal I}
\newcommand{\CF}{\mathcal F}

\newcommand{\CC}{\mathcal C}

\newcommand{\IZ}{\mathbb Z}
\newcommand{\IR}{\mathbb R}
\newcommand{\IC}{\mathbb C}

\renewcommand{\v}{\varphi}
\newcommand{\vt}{\widetilde \varphi}

\newcommand{\ndt}{\noindent}

\renewcommand{\i}{{\rm i}}

\newcommand{\Ge}{\Gamma_\text{e}}
\newcommand{\e}{{\bf e}}
\newcommand{\vth}{\vartheta}
\renewcommand{\th}{\theta}
\newcommand{\half}{\frac12}
\newcommand{\z}{\zeta}

\newcommand{\ret}{\tau_1}
\newcommand{\imt}{\tau_2}

\newcommand{\Seff}{S_\text{eff}}

\newcommand{\uu}{\underline{u}}

\newcommand{\defeq}{\; \coloneqq \;} 

\newcommand{\PsiQ}{\Psi_Q}

\title{Supersymmetric phases of 4d~$\CN=4$ SYM at large $N$}

\author{Alejandro Cabo-Bizet}
\emailAdd{alejandro.cabo\_bizet@kcl.ac.uk}
\author{and Sameer Murthy}
\emailAdd{sameer.murthy@kcl.ac.uk}
\affiliation{Department of Mathematics, King's College London,\\
The Strand, London WC2R 2LS, U.K.}

\abstract{
We find a family of complex saddle-points at large~$N$ of the matrix model for the superconformal index 
of~$SU(N)$ $\CN=4$ super Yang-Mills theory on~$S^3 \times S^1$ with one chemical potential~$\t$.  
The saddle-point configurations are labelled by points~$(m,n)$ on the lattice~$\Lambda_\t = \IZ \t +\IZ$ with~$\text{gcd}(m,n)=1$.
The eigenvalues at a given saddle are uniformly distributed along a string winding~$(m,n)$ times along the~$(A,B)$ cycles 
of the torus~$\IC/\Lambda_\t$. 
The action of the matrix model extended to the torus is closely related to the Bloch-Wigner elliptic dilogarithm, and 
the related Bloch formula allows us to calculate the action at the saddle-points in terms of real-analytic Eisenstein series. 
The actions of~$(0,1)$ and~$(1,0)$ agree with that of pure~AdS$_5$ and 
the supersymmetric~AdS$_5$ black hole, respectively. 
The black hole saddle dominates  the canonical ensemble when~$\t$ is close to the origin, 
and there are new saddles that dominate when~$\t$ approaches rational points. 
The extension of the action in terms of modular forms leads to a simple treatment 
of the Cardy-like limit~$\t\to 0$.
}

\begin{document}
 
\maketitle


\section{Introduction \label{sec:Intro}}

The subject of this paper is a matrix model defined by an integral over~$N \times N$ 
unitary matrices~$U$ of the following type,
\be \label{Uact}
\CI(\vec{\t}) \= \int \, DU\, \exp \biggl( \; \sum_{n=1}^\infty \, \frac{1}{n} \, f(n\vec{\t}) \, \text{tr} \, U^n \, \text{tr} \, (U^\dagger)^n \, \biggr) \,.
\ee
Here~$\vec{\t}$ denotes the vector of coupling constants of the model, and~$DU$ is the invariant measure. 
Upon diagonalizing the matrices, one obtains an integral over the~$N$ eigenvalues 
which experience a purely two-particle interaction governed by the single function~$f(\vec{\t})$.
The particular model that we analyze arises as the generating function of the 
superconformal index~\cite{Romelsberger:2005eg,Kinney:2005ej} of~$\CN=4$ super Yang-Mills (SYM)
theory on $S^3 \times S^1$. In this model~$U$ has the interpretation as the holonomy of the gauge field
around~$S^1$, and there are three coupling constants~$\t$, $\s$, $\v$, which are interpreted, respectively, as the chemical potentials
dual to the two angular momenta on the~$S^3$ and a certain combination of the R-charges of the theory. 
The function~$f$ has the interpretation as the single-letter index of the super Yang-Mills theory~\cite{Romelsberger:2005eg}. 

The matrix model~\eqref{Uact} for the supersymmetric index has generated a lot of interest since the work of~\cite{Kinney:2005ej}. 
The holographically dual 
gravitational theory with~AdS$_5$ boundary conditions admits supersymmetric black hole solutions 
which preserve the same supercharges as the superconformal index~$\CI$ captured by the matrix model. 
This leads to the expectation that the (indexed) degeneracies of states contained 
in~$\CI$ grow exponentially as a function of the charges when the charges become large enough to form a black hole. 
The analysis of~\cite{Kinney:2005ej} seemed to suggest that the shape, or functional form, of~$\CI$ 
does not allow for such a growth of states, thus leading to a long-standing puzzle. 

Following the insightful observation of~\cite{Hosseini:2017mds}, this puzzle has been revisited recently
\cite{Cabo-Bizet:2018ehj,
Choi:2018hmj,Choi:2018fdc,Benini:2018mlo,Benini:2018ywd,Suh:2018qyv,Honda:2019cio,ArabiArdehali:2019tdm,Zaffaroni:2019dhb,
Kim:2019yrz,Cabo-Bizet:2019osg,Amariti:2019mgp,Cassani:2019mms,Larsen:2019oll,Kantor:2019lfo,
Nahmgoong:2019hko,Lezcano:2019pae,Lanir:2019abx}
The emergent picture shows that there is actually an exponential growth of states contained in the index,
which is captured by the saddle-point estimate if one allows the potentials~$\s$, $\t$, $\v$ to take values in the complex plane away 
from the pure imaginary values assumed in~\cite{Kinney:2005ej}.\footnote{If we restrict all three potentials to be purely imaginary, 
these analyses also show that there is no growth of states at large~$N$, as consistent with the analysis of~\cite{Kinney:2005ej}.} 
There are essentially three strands of analysis contained in the recent progress: 
\begin{enumerate}
\item The first strand is an analysis of the bulk supergravity~\cite{Cabo-Bizet:2018ehj} which shows that 
the black hole entropy can be understood as the Legendre transform of the 
regularized on-shell action of a family of Euclidean solutions lying on the surface
\be \label{tsvcons}
\s + \t - 2\v = n_0 \,, \qquad n_0 \= -1 \,.
\ee
It was first noticed in~\cite{Hosseini:2017mds} that this constraint is important to obtain the 
black hole entropy as an extremization of an entropy function. 
The result of~\cite{Cabo-Bizet:2018ehj} gives a physical interpretation to this observation---the entropy function 
is the gravitational action, and the constraint arises from demanding supersymmetry. 
One is thus led to look at configurations on this surface in the field theory index
in order to make accurate saddle-points estimates. 

\item The second strand, referred to as the Cardy-like limit~\cite{Choi:2018hmj,Kim:2019yrz,Honda:2019cio,
ArabiArdehali:2019tdm,Cabo-Bizet:2019osg} 
is the analysis of the index~$\CI$ in the limit when 
the charges are much larger than all other parameters of the theory including~$N$. 
In the bulk gravitational dual, this corresponds to a black hole that ``fills up all of AdS space".
In this limit 
one finds a configuration of eigenvalues clumped near the origin which has an entropy
equal to that of the black hole. This analysis has some subtleties related to convergence 
of infinite sums and the related puzzle that if one literally puts all the eigenvalues at the origin then
the index seems to vanish. The resolution is given by a certain limiting procedure in which 
the eigenvalues are placed close to zero 
and one then takes the limit of the entropy as they go to zero. 
\item The third strand is the Bethe-ansatz analysis for the subspace~$\t=\s$, initiated by~\cite{Benini:2018mlo, Benini:2018ywd}. 
Here the analysis is performed not directly on the matrix model~\eqref{Uact} but on an equivalent 
representation of the index~\cite{Closset:2017bse,Closset:2018ghr},~\cite{Benini:2018mlo} which one can solve 
at large~$N$ by computing the roots of the Bethe-ansatz equations. 
These roots are referred to in~\cite{Benini:2018ywd} as saddle-point-like 
configurations, suggesting that those configurations could correspond to the saddle-points of 
the matrix model~\eqref{Uact}. 
This analysis was shown in~\cite{Benini:2018ywd} to be consistent with that of Point~2 when
one further takes the Cardy-like limit.  
\end{enumerate}

\begin{figure}\centering
\includegraphics[width=6.5cm]{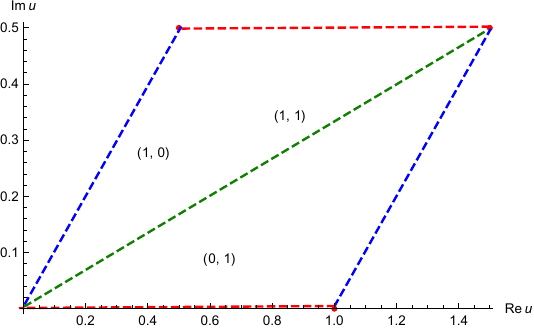}\qquad\qquad\qquad
\includegraphics[width=4.8cm]{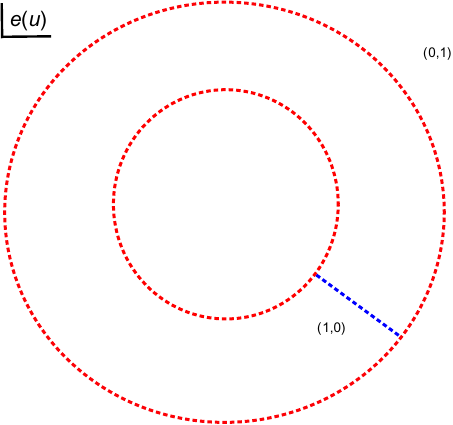}
  \caption{The saddle-point configurations of eigenvalues in the fundamental domain and on the torus for $\tau=\frac{1}{2}(1+\i)$. 
  The saddles shown are $(0,1)$ (red), $(1,0)$ (blue), and $(1,1)$ (green). 
  The original definition of the matrix model is the interval~$[0,1)$ on the left which maps to the outer circle of unit radius on the right.}
  \label{SaddleUp}
\end{figure}

Put together, these three strands of analysis strongly suggest that there should be a complex saddle-point of the 
matrix model~\eqref{Uact} at large $N$ and finite~$\t$, having the same entropy as that of the black hole. 
In this paper we show that this is indeed the case in the subspace of coupling constants~$\s=\t$. Within this 
subspace we find an infinite family of saddle-points labelled by points on the lattice~$\Lambda_\t = \IZ \t +\IZ$. 
More precisely, writing a generic lattice point as~$m\t+n$, $m,n \in \IZ$, the independent saddles are labelled by~$(m,n)$ 
with~$\text{gcd}(m,n)=1$.
As we explain below, our solution is best presented in terms of the 
torus~$\IC/\Lambda_\t$ whose~$A$-cycle is identified with the original circle of unit radius.
The eigenvalues at the~$(m,n)$ saddle are uniformly distributed along a string winding~$m$ and~$n$ times 
along the~$A$ and~$B$ cycles, respectively, of the torus  (see Figure~\ref{SaddleUp}). 
In particular, the saddle~$(0,1)$ is identified with pure~AdS$_5$ and~$(1,0)$ is identified with the 
supersymmetric~AdS$_5$ black hole.

Our method of analysis is as follows. Firstly we need to extend the action to complex values of the fields. 
Denoting the eigenvalues of the matrix~$U$ as~$e^{2 \pi \i u_i}$, $i=1,\dots,N$, 
the variables~$u_i$ of the original unitary matrix model take values in~$\IR/\IZ$, whose representative 
we choose throughout the paper to be~$[0,1)$.
``Complex field configurations" in our context means complex values of~$u_i$ or, equivalently,  
moving away from the original circle of unit radius in the variables~$e^{2\pi \i u_i}$.
A priori there is no canonical manner to extend the action to the complex plane.
A natural first guess may be to use the analytic continuation of the original action away from real values of~$u_i$.
However, the construction of our solution needs the action to be well-defined on the torus, for which the 
analytic continuation is not the right choice as it is only 
quasi-periodic under translations~$u_i \to u_i +\t$ (in a manner that generalizes the 
quasi-periodicity of the Jacobi~$\theta$-function).
The good choice of continuation of the action to the complex~$u_i$ plane turns out to be a doubly periodic 
function, i.e.~one which is strictly periodic under translations of~$1$ and~$\t$. 
This function is not meromorphic but (with a small abuse of terminology) we shall sometimes call it an elliptic function. 
This choice can be thought of 
as a real-analytic (in~$u$) deformation of the analytic continuation which cures the ``small" non-periodicity 
at the expense of meromorphy.

The fact that the extended action is not meromorphic gives rise to new subtleties 
compared to the saddle-point analysis for meromophic functions. Firstly we need to 
find solutions to the variational equations in the real and imaginary directions separately.
It is actually quite easy to see that any periodic uniform distribution of eigenvalues in the 
complex plane solves the saddle-point equations, and we present this as a Lemma in Section~\ref{sec:Saddles}.
Secondly, we need a separate argument to justify the contour deformation so that it passes
through the saddle-points. In Section~\ref{sec:contour} we show that we can replace the 
doubly-periodic integrand by a meromorphic integrand without changing the large-$N$ value of the corresponding integrals
on the original contour as well as on a contour passing through the saddle-point. 
We can thus use the meromorphicity to deform the contour, modulo issues of wall-crossings that we do not analyze here.
Thirdly, we need to consider the space of all contours and analyze the contribution 
of a given saddle-point as a function of the moduli of the theory. We do not address this last issue in this paper. 
We note, however, that, since the contour deformation argument relies on an associated meromorphic 
function, it may be possible to adapt the formalism of Picard-Lefschetz theory 
and the ideas of resurgence to our case (a recent comprehensive review of this subject 
can be found in~\cite{Aniceto:2018bis}).

The particular elliptic function that underlies our extended action can be identified with the \emph{elliptic dilogarithm}, 
a certain single-valued real-analytic extension of the ordinary dilogarithm first studied by D.~Wigner and S.~Bloch~\cite{Bloch}.
This identification immediately places us in a very good position. The main technical tool is a formula due to Bloch~\cite{Bloch} 
which has been studied intensively by number theorists~\cite{ZagierOnBloch,DukeImamoglu}.
Bloch's formula expresses the elliptic dilogarithm in terms of a real analytic lattice sum also known as a 
Kronecker-Eisenstein series~\cite{Weil}, which is not only manifestly elliptic but also 
modular invariant under the usual~$SL_2(\IZ)$ action on~$\t$! 
The Kronecker-Eisenstein series can be thought of as a Fourier expansion along the lattice directions, and this 
makes the calculation of the action at the saddle-points very easy as we just have to calculate the zeroth Fourier 
modes along the direction chosen by a given lattice point.

In Section~\ref{sec:Action} we calculate the resulting actions of the~$(0,1)$ and~$(1,0)$ saddles, 
and find that they agree precisely with that of pure~AdS$_5$ and 
the supersymmetric~AdS$_5$ black hole, respectively. This leads us to the above-mentioned identification 
of these saddles as the duals of the respective gravitational configurations. 
We also calculate the action and the entropy of the generic~$(m,n)$ saddle. 
Comparing the various actions at a given value of~$\t$, we calculate the resulting phase 
diagram for the supersymmetric theory. (The bounding curves of the resulting 
regions are the anti-Stokes lines.)
We find that the dominant phase close to~$\t =0$ is the~AdS$_5$ black hole. 
In addition, when~$\t$ approaches the rational point~$-n/m$, the~$(m,n)$ solution becomes dominant.
These are new supersymmetric phases of the theory.

The form of the Kronecker-Eisenstein series allows us to easily calculate the asymptotics of the action at a 
given saddle-point in the Cardy-like limit~$\t \to 0$. 
Applying this procedure to the black hole saddle immediately yields the leading terms in the limit, 
thus clarifying certain technical issues that arose in the calculations in the Cardy-like limit in~\cite{Cabo-Bizet:2019osg}  
by different methods.\footnote{Preliminary indications of modular invariance in related systems were found 
in~\cite{Razamat:2012uv, DiPietro:2014bca, Shaghoulian:2016gol, Beem:2017ooy, Dedushenko:2019yiw}.
The calculations of~\cite{Cabo-Bizet:2018ehj}, which showed that 
a generalized supersymmetric Casimir energy of~$\CN=4$ SYM on~$S^1 \times S^3$ equals the black hole entropy functional, 
also pointed towards a modular symmetry underlying the system. }
Our formalism thus unifies the three strands of analysis mentioned above and confirms the following satisfying 
physical picture of the Cardy-like limit---it is simply the limit when the black hole saddle-point shrinks to zero size, 
i.e.~the configuration of eigenvalues uniformly distributed between~$0$ and~$\t$ all coalesce to the origin.

Finally, we make contact with the results of the Bethe-ansatz approach. 
In that approach, one directly considers the integral of a product of elliptic gamma-functions 
which is a meromorphic function of~$u_i$. 
The uniform distribution of eigenvalues in the~$(m,n)$ configuration are known to be roots of the Bethe-ansatz 
equations~\cite{Benini:2018ywd}. The effective action of the meromorphic integrand was calculated 
in~\cite{Benini:2018ywd} on the~$(1,n)$ configurations, but had not been calculated previously for $m>1$. 
In order to compare our results for the bigger family, we develop a new method to 
calculate the meromorphic effective action at a generic~$(m,n)$ configuration,
by rewriting the elliptic gamma function as an infinite series, convergent along the direction 
of a given lattice point. 
For all~$(m,n)$ configurations, we find that the results of the elliptic approach and the Bethe-ansatz approach agree. 
In order to illustrate this agreement in a slightly different way, we also calculate the difference between the two effective actions 
and directly show that this (non-periodic) difference vanishes on any~$(m,n)$ configuration  (see Section~\ref{sec:contour}
and Appendices \ref{sec:ExpRep} and~\ref{App:AnalyticAction}).

The plan of the paper is as follows. 
In Section~\ref{sec:Saddles} we present a simple lemma which allows us to find complex saddle-point configurations for 
a class of matrix models defined purely by a two-particle potential. We also briefly review the index for~$\CN=4$ SYM 
on~$S^1 \times S^3$
and its integral representation in terms of elliptic gamma functions~$\Ge$.
In Section~\ref{sec:Extension} we show how to extend the action of the matrix model from the unit circle to a torus using the 
elliptic dilogarithm. 
In Section~\ref{sec:Action} we calculate the action and the entropy of a generic saddle and discuss the phase structure of 
our supersymmetric matrix model. We also make some comments about the Cardy-like limit. 
In Section~\ref{sec:Discussion} we make some brief comments about interesting 
aspects of this problem that we do not discuss in this paper.
In Appendices~\ref{App:theta}--\ref{sec:BA} we collect the definitions and useful properties of 
the various functions that enter in our analysis, discuss the analytic continuation of the integrand 
of the index and its relation with our elliptic extension, and use this discussion to recover our previous 
results and compare them with the Bethe-ansatz approach.

\vspace{0.1cm}

\paragraph{Notation} 
We will use the notation~$\e(x)\coloneqq\rme^{2 \pi \i x}$ throughout the paper:
We will use~$\t \in \mathbb{H}$ as the modular parameter and~$z, u \in \IC$ as elliptic variables. 
We set~$q=\e(\t)$, $\z=\e(z)$. We decompose the modular variable into its real and imaginary parts 
as~$\t=\t_1+\i \t_2$, and the elliptic variables into their projections on to the basis~$(1,\t)$, 
i.e.~$z_2 = \frac{\Im(z)}{\imt}$, $z_1 = \Re(z) - z_2 \, \ret$ so that~$z = z_1 + \t z_2$.
$\mathbb{N}$ denotes the set of natural numbers and~$\mathbb{N}_0 = \mathbb{N} \cup \{ 0\} $.

\section{Complex saddle-points of~$\CN=4$ SYM \label{sec:Saddles}}

In this section we present a simple lemma which allows us to find complex saddle-point configurations for 
matrix integrals of the type~\eqref{Uact}. Upon rewriting these integrals in terms of eigenvalues, the class of models 
that we consider have no single-particle interactions, and only two-particle interactions which depend only 
on the distance between the two particles. With these assumptions the basic lemma states that if the potential 
is periodic in any (complex) direction in field space then the uniform distribution between the origin and the 
point of periodicity is a saddle-point. 
We then review some properties of the~$\CN=4$ superconformal index and 
set up the particular problem of interest to us.

\subsection{Saddle-point configurations for periodic potentials \label{VTSaddles}}

As is well-known, the matrix integral $Z$ in~\eqref{Uact} reduces to an integral over the eigenvalues of~$U$ 
which we denote by $\{ \e(u_i) \}_{i=1,\dots,N}$.
The product of two traces in~\eqref{Uact} leads to an overall~$N^2$ in front of the action, which allows us to 
use large-$N$ methods. In the large-$N$ limit one promotes~$\{u_i\}$ to a continuous variable~$u(x)$, $x \in [0,1]$, 
and replaces the sum over~$i$ by an integral over~$x$. One can further replace each integral over~$x$ in the 
action by an integral over~$u$ with a factor of the eigenvalue density~$\rho(u) = \frac{dx}{du}$ which obeys 
the normalization condition~$\int du \, \rho(u) = 1$. In this limit the integral~\eqref{Uact} can be written in
terms of an effective action~$S(u)$ as
\be \label{ZSeff}
Z \= \int \, [Du] \, \exp \bigl( - S(u) \bigr)  \,, \qquad S \= N^2 \int \, du \, dv \, \rho(u) \, \rho(v) \, V(u-v) \,,
\ee
where the pairwise potential~$V$ is determined by the function~$f(\t)$. 
The paths~$u(x)$ run over the real interval~$(-\half,\half]$ in the unitary matrix model, but 
here we are interested in complex saddle-points of this model. The main observation is that if the pairwise 
potential~$V(z)$ is periodic with (complex) period~$T$, then the uniform distribution~$u(x) = xT$, 
$x \in [0,1]$, extremizes the effective action. To see this we first note that the odd part of the 
potential~$\half(V(u) - V(-u))$ drops out of the integral in~\eqref{ZSeff}), and so~$V(u)$ can be taken 
to be an even function. Now, the variational equations arising from the integral~\eqref{ZSeff} are 
\be \label{SPEqncont}
\int \, dv \, \p_z V(u-v) \, \rho(v) \= 0 \,, \qquad \int \, dv \, \p_{\overline z}V(u-v) \, \rho(v) \= 0\,.
\ee
The partial derivatives here are odd, periodic functions with period~$T$. On the configuration~$u(x) = xT$, 
the left-hand side of this equation equals~$\frac{1}{T} \, \int_0^{T} \, dv \, V'(u-v)$
where the integral is understood to be along a straight line in the complex plane along~$0$ to~$T$. 
Using the periodicity and oddness properties, respectively, we now have (with $'$ denoting 
either the holomorphic or anti-holomorphic derivative),
\be  
\frac{1}{T} \, \int_0^{T} \, dv \, V'(u-v) \= -\frac{1}{T} \, \int_{-T/2}^{T/2} \, dw \, V'(w) \=0 \,.
\ee

This same argument, with minor changes, can be repeated in the discrete variables.  
We present this as a lemma which we will use in the following development. 
Here we set $\underline{u} \equiv \{u_i\}_{i=1,\dots, N}$, and use the 
measure~$[D\underline{u}] = \frac{1}{N!}\,  \prod_{i=1}^{N} \dd u_i$. 
We will work with the~$SU(N)$ gauge theory throughout the paper for which we need 
to include the tracelessness condition~$\sum_{i=1}^N u_i=0$. 
Consider the~$N$-dimensional integral defined by a pairwise potential between the eigenvalues that only 
depends on the difference~$u_{ij} = u_i-u_j$, i.e.,
\be \label{defV}
Z \= \int [D\underline{u}] \exp \bigl(- S (\underline{u}) \bigr) \,, \qquad S(\underline{u}) \= \sum_{i,j=1}^N \, V(u_{ij}) \,.
\ee
We take the potential~$V(z)$ to be any smooth complex-valued function, not necessarily meromorphic. 
A question about contour-deformations immediately arises.\footnote{We thank the referee for emphasizing this point.} 
When the integrand is holomorphic we can freely change the contour so as to pass through complex 
saddle-points. Since we are not demanding holomorphicity, we need an additional argument to be able to 
deform the contour to pass through the saddle-point. We are able to do this in the context of the specific
functions discussed later in this paper, as we elaborate in~Section~\ref{sec:contour}. 
Now we consider the variational equations 
\be \label{varS}
\p_{u_i} S (\underline{u}) \= \p_{\overline{u}_i} S(\underline{u}) \= 0\,, \qquad i = 1, \dots , N \,.
\ee

\paragraph{Lemma}\label{sec:Lemma}
In the above set up, suppose the potential~$V$ is periodic with (complex) period~$T$. Then 
\begin{enumerate}
\item[(i)] The following configuration solves the variational equations~\eqref{varS},
\be \label{uniform}
u_i \= \frac{i}{N} \, T + u_0 \,,
\ee
where~$u_0$ is fixed in such a way the traceless constraint is obeyed. 
\item[(ii)]
The action~$S(\underline{u})$ at the saddle~\eqref{uniform} is given by~$N^2$ times the average value of the 
potential on the straight line joining~$0$ and~$T$ in the large-$N$ limit.
\end{enumerate}

\paragraph{Remark} The distribution~\eqref{uniform} becomes the uniform distribution between~$0$ and~$T$ 
in the large-$N$ limit. 
Since the potential~$V$ is periodic with period~$T$, it admits a Fourier expansion along that direction. 
The second part of the lemma is equivalent to saying that the saddle-point value of the action is~$N^2$ times the 
zeroth Fourier coefficient in the direction~$T$.

To prove the first part of the lemma, we assume, without loss of generality as before, that~$V$ is an even function. 
The saddle-point equations~\eqref{varS} can be written as
\be \label{Vpr}
\sum_{j \neq i} \p_z V (u_{ij}) \= 0  \,, \qquad \sum_{j \neq i} \p_{\overline{z}} \, V(u_{ij}) \= 0  \,.
\ee
On the configuration~\eqref{uniform} we have $u_{ij} =(i-j)T/N$. 
For every point~$j$ in this sum the point~$j'=2i-j \, (\text{mod} N)$ is equidistant from~$i$ and lives on the other side of~$i$, 
that is~$i-j = -(i-j') \, (\text{mod} N)$. 
Since the functions~$\p_z V $, $ \p_{\overline{z}} V $ are odd and periodic with period~$T$, 
these two points cancel against each other and the sums in~\eqref{Vpr} vanish.  
(For even~$N$ the antipodal point~$i+\half N \, (\text{mod} N)$ is its own partner, but the argument 
still applies as~$\p_{z} V(T/2)=\p_{\overline{z}} \, V(T/2)=0$.)

The second part of the lemma follows from a simple calculation of the action 
on the distribution~\eqref{uniform}, which we call the \emph{effective action} of the saddle:
\be \label{SeffT}
\Seff(T) \=
\sum_{i,j=1}^N \, V\Bigl(\frac{i-j}{N}T \Bigr) 
\=  \sum_{i,j=1}^N \, V\Bigl(\frac{i}{N}T \Bigr)
\= N \sum_{i=1}^N \, V\Bigl(\frac{i}{N}T \Bigr)
 \= N^2 \int_0^1 dx \, V(xT) \,.
\ee
In obtaining the second equality we have used the periodicity of the potential,
and in the last equality we have used the large-$N$ limit.
\hfill $\square$

\subsection{The index of~$\CN=4$ SYM}

The superconformal index of~$\CN=4$ SYM theory was calculated in~\cite{Romelsberger:2005eg,Kinney:2005ej}. 
In~\cite{Cabo-Bizet:2018ehj} this index was revisited as a path integral corresponding to the partition function of 
supersymmetric field theories on~$S^3\times S^1$ with complexified angular chemical potentials~$\sigma,\tau$ 
for rotation in~$S^3$, and holonomy for the background R-symmetry gauge field~$A$ given 
by~$\int_{S^1} A = \ii \pi \varphi$, with $\varphi=\ii(\sigma+\tau -n_0)$. The introduction of the integer~$n_0$ 
in~\cite{Cabo-Bizet:2018ehj} corresponds to a shift in the background value of the R-symmetry gauge 
field.\footnote{Equivalently the fermions can be thought of as antiperiodic with the unshifted gauge field values.} 
It was shown in~\cite{Cabo-Bizet:2018ehj} that  
\be\label{RelInd0}
\CI(\s, \t;n_0) \= \CI(\s- n_0,\t;0) \,,
\ee
where 
the function~$\CI(\s,\t;0)$ is the familiar  Hamiltonian definition of the superconformal index~\cite{Romelsberger:2005eg,Kinney:2005ej}
\be \label{InI0rel}
\CI(\s, \t;0) 
\=  \, {\rm Tr}_{\CH_\text{phys}}\,  (-1)^F \rme^{-\beta \{\mathcal{Q},\bar{\mathcal{Q}}\}  
+2 \pi \i \s (J_1+\frac{1}{2}Q)+2 \pi \i \t (J_2+\frac{1}{2}Q)} \,.
\ee
The above two equations make two things immediately clear. On one hand they show 
that the index~$\CI(\t, \s;n_0)$ is protected against small deformations of the parameters of the theory
exactly as the original index.
On the other hand expanding the index as
\be \label{inddeg}
\CI(\s, \t;n_0) \= \sum_{n_1,n_2} d(n_1,n_2; n_0) \, \rme^{2 \pi \i (\s n_1 + \t n_2)} \,
\ee 
in terms of the degeneracies~$d(n_1,n_2)$, we see from~\eqref{InI0rel} that a constant shift of the chemical potentials only 
changes the phase of the indexed degeneracies, so that 
\be
|d(n_1,n_2; n_0)| \= |d(n_1,n_2; 0)| \,.
\ee

Thus we have that the index~$\CI$, i.e.~the canonical partition function, depends on~$n_0$ whose correct value is dictated 
by the holographic dictionary. For the degeneracies~$d$, i.e.~the microcanonical observables, the value of~$n_0$  
does not change the absolute value of the degeneracies. It does, however, play a role in the 
calculation of the saddle-point value of the degeneracies as a function of large charges, since a change of~$n_0$ 
corresponds to a change in the~$\t$-contour of integration used to define the saddle-point value, and for 
the purpose of this calculation one 
should use a value of~$n_0$ that maximizes the answer. 
In our situation, since the~$R$-charges of the~$\CN=4$ SYM theory are quantized in units of~$\frac13$, the 
degeneracies only depend on~$n_0$~(mod~3). We can choose the representatives to be~$n_0=0,\pm 1$. 
As it turns out, the natural contours associated to~$n_0=\pm 1$ 
both have leading entropy at large charges equal to that of the black hole, which dominates the natural contour associated to~$n_0=0$. 
We note that one can map $n_0= +1 \mapsto n_0 = - 1$ by using the transformation $(\Re\, \s, \Re \,\t)\to-(\Re\, \s, \Re\, \t)$.
All this is consistent with the supergravity calculation of the Euclidean black hole~\cite{Cabo-Bizet:2018ehj}.
We will use the value~$n_0=-1$ in the rest of the paper and suppress writing it 
explicitly.

The superconformal index of~$SU(N)$~$\CN=4$ SYM 
can be written in the form~\eqref{Uact} with 
\be
f(\s,\t) \= 1\,-\,\frac{\bigl(1-\e\bigl(\frac13 (\s+\t+1) \bigr)\bigr)^3}{\bigl(1-\e(\t)\bigr)\bigl(1-\e(\s)\bigr)} \,.
\ee 
One can recast the exponential of the infinite sum expression in~\eqref{Uact} 
in terms of an infinite product to obtain the following~$N$-dimensional integral, 
\cite{Dolan:2008qi}, \cite{Cabo-Bizet:2018ehj} 
\bea\label{SYMIndexst} 
\CI(\s,\t)  \=  && \bigl(\e(-\tfrac{\s}{24}-\tfrac{\t}{24})  \, \eta(\s) \, \eta(\t) \bigr)^{N} \times   \\
&& \qquad  \int [D\uu]   \, 
\prod_{i\neq j} \, \Ge \bigl(u_{ij} + \s+\t; \s,\t \bigr) \, \prod_{i,j} \, \Ge \Bigl(u_{ij} + \frac13 (\s +\t -n_0); \s,\t \Bigr)^3 \,, \nn
\eea
where the measure factor is $[D\uu] = \frac{1}{N!}\,  \prod_{i=1}^{N} \dd u_i\, \delta\left(\sum_i u_i\right)$ 
and the integral over each~$u_i$ runs over the real range~$(-\frac12,\frac12 ]$ unless otherwise indicated.
Here the elliptic gamma function\footnote{See \cite{Felder,Spiridonov:2010em,Spiridonov:2012ww} for a development 
of the theory and a discussion of the properties of this function.} 
is defined by the infinite product formula~\cite{Felder},
\be\label{GammaeDef}
\Ge(z;\s,\t) \=  \prod_{j,k=0}^{\infty}
\frac{1- \e \bigl(-z+\sigma (j+1)+ \tau(k+1) \bigr)}{1-\e( z +\sigma j + \tau k)} \,.
\ee
The interpretation of the various pieces in~\eqref{SYMIndexst} in its derivation using supersymmetric localization 
of the partition function~\cite{Cabo-Bizet:2018ehj} are as follows. 
The gauge holonomies~$u_i$ label the localization manifold, and the classical action vanishes at any point on this manifold.
The infinite products in the integrand comes from the one-loop determinant of the localization action. 
The first elliptic gamma function arises from the vector multiplet which has R-charge~$2$, and the other three 
elliptic gamma functions arise from the chiral multiplets which carry R-charge~$\frac23$. 
The product over~$i,j$ in the integrand of~\eqref{SYMIndex} reflects the fact that all the fields are in the adjoint 
representation of the gauge group. 
For the vector multiplets one needs to remove the zero modes which arise for the Cartan elements $i=j$ for the vector field, 
and the non-zero modes of the Cartan elements form the pre-factor in front of the integral~\eqref{SYMIndexst}.

In the rest of the paper we analyze the model with~$\s=\t$ and~$n_0=-1$ so that the SYM index is
\be\label{SYMIndex} 
\CI(\t) \=  q^{-\frac{N}{12}} \, \eta(\t)^{2N} \; 
 \int [D\uu]   \, 
\prod_{i\neq j} \, \Ge \bigl(u_{ij} + 2\t; \t,\t \bigr) \, \prod_{i,j} \, \Ge \Bigl(u_{ij} + \frac13 (2\t +1); \t,\t \Bigr)^3 \,.
\ee
The product expression~\eqref{GammaeDef} makes it clear that~$\Ge(z;\t,\s)$ is separately periodic 
under the translations~$z \to z+1$, $\s \to \s+1$, and~$\t \to \t+1$. Thus we see from~\eqref{SYMIndex} that 
the~$\CN=4$ SYM index~$\CI(\t)$ manifestly has the symmetry~$\t \mapsto \t+3$. 
It is also clear from~\eqref{SYMIndex} that by shifting~$\t$ by~1 and~2, respectively, we reach the other two independent 
values of~$n_0$, thus explicitly showing the relation of~$n_0$ with the contour of integration.

\vspace{0.1cm}

Our goal now is to find complex saddle-points of the integral~\eqref{SYMIndex}. 
In order to do this we need 
to extend the integrand to the complex~$u$-plane. One natural guess would be to use analytic continuation 
of the elliptic gamma function~\cite{Felder}. 
This function is periodic with period~$1$, and is almost---but not quite---periodic\footnote{As we review below, the quasi-periodicity of 
the~$\Ge$-function is similar to that of the Jacobi~$\vth$-functions. Relatedly, it admits an interpretation as ``higher degree 
automorphic forms"~\cite{Felder}, but we will not use this cohomological interpretation here.} with period~$\tau$: it obeys 
\be \label{Geperiod}
\frac{\Ge(z+\t;\t,\t)}{\Ge(z;\t,\t)} \= \th_0(z;\t) \,,
\ee
where the function~$\th_0$ is related to the odd Jacobi theta function~$\vth_1$ as
\be \label{defth0}
\th_0(z;\t) \defeq -\z^\half \, q^{\frac{1}{24}}  \, \frac{\vth_1(\t,z)}{\eta(\t)} \,.
\ee
In order to use the method of the previous subsection we need a strictly periodic function, and so we need to 
find a periodic extension.

Our idea is simple in principle---we deform the integrand away from the real axis such that
the resulting function is periodic with period~$\t$. 
Of course the result will not be meromorphic in~$u$ and the best we can hope for a real-analytic function of~$\Re(u)$, $\Im(u)$. 
(We still demand that the action is meromorphic in~$\t$.)
The technical goal is thus to replace the integrand of~\eqref{SYMIndex} by a function that 
\begin{enumerate}
\item[(a)] is periodic under~$u \mapsto u+ m\t +n$, $m,n \in \IZ$ (an \emph{elliptic function}),
\item[(b)] is real-analytic  (apart from perhaps a finite number of points on the torus), and 
\item[(c)] reduces to the product of elliptic gamma functions in~\eqref{SYMIndex} when~$u \in \IR$. 
\end{enumerate}
The properties~$(a)$--$(c)$ above do not uniquely fix such a function. We will make a particular
choice which has three nice properties.
Firstly, our extended action turns out to be closely related to the elliptic dilogarithm function 
which has, quite remarkably, modular properties under the~$SL_2(\IZ)$ action on~$\tau$. 
As mentioned in the introduction, these properties are made manifest by a formula of 
Bloch~\cite{Bloch}, which allows us to calculate the saddle-point action in one simple step.
Secondly, the action of the black hole configuration agrees precisely with the supergravity action of the 
black hole. 
The third property looks to be an aesthetic or mathematical one at the moment---the 
elliptic dilogarithm obeys interesting equations, and its values at special points 
are interesting from a number-theoretic point of view.

We present the details of the extension of the action to the torus in Section~\ref{sec:Extension}, 
and continue for now by assuming that there is an extension obeying Properties (a)--(c). 
The periodicity of the extension under translations of the lattice~$\Lambda_\t = \IZ\t+\IZ$
can be restated as saying that the function is well-defined on the torus~$\IC/\Lambda_\t$. 
Now, our lemma about complex saddle-points in the previous subsection means that 
there is one saddle-point for every lattice point in~$\IZ\t + \IZ$. If we think of the uniform 
distribution of eigenvalues~$u(x)$ as a string, then a solution corresponding to the point~$m\t+n$ can be 
thought of as a closed string that winds~$(m,n)$ times along the~$(A,B)$ cycles of the torus.
In terms of the original matrix model, the eigenvalues have moved off the original circle of 
unit radius into the complex plane, as shown in Figure~\ref{SaddleUp}.

\vspace{0.4cm}
\noindent {\bf Comment about other discrete saddles and gcd condition}
\vspace{0.1cm}

\noindent In the context of our solution-generating lemma in Section~\ref{VTSaddles}, 
suppose~$V(x)$ has \emph{minimal} period~$T$. 
Then clearly the (discrete) uniform distribution spread between~$0$ and~$pT$, for any non-zero~$p \in \IZ$ 
also solves the saddle-point equations by the same arguments as those given in the lemma.  
These solutions all give rise to the same distribution of points on the torus if and only if~$\text{gcd}(p,N)=1$.
When~$(p,N)=(dp',dK)$ with~$d>1$ and~$\text{gcd}(p',K)=1$, then we have a new solution which is a uniform
distribution of~$K$ ``stacks" of eigenvalues, each stack containing $d=\frac{N}{K}$ eigenvalues. 

Applying these considerations to our problem on the torus, we find a family of saddles that is 
classified by the following three-label notation: $(K| \,m,n)$ with~$\text{gcd}(m,n)=1$ and~$K$ 
is a divisor of~$N$. The~$m\t+n$ saddles discussed above correspond to~$K=N$, and we will continue to use that 
notation in that case. 
If $N$ is prime, saddles are essentially isomorphic to lattice points~$(m,n)$ 
with~$\text{gcd}(m,n)=1$.\footnote{The relation of these saddles to the Bethe-ansatz configurations will be 
discussed in Appendix~\ref{sec:BA}.} 
The one solution which needs to be discussed separately is~$(1|\,m,n)\equiv (0,0)$.
This saddle corresponds to the distribution where all eigenvalues are placed at the origin $u=0$.
As we discuss at the end of Section~\ref{App:AnalyticAction}, this distribution of eigenvalues is 
highly suppressed with respect to the dominant one at finite values of~$\tau$. However, one has to 
be careful to take the limit~$\t \to 0$ because other effects start to appear. 
From now on, we focus on $(N|\, m,n)\equiv(m,n)$ saddles.

\section{Extension of the action to the torus \label{sec:Extension}}

In this section we show how to extend the matrix model to the torus using elliptic functions.
We collect the definitions and properties of some standard functions that appear in our 
presentation in Appendix~\ref{App:theta}.

\subsection{Elliptic functions, Jacobi products, and the elliptic dilogarithm}

In order to define our deformation, and as a warm-up, we begin with a brief discussion of the 
quasi-elliptic function~$\th_0(z)=\th_0(z;\t)$ defined in Equation~\eqref{defth0}. 
It has the following periodicity properties,
\be\label{periodth0}
\th_0(z) \= \th_0(z+1) \= -\e(z) \, \th_0(z+\t) \,,
\ee
and a product representation which follows from the Jacobi product formula,
\be
\th_0(z;\t) \=  (1-\z) \prod_{n=1}^\infty (1-q^n \z) \, (1-q^n \z^{-1}) \,.
\ee
The function~$\th_0$ can be modified slightly in order to obtain a function that is elliptic. 
Define the related function (recall~$z=z_1+\t z_2)$
\be \label{defP}
P(z) \= P(z;\t)  \=  \e(\a_P(z)) \, q^{\half B_2(z_2)} \, \th_0(z;\t) \,,
\ee
where~$B_2$ is the second Bernoulli polynomial~$B_2(x) = x^2 - x + \frac16$.
Here the function~$\a_P$ is chosen to obey
\be
\a_P(z) \in \IR \,, \qquad  \a_P(z) \= 0  \; \, \text{when~$z_2=0$} \,,
\ee
and will be specified below.

It is easy to check that the function~$q^{-\frac{1}{12}} P(z)$ agrees with~$\th_0(z)$ on the real axis, i.e.,
\be
q^{-\frac{1}{12}} P(z;\t) \= \th_0(z;\t)  \quad \text{when~$z_2=0$} \,. 
\ee
The elliptic transformations~$z \to z+1$, $z \to z+\t$ are simply shifts of the real variables~$z_1 \to z_1 +1$, $z_2 \to z_2+1$, 
respectively.  Under~$z\to z+1$, the absolute value of each factor 
in~\eqref{defP}, and therefore~$|P|$, is invariant.
Under~$z\to z+\t$, the shift of the theta function 
(Equation~\eqref{periodth0}) is cancelled by the corresponding shift of~$B_2$ (Equation~\eqref{BernPeriod}).  
Therefore~$|P|$ is a doubly periodic function. 
The phase~$\a_P$ is chosen so that~$P$ itself is doubly periodic. 
This criterion fixes~$\a_P$ in terms of~$\th_0$ up to an additive ambiguity of a periodic 
real function. This ambiguity can be fixed by making a particular choice for~$\a_P$, 
for example, by defining it\footnote{This is equivalent to defining~$P$ to be~$q^{\half B_2(\{z_2\})} \, \th_0(z_1+\{z_2\}\t;\t)$.} 
 be to zero in the first fundamental domain~$0 < z_2 <1$, and extending 
it by periodicity of~$P$. Defined in this manner, the phase~$\a_P(z)$ is locally constant in~$z_2$, 
but exhibits a discontinuity when~$z_2$ hits an integer. 
We can smoothen this discontinuity over a small range~$\epsilon$ (for instance we can locally replace the Heaviside step
function by~$\frac12(1+\text{erf}(x))$), which is taken to zero at the end of the calculations.  
As we explain below, the calculations in the following sections are insensitive to the details of the smoothening.

A non-trivial fact is that~$|P|$ is also invariant under the modular transformations~$\t \to \frac{a\t+b}{c\t+d}$, 
$z \to \frac{z}{c\t+d}$. In other words, it is invariant under the full Jacobi group. These properties are immediately 
demonstrated by the second Kronecker limit formula~\cite{Weil},
\be \label{KroneckerLimit}
- \log|P(z;\t)| \= \underset{s\to 1}{\text{lim}} \; 
\frac{\imt^s}{2 \pi} \!  {\underset{m,n \in \IZ \atop (m,n) \neq (0,0)}{\sum}} \; \frac{\e(n z_2 - m z_1)}{|m\t+n|^{2s}} \,.
\ee
The right-hand side of~\eqref{KroneckerLimit} is a real-analytic Kronecker-Eisenstein series which is manifestly 
invariant under the Jacobi group.

\vspace{0.2cm}

These properties of~$\th_0$ and~$P$ are classical facts known for over a century (see~\cite{Weil} for a beautiful exposition). 
Interestingly there is a similar story for the elliptic gamma function, which is much more recent. 
Following~\cite{DukeImamoglu, Pasol:2017pob}, we construct the function, for~$z\in \IC$,
\be \label{defQtl}
Q(z) \= Q(z;\t) \= \e(\a_Q(z)) \, q^{\frac13 B_3(z_2) - \frac12 z_2 B_2(z_2)} \, \frac{P(z;\t)^{z_2} }{\Ge(z+\t;\t,\t)} \,,
\ee
where~$B_3$ is the third
Bernoulli polynomial~$B_3(x) = x^3 - \frac32 x^2 + \frac12 x$. 
Here we have introduced the phase function~$\a_Q$ obeying the properties 
\be
\a_Q(z) \in \IR \,, \qquad \a_Q(z) \= 0  \quad \text{when~$z_2=0$} \,,
\ee
which we discuss more below.
It is easy to check that $Q$ agrees with the~$\Ge$-function on the real axis, i.e.,
\be \label{QtlGamrel}
Q(z;\t)  \= \Ge(z+\t;\t,\t)^{-1} \quad \text{when~$z_2=0$} \,. 
\ee 
The quasi-periodicity relation of the elliptic Gamma-function 
\be \label{Geperiod1}
\Ge(z;\t,\t) \= \Ge(z+1;\t,\t) \= \th_0(z;\t)^{-1} \, \Ge(z+\t;\t,\t) \,
\ee
implies that the function~$|Q|$ is periodic under translations of the lattice~$\IZ\t+\IZ$. 

Remarkably, the function~$|Q|$ also satisfies a relation similar to, but more complicated than, 
the Kronecker limit formula~\eqref{KroneckerLimit} for~$|P|$. 
To see the relation, one first notices that the function~$|Q|$ is very closely related to the Bloch-Wigner 
elliptic dilogarithm. 
The Bloch-Wigner dilogarithm function is the single-valued non-holomorphic function 
(we follow the treatment of~\cite{ZagierOnBloch}),
\be
D(x)  \defeq \Im(\text{Li}_2(x)) + \arg (1-x) \, \log|x| \,.
\ee
Its elliptic average, which is manifestly invariant under lattice translations, is defined as
\be
D(q,x) \defeq \sum_{\ell \in \IZ} D(q^\ell x) \,.
\ee
The function~$D(q,x)$ turns out to have a natural imaginary partner 
\be
J(q,x)\defeq \sum_{\ell=0}^\infty J(q^\ell x) - 
\sum_{\ell=1}^\infty J(q^\ell x^{-1}) + \frac12 \log^2|q| \, B_3\Bigl(\frac{\log|x|}{\log|q|} \Bigr)\,,
\ee
which is itself an elliptic average\footnote{The definition of~$J(q,x)$ is a little more complicated than that 
of~$D(q,x)$ because the function~$J(x)$ does not decay as rapidly as~$D(x)$ and consequently one needs to  
regulate the infinite sum.} of the function
\be
J(x) \defeq \log|x|\, \log|1-x| \,,
\ee
so that the most elegant formulas are written in terms of the combination
\be\label{FDJrel}
F(z;\t)\defeq \frac{1}{2\pi} \bigl( D(q,\z) + \i \, J(q,\z) \bigr)\,.
\ee

The relation of~$|Q|$ to~$F$ is as follows~\cite{DukeImamoglu}, \cite{Pasol:2017pob} 
\be \label{QtlFrel}
\log\,\Bigl|Q\Bigl(\frac{z}{\t}; -\frac{1}{\t} \Bigr) \Bigr| - \t \log|Q(z;\t)| \= F(z;\t) \,.
\ee
The relation to Kronecker-Eisenstein series follows from the following formula~\cite{Bloch,ZagierOnBloch},
\be \label{Bloch}
F(z;\t) \= \frac{\imt^2}{2\pi^2} \! 
{\underset{m,n \in \IZ \atop (m,n) \neq (0,0)}{\sum}} \; \frac{\e(n z_2 - m z_1)}{(m\t+n) (m\overline{\t} +n)^2} \,.
\ee
The function~$F$ is manifestly invariant under shifts of~$u$ by lattice points. The form of~$F(\t,z)$ as a lattice sum 
also makes it clear that under modular transformations
\be
\t \mapsto \frac{a\t+b}{c\t+d} \,, \qquad z \mapsto \frac{z}{c\t+d} \,, \qquad \begin{pmatrix} a & b \\ c & d \end{pmatrix} \in SL_2(\IZ) \,,
\ee 
it transforms as a Jacobi form of weight~$(0,1)$. 
The real and imaginary parts of Equation~\eqref{QtlFrel} give us formulas for~$\log|Q|$ as well as its 
modular transform in terms of the real and imaginary parts of~$F$,
\be \label{QFrel}
-\t_2  \log|Q(z;\t)| \= \frac{1}{2\pi} J(q;\z) \,, \qquad 
\log\Bigl|Q\Bigl(\frac{z}{\t}; -\frac{1}{\t} \Bigr) \Bigr|  \=  \frac{1}{2\pi}  \Bigl( D(q;\z) + \frac{\t_1}{\t_2} J(q;\z) \Bigr)\,,
\ee
for which we can write Fourier expansions using the expansion~\eqref{Bloch}.

The above discussion completely defines the function~$\log|Q|=\Re \log Q$, but 
the phase function~$\a_Q$, and therefore~$\Im \log Q$, still needs to be defined 
properly.\footnote{The function~$Q$ should be related to a version of the elliptic dilogarithm holomorphic in~$\t$
studied in~\cite{Levin,BrownLevin}, we leave a detailed study of this to the future.}
We want to choose the phase~$\a_Q$ such that the whole function~$Q$ is doubly periodic.
As in the case of~$P$, this criterion determines~$\a_Q$ in terms of the function~$\Ge$ up to an additive 
ambiguity of a periodic real function. Unlike in the case of~$P$, we do not fix this ambiguity 
completely yet. Instead, we write~$\a_Q (z) \= \a^0_Q (z) + \PsiQ (z)$ where~$\a^0_Q$ 
is fixed by a particular choice, e.g.~by demanding that it vanishes in the first fundamental region
and extending it by demanding periodicity of~$Q$.\footnote{This choice leads to the 
value~$ \a^0_Q(z)=-\frac14 (1+2 \{z_1\}) \, \lfloor z_2 \rfloor \, (1+\lfloor z_2 \rfloor)$. \label{a0value}}.
We have explicitly parameterized the ambiguity by a doubly-periodic real function~$\PsiQ(z)$ which 
vanishes at~$z_2=0$, which we fix in Section~\ref{sec:contour}.

Alternatively,  we can define~$\Im \log Q$ 
to be the harmonic dual of~$\Re \log Q$ with respect to~$\t$.
In other words we construct~$\log Q$ by demanding that it is holomorphic in~$\t$ 
and has the same real part as~$\Re \log Q$. 
Using the relations~\eqref{QFrel}, \eqref{FDJrel}, and the Fourier expansion~\eqref{Bloch},
we can write a Fourier expansion for~$\log Q$ as follows,
\bea \label{QKron}
\log Q(z;\t) & \= & - \frac{1}{4\pi^2} \! 
{\underset{m,n \in \IZ \atop m \neq 0}{\sum}} \; \frac{\e(n z_2 - m z_1)}{m(m\t+n)^2} + 
\frac{\t }{2\pi^2} \, 
{\underset{n \in \IZ \atop n \neq 0}{\sum}} \; \frac{\e(n z_2)}{n^3} + 
\frac{\pi\i}{4}  {\underset{m,n \in \IZ \atop m \neq 0}{\sum}} K(m,n) \, \e(n z_2 - m z_1) \,, \nn \\
& \= & - \frac{1}{4\pi^2} \! 
{\underset{m,n \in \IZ \atop m \neq 0}{\sum}} \; \frac{\e(n z_2 - m z_1)}{m(m\t+n)^2} 
+\frac{2 \pi \i \t}{3}  B_3\bigl( \{z_2\} \bigr) + \pi\i \,\Psi(z) \,.
\eea
In defining the harmonic dual there is an additive ambiguity in~$\Im \log Q$ of a~$\t$-independent doubly periodic real function, 
which we have denoted by~$\Psi(z)$, and whose Fourier coefficients we have denoted by~$\frac{1}{4}K(m,n)$. 
By comparing the expression~\eqref{QKron} to the expansion~\eqref{ExpRepG} of~$\Ge$, 
and using the constraint~\eqref{QtlGamrel}, we obtain that~$\Psi(z)|_{z_2=0}=0$. 
The ambiguity~$\Psi$ in the Fourier expansion is in one-to-one correspondence with the ambiguity~$\PsiQ$ 
in the definition~\eqref{defQtl} which we discussed in the previous paragraph. 
The difference~$\Psi(z)-\PsiQ(z)$ is an unambiguous doubly periodic function which we do not calculate here.

The function~$\Im \log Q$ as defined above is not smooth, in particular, it is continuous but 
its first derivative is discontinuous at integer values of~$z_2$. We can smoothen this discontinuity, 
as we did for~$\log P$, over a small range~$\epsilon$. In the following sections we apply the lemma in 
Section~\ref{VTSaddles} to potentials which are linear combinations of~$\log P$ and~$\log Q$. 
The first part of the lemma states that periodic configurations of eigenvalues 
solve the saddle point equations for smooth periodic functions, 
this part clearly goes through in a straightforward manner for the smoothened functions. 
The second part of the lemma states that the action of the periodic configurations is the 
zeroth Fourier coefficient of the potential~$V$. Here, we note that the region of smoothening, and therefore 
the contribution of the integrals of~$\log P$ and~$\log Q$ from the region near the discontinuities, vanishes 
as~$\epsilon \to 0$ (note that there is no derivative acting on~$V$ in the calculation of the action).
In other words, the zeroth Fourier coefficients of the smoothened functions~$\log P$ and~$\log Q$ can  
still be calculated from the Fourier expansions~\eqref{KroneckerLimit}, \eqref{Bloch}, and~\eqref{QKron}.

\subsection{Elliptic extension of the action \label{sec:ellext}}

Having set up the basic formalism, we can now connect to the action of~$\CN=4$ SYM. 
In order to do so, we define the related function
\be\label{Qab}
Q_{a,b}(z) \= Q_{a,b}(z;\t) \defeq  q^{\frac{a^3}{6} - \frac{a}{12} } \, \frac{Q(z+a\t+b)}{P(z+a\t+b)^a} \,, \qquad a,b\in \IR \,.
\ee
This function is doubly periodic since all its building blocks are, and it obeys the property  
\be \label{QGamrel}
Q_{a,b}(z) \=  \Ge(z+(a+1)\t+b;\t,\t)^{-1}  \quad \text{when~$z_2=0$} \,. 
\ee 

The index~\eqref{SYMIndex} can now be written as
\be\label{SYMIndex1again}
\CI(\t) \=  \bigl(q^{-\frac{1}{24}} \eta(\t) \bigr)^{2N}
 \int [D\underline{u}]   \, \prod_{i\neq j} \, Q_{1,0} (u_{ij})^{-1} \, \prod_{i, j} Q_{-\frac13,\frac13} (u_{ij})^{-3} \,.
\ee
We can write this in terms of an effective action as
\be
\CI(\t) \=  \int [D\underline{u}]   \, \exp(-S(\underline{u})) \,,
\ee
with
\be \label{EffActQ}
S(\underline{u})\=  -2N \log\bigl(q^{-1/24} \, \eta(\t) \bigr) + \sum_{i\neq j} \,\log  Q_{1,0} (u_{ij}) 
+ 3 \sum_{i,j}  \log Q_{-\frac13,\frac13} (u_{ij}) \,   \,. 
\ee
In terms of the functions~$Q$ and~$P$ we have 
\be \label{SPQ}
\begin{split}
S(\underline{u})& \=  -2N \log\bigl(q^{-1/24} \, \eta(\t) \bigr)  - \frac{1}{6}  N \pi \i \t + \frac{8}{27} \pi \i \t  N^2  \\
& \qquad + \sum_{i \neq j}  \log Q \bigl(u_{ij} + \t  \bigr) + 3 \sum_{i,j}  \log Q \bigl(u_{ij} - \tfrac13 \t +\tfrac13 \bigr) \\
& \qquad - \sum_{i\neq j}   \log P \bigl(u_{ij} +\t \bigr) + \sum_{i, j}   \log P \bigl(u_{ij} - \tfrac13 \t +\tfrac13 \bigr) \,.
\end{split}
\ee
We have thus reached our goal of extending the action to the complex plane in terms of doubly periodic functions. 
The action is a real-analytic function on the torus except for a finite number of points where it has singularities,
we comment on this in Appendix~\ref{App:zeropole}.

The fact that the expansions~\eqref{Bloch} and~\eqref{QKron} are written as a double Fourier series in~$z_1, z_2$ 
means that we can read off the average value in any desired direction in the~$z$-plane. 
Secondly this expression in terms of a lattice sum makes the modular properties completely manifest.
We will use these properties in the following section to calculate the action at the saddle-points,
and to calculate asymptotic formulas in the Cardy-like limit.

\subsection{Deforming the contour \label{sec:contour}}

We now turn to the evaluation of the integral~\eqref{SYMIndex1again}. 
By construction the integrand is a doubly periodic function with periods~$1$ and~$\t$.
Applying the lemma of Section~\ref{VTSaddles}, we conclude that the uniform distribution of the eigenvalues
between~$0$ and the lattice point~$m\t+n$, $m,n \in \IZ$ solves the variational equations. 
However, in order for this configuration to be a genuine saddle-point of the integral, the contour of 
integration needs to pass through it. The original contour in~\eqref{SYMIndex1again} 
runs over~$u_i \in [0,1]$, $i=1, \dots , N$, while the 
value of~$u_i$ at the saddle~$(m,n)$ is
\be \label{uisadval}
u_i \= \frac{i}{N} (m\t+n) \; \Longleftrightarrow \; u_{ij} \= \frac{i-j}{N} (m\t+n) \,,
\ee
which is not on the original contour.

To our knowledge, two approaches have been followed to deal with this issue in similar problems with meromorphic integrands.
The first approach can be seen in the problem of counting black hole microstates in~$\CN=4$ string 
theory in asymptotically flat space,
wherein the degeneracy of microstates is given by a contour integral of a meromorphic function. 
In this case one deforms the contour so as to pass through the saddle, and keeps track of any 
residues that are picked up in the process~\cite{Sen:2007vb,Dabholkar:2012nd}. 
The other approach is Picard-Lefschetz theory wherein one constructs a basis 
of the homology of the manifold (or relative homology for non-compact spaces),
and then decomposes a given contour in terms of these basis 
elements, see e.g.~\cite{Witten:2010cx},\cite{Aniceto:2018bis}.
Since the integrand in our discussion is not meromorphic, we cannot apply either of these methods 
directly. As we do not know of a rigorous mathematical formalism to approach this problem,  
we only make some observations in this subsection, 
and leave a complete analysis of this important issue to future work. 
In particular, we make two related points, pertaining to the two methods mentioned above.

The first point is that there is a close relation between the doubly periodic integrand~\eqref{SYMIndex1again} 
that is our focus, and the meromorphic integrand of~\eqref{SYMIndex}. 
As we saw in the previous subsection, these two integrands are equal to each other 
when all the~$u_i$ are real because of the identity~\eqref{QGamrel}. 
In fact, as we show below, they are also equal to each other when evaluated on the 
saddle-point configuration~\eqref{uisadval}.
Thus, one way to deal with the contour deformation could be to apply the Picard-Lefschetz theory
to the meromorphic integrand, and use its equality to the doubly-periodic integrand on the saddles 
at the end of the process.\footnote{Another possibility, suggested by a referee whom we thank, is to use real-harmonic 
functions. We have not been able to make this interesting suggestion more precise.}

In order to show this equality we first note that, as a direct consequence of the 
definitions of the various functions, we can write, for all~$z$,  
\be \label{QabPgamrel}
Q_{a,b}(z) \=  e^{2\pi \i(\PsiQ(z+a\t+b)+\a^0_Q(z+a\t+b))} \, q^{-A_{a}(z_2)} \,
\frac{P(z+(a+1)\t+b;\t)^{z_2}}{\Ge(z+(a+1)\t+b;\t,\t)}  \,,
\ee
where the cubic polynomial~$A_a$ is defined as 
\be
A_{a}(x) \=  \tfrac16  \, x^3  \, +\, \tfrac12 a\, x^2\,+\, \bigl(\tfrac{1}{2} a^2 - \tfrac{1}{12} \bigr)\, x\,.
\ee
Now, the integrands of~\eqref{SYMIndex1again} and~\eqref{SYMIndex} involve a product  
over~$a=1,-\frac13,-\frac13,-\frac13$ (from the product over vector and chiral multiplets), and 
over all pairs~$(i,j)$. The fact that we sum over the pairs~$(i,j)$ and~$(j,i)$ for a given~$i,j$ 
means that only the quadratic term in the above polynomial survives in the full integrand. 
This term is proportional to~$a$, which vanishes after summing over the four values that it takes. 
Thus the contribution of the cubic polynomial~$A_{a}$ to the integrand vanishes.
The contribution of the function~$|P|$ to the integrand can be 
written as the exponential of
\be \label{SSdiff}
\sum_{i,j=1}^N \, (u_{ij})_2 \,  \Bigl(\log \big|P \bigl(u_{ij}+ 2 \t+ 1 \bigr) \big| 
+ 3 \log \big|P \bigl(u_{ij}+ \tfrac23\, (2\t+1)\bigr) \big| 
\Bigr)\,.
\ee
We can evaluate this expression on the saddle point~$u_i=\frac{i}{N} (m\t+n)$ 
using the double Fourier expansion~\eqref{KroneckerLimit} for the function~$\log|P|$.
In this manner we obtain a sum over the integers~$\wt n, \wt m$ of two terms corresponding to the 
two terms in~\eqref{SSdiff}. 
Now, each term contains the factor
\be \label{ijsumzero}
 \sum_{i,j=1}^N \, (i-j) \, \e \bigl(\tfrac{i-j}{N}(\wt n m - \wt m n) \bigr) \,,
\ee
which actually vanishes, as proved in Equation~\eqref{Identity1}.\footnote{The fact that the full function~$P$ 
is doubly periodic implies that it has a double Fourier expansion similar 
to~\eqref{KroneckerLimit}. The identity~\eqref{ijsumzero} then also implies that the contribution of~$P$ to
the action also vanishes.}
Thus we reach the conclusion that the integrands of~\eqref{SYMIndex1again} and~\eqref{SYMIndex}
are equal in magnitude on the saddle point configurations. We had left the phase~$\PsiQ(z)$ 
ambiguous until now, and we fix it by demanding that the phases of the 
two integrands are also equal at each saddle point.\footnote{\label{footPhidef} Here a question arises as to whether this
prescription for~$\PsiQ$ is well-defined. 
In particular, it could happen that a certain point~$z$ on the torus lies on the string of eigenvalues 
for two different saddles~$(m,n)$ and~$(m',n')$. The point~$z$ would correspondingly lift to 
two different points in the complex plane which differ by a lattice translation. 
The question then is whether the value of the phase of~$\Ge$ and in particular the value of~$\a^0_Q$ 
agrees at these two points. 
This is a subtle question whose complete analysis will be posted elsewhere. For our 
purposes here, we restrict our analysis to a set of saddles with an upper cutoff on~$m$.
In this situation if we take~$\a^0_Q$ to be defined as in Footnote~\ref{a0value}, the difference 
in~$\a^0_Q$ between two points differing by a lattice translation is a rational number with a bounded
denominator. We can then lift our discussion to a larger torus (which is still finite) 
on which~$\PsiQ$ is well-defined. We note that all the calculations of the action are done by considering 
points on the complex plane, so that they are not affected by this cutoff.
}
In Appendix~\ref{App:AnalyticAction} we have another proof of the equality of the two 
forms of the action using a series representation of the various functions.

\begin{figure}\centering
\includegraphics[width=6cm]{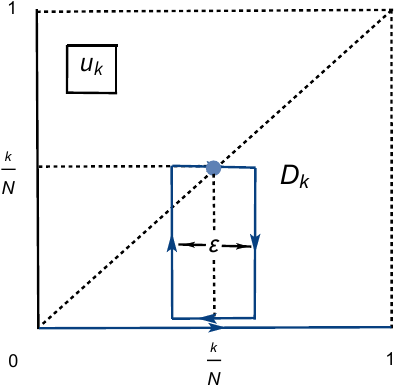} 
 \caption{  The new contour of integration that passes through the saddle-point is the sum of the contour along~$[0,1]$ 
  and the closed contour~$D_k$ in each coordinate~$u_k$. The figure shows the~$u_k$ plane for one~$k$ with the axes
  denoting the two components of~$u_k$ (decomposed in our notation~$z=z_1+\t z_2$) running from~0 to~1 so as
  to cover the fundamental parallelogram.  }
  \label{fig:Contour}
\end{figure}

The second point is the construction of a new contour that passes through the saddle-point,
and the associated analysis of the effect of the change in contour. 
The new contour~$\CC$ of integration for the integral~\eqref{SYMIndex1again} is constructed as follows. 
In each~$u_k$-plane, we cut a small interval from the real axis of width~$\ve$ (which we eventually take to zero), 
and lying directly below the saddle-point value~$u_k=\frac{k}{N}(m\t+n)$, and parallel transport it upwards 
so that it goes through the saddle point value.
Then we complete the contour by adding two vertical lines with opposite orientation. 
The new contour~$\CC$ is the sum of the original contour~$[0,1]$ and the closed contour~$D_k$
in each~$u_k$-plane. 
This is shown in Figure~\ref{fig:Contour}. 
We now want to calculate the change in the value of the integral~\eqref{SYMIndex1again} when 
we change the contour from~$[0,1]^N$ to~$\CC$ as~$N \to \infty$, or, equivalently, the sum of the  
integrals around the closed contour~$D_k$. The assertion is that we can essentially replace the integrand 
of~\eqref{SYMIndex1again} by the meromorphic integrand of~\eqref{SYMIndex} everywhere along~$D_k$,
which we can calculate using the method of residues. 

How do we justify the replacement? 
On the bottom horizontal piece of~$D_k$ this is because of the equality~\eqref{QGamrel}. The two vertical 
sides have opposite orientations and therefore cancel out as~$\ve \to 0$. 
In this limit of vanishing width, we can also 
replace the integrand of~\eqref{SYMIndex} by the integrand of~\eqref{SYMIndex1again}
on the top horizontal part of the contour as we argued above. 
Having reached the contour~$\CC$ we can now use the saddle-point approximation to calculate 
the integral. 
There is one point in this argument where we have to be careful. On the one hand, we want~$\ve \to 0$ so that the 
replacement of the doubly periodic integrand by the meromorphic integrand on the top strip is a 
good approximation. On the other hand, we recall that in the saddle-point method we essentially
approximate the function by a Gaussian near its saddle, and most of the value of the integral comes from
a region close to the saddle-point. In order for the method to be valid 
we should not take~$\ve \to 0$ too fast. In our case the large factor in front of the exponent for each eigenvalue is~$N$,  
and so if we take~$\ve = \frac{1}{\sqrt{N}}$ we safely pick up most of the value of the Gaussian. 

We can now check that the error caused by the replacement on the interval at the top of~$D_k$ is sub-leading in~$N$. 
The configuration~$u_k=\frac{k}{N}(m\t+n)$ is a saddle-point for the doubly periodic action and we can use
the Gaussian formula including the first subleading term to estimate the integral along the top interval. 
On the other hand it is regular point for the meromorphic action
and the value of its integral along the top interval is the value at the saddle point
multiplied by~$\ve$ up to an order one constant. 
Thus, the ratio of the two integrals on the top strip equals $a \times \ve \sqrt{N} = a$,
where~$a$ is an order one constant, for each variable, so that the total error in the full effective action is~$N \log a$.
In the next section we estimate the saddle-point value of the action and we find that for an infinite family of saddles, 
including the black hole saddle, the action is proportional to~$N^2$. For these saddles the error term~$N \log a$ 
is a sub-leading contribution. 
Thus we reach the conclusion that the change in the integral~\eqref{SYMIndex1again} 
upon deforming the contour from~$[0,1]^N$ to~$\CC$ as~$N \to \infty$ equals the sum of the residues of the integrand 
of~\eqref{SYMIndex} inside the various~$D_k$. 
We postpone a complete analysis of the contours and residues to the future.

Upon putting together the two observations above, we reach the conclusion that  
the~$(m,n)$ solutions to the variational equations are genuine saddle-point configurations contributing to the integral 
expression for the index. We emphasize again that the above discussion only 
applies locally near the saddles, i.e.~it only shows that the contour can pass through the saddle-point. In particular, 
it does not tell us what the exact contour is, and a rigorous global analysis of the contour remains to be done.
We proceed by writing the full integral as a sum over the various saddles (or phases) 
with an effective action at each saddle. This sum should be regarded in the same sense as the sum over 
instantons in quantum field theory and, in particular, we do not treat issues of convergence. 
We are, however,  in a position to compare the effective actions at the various saddles and  
find the phase structure of the theory at leading order in the large-$N$ approximation.  
In the following section we calculate this effective action and the consequent phase structure.

\section{Action of the saddles, black holes, and the phase structure \label{sec:Action}}

The considerations of the previous two sections have shown that the complex saddles for the 
supersymmetric index~\eqref{SYMIndex} of~$\CN=4$ SYM on~$S^3 \times S^1$ in the 
large-$N$ limit are labelled by the lattice points~$m\t+n \in \IZ \t+\IZ$ with~$\text{gcd}(m,n)=1$.\footnote{The expressions 
in the rest of this section are valid for points obeying the condition~$\text{gcd}(m,n)=1$.}
After taking into account the caveats in the discussion of Section~\ref{sec:contour},  
we can write the index as a sum over saddles as follows 
\be\label{ISaddles}
\CI(\t) \; \sim \;  \sum_{m,n \in \IZ \atop \text{gcd}(m,n)=1}  \, \exp \bigl(-\Seff(m,n;\tau)\bigr) \,.
\ee
The effective action~$\Seff(m,n;\tau)$ is the classical 
action~\eqref{EffActQ} evaluated at the saddle-point configuration~$(m,n)$ plus the quantum corrections 
induced by loop effects around the saddle. 
In this paper we only calculate the classical part. In this section we show that the action of the~$(1,0)$ saddle 
agrees with that of the supersymmetric black hole in the dual~AdS$_5$ theory. 
We then compare the real parts of the actions at the different saddle-points as a function of~$\t$ which 
leads to the phase structure of the theory.

\subsection{Action of the saddles \label{subsec:Action}}

In the large-$N$ limit the effective action~\eqref{EffActQ} of the SYM theory extended to the torus can 
be written in terms of a potential as in \eqref{defV}, 
\be \label{defVSYM}
\begin{split}
V(z) & \= \log  Q_{1,0} (z) + 3 \log Q_{-\frac13,\frac13} (z) \\
& \=  \frac{8}{27} \pi \i \t   +  \log Q \bigl(z + \t  \bigr) + 3 \log Q \bigl(z - \tfrac13 \t +\tfrac13 \bigr)  
	-   \log P \bigl(z +\t \bigr) +  \log P \bigl(z - \tfrac13 \t +\tfrac13 \bigr) \,.
\end{split}
\ee
The formula~\eqref{SeffT} says that the value of the action at the saddle-point~$m\t+n$ 
is given by the average value of the action (as a function of~$z$) on the straight line 
between the origin and that point.
In terms of the coefficients of the double Fourier series
\be
V(z) \= \sum_{m,n \in \IZ} \, \e(n z_2 - m z_1) \, V_{m,n} \,,
\ee
the average value of~$V$ between~$0$ and~$m \t +n$ ($\text{gcd}(m,n)=1$) is
\be \label{Vavg}
\int_0^1 dx \, V(x(m\tau+n)) \= \sum_{m',n' \in \IZ} \, V_{m',n'} \, \int_0^1 dx \, \e\bigl((n' m - m' n)x\bigr)
\= \sum_{p \in \IZ} \, V_{pm,pn} \,.
\ee
Here in the second equality we have used that the integral of~$\int_0^1\e(nx) dx = \delta_{n,0}$.

We will now use the formula~\eqref{Vavg} to calculate the average value of the action.  
In preparation for the result we first calculate the average values of the functions~$\log P$ and~$\log Q$. 
As discussed below Equation~\eqref{QFrel}, it is easier to calculate the average value of the real 
parts~$\log |P|$ and~$\log |Q|$. To reach the complex function we construct an expression 
holomorphic in~$\t$ whose real part agrees with these calculations. (We only present that 
after having assembled the full action, but it is easy to do this at every step if required.)
This still leaves the ambiguity of a~$\t$-independent purely imaginary term, we shall discuss 
this in the context of the~$(1,0)$ saddle in detail in the following subsection, and more generally in 
Appendix~\ref{App:AnalyticAction}.

\paragraph{Average value of~$\log |P|$.}
The double Fourier series given by the Kronecker limit formula~\eqref{KroneckerLimit} implies that, for~$(m,n) \neq (0,0)$,
\be
\begin{split}
\int_0^1 dx \, \log |P\bigl(x(m\tau+n) +a \t + b \bigr)|  
& \= - \frac{\imt}{2\pi} \, \frac{1}{|m\t+n|^{2}} \, \sum_{p \in \IZ \atop p \neq 0} \, \frac{1}{p^2} \, \e\bigl( (n a - m b) p \bigr) \\
& \=  - \pi  \imt  \, \frac{1}{|m\t+n|^{2}} \, B_2 \bigl( \{m b-na\} \bigr) \,.
\end{split}
\ee

\vspace{0.2cm}

\paragraph{Average value of~$\log |Q|$.}
The double Fourier series given by the Bloch formula~\eqref{Bloch} implies that, for~$(m,n) \neq (0,0)$,
\be
\begin{split}
\int_0^1 dx \, \log |Q\bigl(x(m\tau+n) +a \t + b \bigr)|   
& \=  \i \, \frac{\imt}{2\pi^2} \,  \frac{(m \ret+n)}{|m\t+n|^{4}}\, \sum_{p \in \IZ \atop p \neq 0} \,  \frac{1}{p^3} \, \e\bigl( (n a - m b) p \bigr) \\
& \= \frac23 \pi \imt  \,  \frac{(m \ret+n)}{|m\t+n|^{4}}\,  B_3 \bigl( \{m b-na\} \bigr) \,.
\end{split}
\ee
In reaching the second lines in the above formulas we have used the Fourier 
expansions~\eqref{B2Four}, \eqref{B3Four} for the Bernoulli polynomials.

\vspace{0.2cm}

\paragraph{Average value of~$\log |Q_{ab}|$.} 
Putting these two together we can calculate the average value of~$\log Q_{ab}$ to be
\bea
&& \int_0^1 dx \, \log |Q_{ab}\bigl(x(m\tau+n) \bigr)|   \= \\
&&\qquad  \= \pi \imt  \, \Bigl( \frac{a(1-2a^2)}{6} + \frac23 \,\frac{(m \ret+n)}{|m\t+n|^{4}}\,  B_3 \bigl( \{m b-na\} \bigr) 
+ a \, \frac{1}{|m\t+n|^{2}} \, B_2 \bigl( \{m b-na\} \bigr)  \Bigr) \,. \nn
\eea

\vspace{0.4cm}

\paragraph{Vector multiplet action.}
The real part of the vector multiplet contribution to the saddle-point action is given by
\be
\begin{split}
\Re \, \Seff^\text{vec}(m,n;\t) 
& \= N^2 \int_0^1 dx \, \log |Q_{1,0}\bigl(x(m\tau+n)\bigr)|   \\
&  \= N^2 \pi \imt \Bigl( -\, \frac{1}{6} 
 + \frac23   \,  \frac{(m \ret+n)}{|m\t+n|^{4}}\,  \, B_3 (0) 
 +  \, \frac{1}{m(m\t+n)}  \, B_2 (0) \Bigr) \, , \\
&  \=  N^2 \pi \t_2 \Bigl( - \frac{1}{6} 
 +  \,  \frac{1}{6} \, \frac{1 }{|m\t+n|^{2}} \,  \Bigr) \,.
\end{split}
\ee

\vspace{0.2cm}

\paragraph{Chiral multiplet action.}
The real part of the chiral multiplet contribution to the saddle-point action is given by
\bea
&& \Re \, \Seff^\text{chi}(m,n;\tau) \nn \\
&& \qquad \= N^2 \int_0^1 dx \, \log |Q_{-\frac13,\frac13}\bigl(x(m\tau+n)\bigr)|   \\
&& \qquad \=  N^2 \pi \t_2 \Bigl( - \frac{7}{162}
 + \frac23   \,  \frac{(m \ret+n)}{|m\t+n|^{4}}\,  B_3 \bigl( \{\tfrac13(m+n)\} \bigr) 
 -  \, \frac13 \frac{1}{|m\t+n|^{2}} \, B_2 \bigl(  \{\tfrac13(m+n)\} \bigr) \Bigr) \,. \nn
\eea

\vspace{0.4cm}

\paragraph{$\CN=4$ SYM action.}
Putting everything together, we reach the following concise formula for 
the real part of the total action at the~$(m,n)$ saddle-point,
\bea \label{StotRe}
&& \Re\, \Seff(m,n;\tau)  \nn \\
&& \= \Re\, S^\text{vec}(m,n;\t) + 3 \, \Re\, S^\text{chi}(m,n;\t) \\
&& \= N^2 \pi \imt \biggl( - \frac{8}{27} 
  +  2 \, \frac{(m \ret+n)}{|m\t+n|^{4}} \, B_3 \bigl( \{\tfrac13(m+n)\} \bigr) 
 +  \,  \frac{1}{|m\t+n|^{2}}  \,  \Bigl(\frac16 -B_2 \bigl(  \{\tfrac13(m+n)\} \bigr)  \Bigr)\biggr) \,. \nn
\eea
The coefficients of the second and third terms in the above formula are periodic in their arguments 
and take the values given in Table~\ref{Table:B2B3vals}.
\begin{table}[h!]
\centering
\begin{tabular}{|c | c c c  |} 
\hline
$\ell \; (\text{mod} \, 3)$ & 0 & 1& $2$    \\ [0.5ex] 
\hline  
 $B_3(\bigl\{\frac{\ell}{3} \bigr\})$  & 0 & $1/27$& $-1/27$   \\ 
 [0.5ex] 
 $\frac16 - B_2(\bigl\{\frac{\ell}{3}\bigr\})$ & 0 & $2/9$ & $2/9$   \\
 [0.5ex] 
\hline
\end{tabular}
\caption{The coefficients of the second and third terms in Equations~\eqref{StotRe}}
\label{Table:B2B3vals}
\end{table}
The symmetry~$\t \mapsto \t+3$ is realized in this formula as~$(m,n) \mapsto (m,n+3m)$. Under this operation
the combination~$\frac13(m+n) \mapsto \frac13(m+n) + m$, so that the fractional part of this number in the argument of the 
Bernoulli polynomials does not change. 
It is clear from this formula that 
\be
\Re \, \Seff(m,n;\t) \=  - \frac{8}{27} N^2 \pi \imt  \qquad \text{when~$m+n =0$ mod 3}\,.
\ee

Using the values of the Bernoulli polynomials in Table~\ref{Table:B2B3vals} we can rewrite the real part of the action as
\be \label{StotRe1}
\Re\, \Seff(m, n; \t)  
\= \frac{2}{27} N^2 \pi \, \imt \biggl( - 4 +  \frac{(m \ret+n)}{|m\t+n|^{4}} \,  \chi_1(m+n) 
 +   \frac{1}{|m\t+n|^{2}}  \,  3 \chi_0(m+n) \biggr) \,,
\ee
where~$\chi_i$, $i=0,1$ are the two Dirichlet characters of modulus~$3$ (see Table~\ref{Table:chi01}).
\begin{table}[h!]
\centering
\begin{tabular}{|c | c c c  |} 
\hline
\diagbox[width=1cm,height=0.6cm]{$\chi$}{$\ell$} & 0 & 1& $2$    \\ [0.5ex] 
\hline  
 $\chi_0$  & 0 & $1$& $1$   \\ 
 [0.5ex] 
 $\chi_1$ & 0 & $1$ & $-1$   \\
 [0.5ex] 
\hline
\end{tabular}
\caption{The two Dirichlet characters modulo 3.}
\label{Table:chi01}
\end{table}

Upon constructing the holomorphic part of the action as explained above, we obtain $S(0,1;\tau) = 0$ 
and a short calculation shows that, for~$m\neq 0$, 
\be\label{actionEll}
\Seff(m,n; \tau) \= \frac{N^2 \pi \i }{27\,m} \,\frac{  \bigl(2 (m\t+n)  +  \chi_1(m+n) \bigr)^3}{(m\t+n)^2} + N^2 \pi \i  \, \v(m,n) \,,
\ee
where the~$\t$-independent real phase~$\v$ is 
\be\label{phase}
 \v(m,n) \=  \wt K(m,n)  - \frac{1}{27}  \Bigl( 8 \frac{n} {m}  + \frac{12}{m}    \, \chi_1(m+n) \Bigr) \,,
\ee
where in terms of the function~$K$ in~\eqref{QKron}, we have 
\be
\wt K(m,n) \= \sum_{p\in \IZ} K(pm,pn) \,.
\ee

It is interesting to ask what is the action of a given saddle-point configuration if we use the 
analytic continuation of the elliptic gamma function instead of the elliptic continuation presented here. 
We address this question in Appendix~\ref{App:AnalyticAction} and find that the results coincide, 
up to a $\tau$-independent imaginary constant.
The phenomenon that the action of a given saddle essentially does not change if we use the 
analytic continuation of the elliptic gamma function instead of the elliptic continuation is interesting. 
At the practical level, the meromorphic analysis of Appendix~\ref{App:AnalyticAction}, 
can be used to extend the current results of the Bethe ansatz approach of~\cite{Benini:2018mlo,Benini:2018ywd} and, 
to relate those with the results of the elliptic approach introduced here.

\subsection{Entropy of supersymmetric AdS$_5$ black holes \label{sec:BHent}}

A particularly interesting saddle-point is the black-hole~$(m,n)=(1,0)$, whose action is
\be \label{SBH}
\Seff(1,0;\t)    
\=\frac{\pi \i\,  N^2\, (2\tau\, +\,1)^3}{27\, \tau ^2} +\pi\i \,N^2 \v(1,0)\,.
\ee
Up to a~$\t$-independent additive constant that we discuss below, 
this action is equal to the action of the supersymmetric~AdS$_5$ black hole, 
calculated in~\cite{Cabo-Bizet:2018ehj} from supergravity after explicitly solving 
the constraint~\eqref{tsvcons} for~$\v$ and subsequently setting~$\t=\s$. 
In~\cite{Cabo-Bizet:2018ehj} the black hole entropy was derived by performing a 
Legendre transform which includes a Lagrange multiplier that enforces the constraint. 
Here we verify that one can equivalently obtain the result for the black hole entropy 
by explicitly solving the constraint. As we see below  
the Legendre transform of the action~\eqref{SBH} combined with a set of reality conditions on the 
expectation values of charges and the entropy
leads to the Bekenstein-Hawking entropy of the BPS black hole \cite{Hosseini:2017mds,Cabo-Bizet:2018ehj}.

The entropy of the black hole comes from the $(1,0)$ saddle-point as
\be\label{ContourInt}
\rme^{ \mathcal{S}_{BH}(Q,\,J)}\=\int d\tau\,  \exp \bigl(\mathcal{E}(\tau) \bigr)\, , \qquad 
\mathcal{E}(\tau) \= - \Seff(1,0;\tau)-2 \pi \i \,\tau \bigl( 2 J+Q \bigr)\,-\, \pi  \i \, Q\,,
\ee
with $J= \frac12(J_1+J_2)$. 
There are two $\t$-independent terms in the exponent of the right-hand side which deserve a comment.
One is the last term on the right-hand side which appears from a similar term in the Hamiltonian definition
of the index~\eqref{RelInd0} with~$n_0=-1$. 
The other is the value of the constant~$\v(1,0)$ that we fix to zero by the 
choice~$\wt {K}(1,0)=\frac{12}{27}$ in~\eqref{SBH}, so that 
\be \label{Sefftl}
 \Seff(1,0;\tau)\=\frac{\pi \i  \, N^2}{27} \frac{(2 \,\tau\, +\,1)^3}{\tau ^2}\,. 
\ee
In the large-$N$ regime, the integral \eqref{ContourInt} is approximated by its saddle-point value, and 
as usual we assume that the integration contour goes through the saddle-point. 
Thus we have to extremize the functional~$\mathcal{E}(\tau)$.

We find that, as a result of this extremization procedure, 
the black hole entropy obeys $P(\frac{\i}{2 \pi}\mathcal{S}_{BH})=0$,  
 where~$P$ is the cubic polynomial
\be\label{PPol}
\begin{split}
P(x)& \=\Bigl(- \frac{Q}{2}+ x \Bigr)^3-\frac{N^2}{2} \bigl( J + x \bigr)^2, \\ 
&\; \equiv \; p_0+ p_1\, x + p_2 \, x^2+x^3 \,.
\end{split}
\ee
This is precisely the polynomial that was found to govern the gravitational entropy of the 
supersymmetric extremal black holes in~\cite{Cabo-Bizet:2018ehj}. 
In addition if we impose the reality conditions
\be\label{Reality}
\text{Im}(\mathcal{S}_{BH})\=\text{Im}(J)\=\text{Im}(Q)\=0 \,,
\ee
we find that the charges $J$ and $Q$ obey the following constraint
\be\label{Constraint}
p_0\=p_1\, p_2\,,
\ee
where
\be
p_0\=-{N}^2\frac{J^2 }{2}-\frac{Q^3}{8}\,, \qquad 
p_1\=\frac{3\, Q^2}{4}\,-\, {N}^2J \,, \qquad 
p_2\=\,-\,\frac{3 Q}{2}\,-\,\frac{{N}^2}{2} \,,
\ee
are the coefficients of~$P$ defined in~\eqref{PPol}.
With these reality conditions, we obtain the following result for the entropy,
\be\label{BPSEntropy}
\mathcal{S}_{BH}\= \pi \sqrt{{3} Q^2 \,-\,4{N^2} J}\, ,
\ee
which agrees with the Bekenstein-Hawking entropy of the corresponding BPS black holes. Based on 
the black hole properties,
we shall call the SYM saddle-points with real expectation value of charges and entropy as \emph{extremal}.

The above results can be obtained as follows. Upon extremising $\mathcal{E}(\tau)$ one obtains the following relation
\be
\label{Jtau}
2J + Q \= - N^2 \,\frac{(\tau -1) (2 \tau +1)^2}{27 {\tau}^3} \,.
\ee
Plugging this relation into $\mathcal{E}(\tau)$ one obtains, at the saddle point
\be\label{parametS2}
\mathcal{S}_{BH} \= \mathcal{E}(\t)  \= 
-  \pi\, \i \, N^2 \frac{(2 \tau +1)^2}{9 {\tau}^2}\, - \pi\,\i \, Q\,.
\ee
The above two equations imply that the quantity~$s=\frac{\i}{2 \pi}\mathcal{S}_{BH}$ obeys, at the saddle point, 
\be 
s-\frac{Q}{2}  \=  \frac{N^2}{18} \frac{(2 \tau +1)^2}{{\tau}^2} \,, \qquad 
s+J \=  \frac{N^2}{54} \frac{(2 \tau +1)^3}{{\tau}^3} \,.
\ee
Putting these two equations together we find that~$P(s)=0$.

Upon imposing the reality condition of $J$ and $Q$ in \eqref{Jtau} we obtain that the extremal solutions lie on the 
following real one-dimensional locus in the complex~$\t$-plane 
(we denote the corresponding quantities by~$^*$), 
\be\label{CurveExt}
\tau^*_2\=\sqrt{-\frac{3 {\tau_1}^2 (2 \tau_1+1 )}{6 \tau_1-1}} \,.
\ee
Now imposing the reality of $\mathcal{S}_{BH}$ and $Q$ in \eqref{Jtau}, \eqref{parametS2},  we obtain
\be\label{JQReal}
J^* \= \frac{N^2 \,(2\ret+1)^2 \, (7 \ret -1)}{432 \,{\tau_1}^3}\,,\qquad 
Q^* \=  -\,\frac{N^2\, \left( 2\t_1+1\right)\left(10\t_1-1\right)}{72 \, {\tau_1}^2} \,.
\ee
One can now check that the expressions \eqref{JQReal} satisfy the non linear constraint~\eqref{Constraint}, 
and that the right-hand side of~\eqref{parametS2} equals the expression~\eqref{BPSEntropy}. 
One can ask what happens if we choose differently the~$\t$-independent term $\v(1,0)$ in the 
extremization problem~\eqref{ContourInt}.
This is answered by noticing that without these terms the exponent has the symmetry which 
shifts~$J$ and~$\frac{Q}{2}$ equally in opposite directions keeping~$2J+Q$ invariant. 
The actual minimization procedure (e.g.~in Equation~\eqref{Jtau}) only depends on this combination.
However, the reality conditions that we impose break this symmetry. 
Now it is clear that if we change these terms, e.g.~if we change the value of the 
constant~$\wt K= \wt K(1,0)$ away from~$\frac{12}{27}$, then
the whole procedure that we described above still goes through, but with the redefinitions~$Q \to Q+N^2(\wt K-\frac{12}{27})$, 
$J \to J - \frac{N^2}{2}(\wt K-\frac{12}{27})$. 
However, we note that the value of~$\wt K=\frac{12}{27}$ that agrees with the supergravity 
entropy~\emph{function} is precisely the one given by the analytic continuation of 
the original action~\eqref{SYMIndex} (see Section~\ref{sec:contour} and Appendix~\ref{App:AnalyticAction}). 
It will be nice to understand this observation at a deeper level.

\begin{figure}[h]\centering
  \includegraphics[width=7cm]{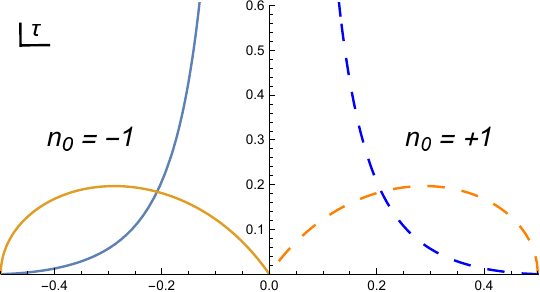}
  \caption{ The orange curves are plots of extremal curves $\tau^*$. The blue curves are plots of their entropy as a function of 
  the worldline parameter $\tau_1$. These curves agree with the respective gravitational curves with the same labels. 
  In the gravitational theory the extremal curves for~$n_0=-1$ (solid) and for~$n_0=+1$ (dashed) have the same 
  metric and gauge field configurations. }
  \label{BlackholeUnicity}
\end{figure}

We move on to compare these results with the gravitational picture.
In~\cite{Cabo-Bizet:2018ehj}, a family of complex supersymmetric asymptotically $AdS_5$ solutions 
of the dual supergravity theory was studied, using earlier results of~\cite{Chong:2005hr,Gutowski:2004ez}
on general non-supersymmetric black hole solutions. 
Upon imposing reality conditions on the solutions within this family, 
one reaches the extremal supersymmetric black hole. 
The BPS angular velocities of the general complex solutions was defined in~\cite{Cabo-Bizet:2018ehj},
using a certain limiting process to the supersymmetric locus,  (in the present context of equal angular 
potentials) as 
\be\label{Inver}
\omega \= 2 \pi \i\, \tau \= 2 \pi \i\, \frac{(a-1) (-(a+2)n_0-3 \i r_+)}{(a+2)^2+9\, r_+^2}\,,
\ee
where $a\in[0,1)$ and $r_+>0$. At extremality, i.e.~for $r_+=\sqrt{2a+a^2}$, $\tau$ reduces to \cite{Cabo-Bizet:2018ehj}
\bea\label{inverExtre}
 \tau^*(a)&\=&\frac{(a\,-\,1) \bigl(-(a+2) n_0-3 \i \sqrt{a(a+2)} \bigr)}{2(a+2) (5 a+1)}\,.
 \eea
The two values $n_0=\pm1$ in \eqref{Inver} are related in field theory by~$\ret \to - \ret$, and 
label two complex conjugated solutions in the gravitational theory related by~$\Im \, \omega \to -\Im \, \omega$. 
At extremality the metric and gauge field configurations of two gravitational solutions  
are real, and they coincide which means that the two values~$n_0=\pm1$ 
are simply two descriptions of the very same solution.
One can check that for $n_0=\pm 1$ the gravitational extremal curve~$\tau^*(a)$ in~\eqref{inverExtre} agrees with the 
field theory extremal curve $\tau^*$, obtained from \eqref{CurveExt} as one expects.\footnote{We have checked that 
there are no~$(m,n)$ saddles in the $n_0=-1$ field theory index that matches the $n_0=+1$ extremal curve. 
Indeed, once the holographic dictionary is established, it would have been surprising to have two different 
field-theory saddles corresponding to the same BPS black hole solution.}
We plot these curves in Figure \ref{BlackholeUnicity}. The dashed curve in this figure is the same as the~$(1,0)$ curve in~\cite{Benini:2018ywd}.
\begin{figure}[h]\centering
  \includegraphics[width=6cm]{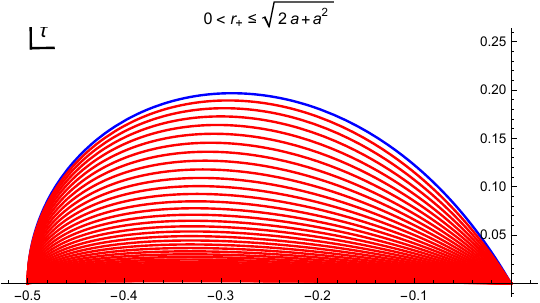}
    \includegraphics[width=6cm]{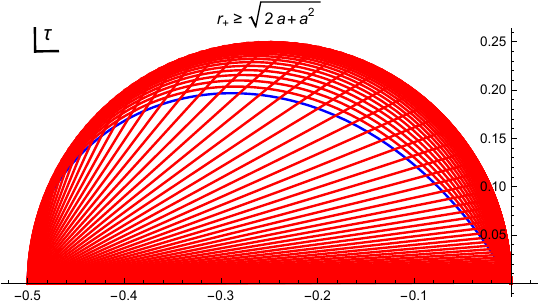}
  \caption{  The blue curve denotes the extremal black hole-locus with the horizon radius~$r_+=\sqrt{2 a+a^2}$. 
  The red curves correspond to different values of $r_+$ (where the solution is complex). 
  Notice that the envelope of these complex solutions in~$\mathbb{H}$ is bounded by the semi-circle on the right. 
}
  \label{NonExtremalBHs}
\end{figure}

In Figure~\ref{NonExtremalBHs} we have plotted the regions covered by such solutions as one varies the parameters $a$ and $r_+$. 
We note that the known non-extremal supersymmetric solutions discussed in~\cite{Cabo-Bizet:2018ehj} (which have~$0 \le a<1$
and~$r_+>0$)  do not cover the full range of complex coupling $\tau \in \mathbb{H}$, and, in particular, the regions with large~$\imt$. 
As we will see in the following subsections, there are other saddles~$(m,n)$ which exist (and dominate) other regions.

\subsection{Entropy of $(m,n)$ solutions \label{sec:mnent}} 

In this subsection we extremize the entropy functional of the ensemble described by $(m,n)$ saddles,
and calculate their entropy. 
The procedure is exactly as in the~$(1,0)$ case, namely that the extremization combined with 
the imposition of reality of the expectation values of field-theory charges and entropy forces the 
chemical potential $\tau$ to live on a specific one-dimensional curve in the upper half plane.

We start by rewritting \eqref{actionEll} in terms of the variable $\wt\t=m\t+n$,
 \be\label{GeneralAction}
    \Seff(m,n;\wt\t)\= \frac{1}{m}\, \frac{\pi\i}{27}\, N^2 \, \frac{\bigl(2\widetilde{\tau}+\chi_1(m+n)\bigr)^3}{\widetilde{\tau}^2}
    + N^2 \pi\i \,\varphi(m,n)\,.
\ee
We suppress~$\v(m,n)$ in the following discussion, keeping in mind that it can be reinstated 
following the discussion in the previous section.
To ease presentation we also suppress the dependence of $\chi$ on $m$ and $n$. 
The entropy of $(m,n)$ saddles is obtained by the following 
extremization, 
\be\label{ExtremizationEntropyF}
\mathcal{S}_{(m,n)}\=\underset{\t}{\text{ext}}\, \mathcal{E}(\tau,J,Q)\,.
\ee
where the entropy functional~$\mathcal{E}$ is 
\be\label{LOp}
\mathcal{E}(\tau,J,Q)\=- \Seff(m,n;\tau)-2 \pi \i \,\tau \bigl( 2 J+Q \bigr)\,-\, \pi  \i \, Q\,.
\ee
From the chain rule it follows that
\bea\label{ChainRule}
\underset{\t}{\text{ext}}\, \mathcal{E}(\tau,J,Q)&=&\underset{\wt\tau}{\text{ext}}\, \mathcal{E}(\wt\tau;\wt J,\wt Q)\,,
\eea
where we have defined useful auxiliary charges in terms of the physical charges as
\be\label{JQtilde}
\widetilde{Q} \= Q\,-\, 2\,\frac{n}{m} \,(2J+Q)\,, \qquad 
2\widetilde{J}+\widetilde{Q}\=\frac{2J+Q}{m}\,.
\ee
It should be clear from the above considerations that the entropy of extremal $(m,n)$ solutions has a similar form as
that of the~$(1,0)$ solution~\eqref{BPSEntropy} with~$Q$, $J$ replaced by~$\wt Q$, $\wt J$, respectively.
We now spell out some of the details. 
 
We have that $\mathcal{S}_{(m,n)}$ is a root of the cubic polynomial $\wt P(\frac{x}{2\pi\i})$, where
\bea\label{PolynomMN}
\widetilde{P}(x)& \= &\Bigl(-\frac{\chi_1  \widetilde{Q} }{2}\,+\,x \,\Bigr)^3\,-\,
 \frac{\chi_1\,N^2}{2 m} \bigl( \chi_1 \widetilde{J} \,+ \, x \bigr)^2\,, 
\\
&=&\widetilde{p}_0\,+\,\widetilde{p}_1\, x\,+\,\widetilde{p}_2\, x^2+x^3 \,,
\eea
with
\be  \label{Definitionp} 
\widetilde{p}_0\= -\chi_1 {N}^2\frac{\wt J^2 }{2m}-\chi_1\frac{\wt Q^3}{8}\,, \qquad 
\widetilde{p}_1\= \frac{3}{4} \wt{Q}^2\,-\,\frac{N^2 \wt{J} }{m}\,, \qquad
\widetilde{p}_2\=-\frac{3}{2} \, \chi_1\, \wt{Q} \,-\, \chi_1\,\frac{N^2}{2 m}\, .
\ee

Next we impose reality conditions. Notice that reality of~$(J,Q)$ is equivalent to reality of~$(\wt J, \wt Q)$. 
Imposing the reality conditions on the charges as well as the entropy~$\mathcal{S}_{(m,n)}$, 
we obtain 
\be\label{E1E2}
\widetilde{J}^*(\widetilde{\t}_1)
\=\frac{N^2 \, (2 \wt\t_1+\chi_1 )^2 (7 \wt\tau_1-\chi_1 )}{432\, m\, \wt\tau_1^3} \,, \qquad 
\widetilde{Q}^*(\widetilde{\t}_1)
\= -\,\frac{N^2\, ( 2\wt\t_1+\chi_1 ) (10\wt\t_1-\chi_1 )}{72 \,m\, \wt\tau_1^2} \,,
\ee
and the extremal locus 
\be\label{curves}
\wt\tau_2^*\=\sqrt{\frac{3 \wt\tau_1^2\, (2 \,\wt\tau_1+\chi _1 )}{\chi _1- 6 \wt\tau _1}} \,.
\ee
Notice that~$\wt \ret =0$ is a cusp of the curve.   
The corresponding singularity is related to the phenomenon of the 
solution at this point becoming infinitely large, which can be seen 
from the fact that the charges~\eqref{E1E2} become infinite. 
Excluding this point, and demanding that the expression under the square root in~\eqref{curves} 
is positive gives us the following allowed values of~$\wt\ret$:
\bea\label{domaincurves1}
\left. 
\begin{array}{cc}
-\,\frac{1}{2}\,<\,\wt\t_1\,<\,0 & \\
\quad 0\,<\,\wt\t_1\,<\,\frac16 & \\ 
\end{array}
\right\}
\qquad \text{if} \quad(m+n)\,\text{mod}\,3\,=\,1\,, \\
\label{domaincurves2}
\left. 
\begin{array}{cc}
\quad 0\,<\,\wt\t_1\,<\,\frac12 & \\ 
-\,\frac{1}{6}\,<\,\wt\t_1\,<\,0 & \\
\end{array}
\right\}
\qquad \text{if} \quad (m+n)\,\text{mod}\,3\,=\,2\,.
\eea
Note that the real part of the action~$\Seff$ is less than (greater than) the limiting value of all saddles as~$\imt \to \infty$ (which 
we will discuss in the next section) for the first lines (second lines) in~\eqref{domaincurves1},\eqref{domaincurves2}.
Relatedly, the first lines reproduce the action and entropy of the known black hole solutions, while 
the second lines do not correspond to anything that we know of in gravity.\footnote{From now on we only refer 
to the regions in the first lines when we discuss extremal curves.}
For this reason we define the Cardy-like limit (discussed later in more detail) as~$\wt \t \to 0$, $-\chi_1 {\wt \t}_1 >0$.

In terms of the physical chemical potential $\tau$, \eqref{curves} takes the forms
\bea\label{curves2}
\tau _2^* \=\frac{1}{m} {\sqrt{\,\frac{3\left(m\, \tau _1\,+\,n\right)^2 \left(2\, m \,\tau _1\,+\,2
\,   n\,+\,\chi _1\right)}{\chi _1\,-\,6\, m\, \tau _1\,-\,6\, n}}}\,,
\eea
and the first lines of \eqref{domaincurves1}, \eqref{domaincurves2} take the form,
\bea
\begin{cases}\label{rangetau1}
-\,\frac{1\,+\,2\,n}{2m}<\,\t_1\,<\,-\,\frac{n}{m} & \text{if} \quad(m+n)\,\text{mod}\,3\,=\,1\,,  \\
\quad\,\,-\,\frac{n}{m}\,\,\,<\,\t_1<\,\frac{1\,-\,2 n}{2 \,m} & \text{if} \quad (m+n)\,\text{mod}\,3\,=\,2\,.  \\
\end{cases}
\eea

\begin{figure}[h]\centering
  \includegraphics[width=6cm,height=5cm]{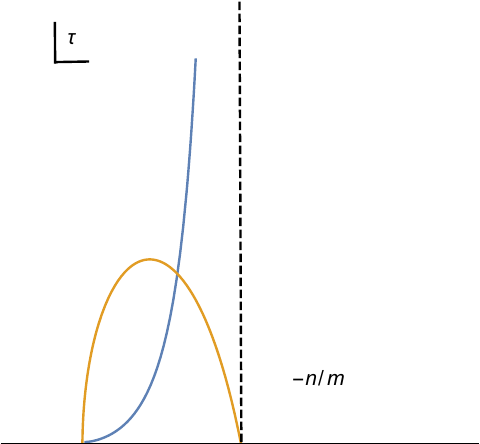}\qquad \text{or}\qquad
  \includegraphics[width=6cm,height=5cm]{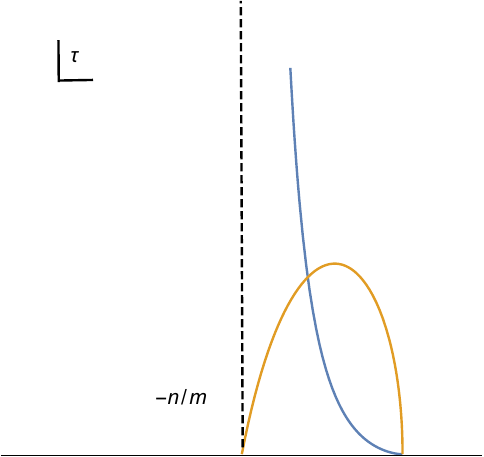}
  \caption{The orange lines are extremal curves in the $\tau$-plane for the $(m,n)$ saddle (where~$J$,~$Q$, and~$\mathcal{S}_{(m,n)}$ 
  are all real).  The blue line plots the normalized entropy~$\frac{\mathcal{S}_{(m,n)}}{N^2}$ along the corresponding extremal curve. 
  For every~$\frac{n}{m}$, with $m>0$, $\text{gcd}(m,n)=1$, and such that $(m+n)  \,\text{mod}\,3 \neq 0 $, the normalized entropy grows 
  to~$+\infty$ when the extremal curve approaches $-\frac{n}{m}$ from the left ($(m+n)\,\text{mod}\,3  = 1$), or from the 
  right ($(m+n) \,\text{mod}\,3 = -1$).}
  \label{EveryRational}
\end{figure}

From Equation~\eqref{PolynomMN} it follows that $\mathcal{S}_{(m,n)}=0$ for solutions such 
that $\chi_1(m+n)=0$. From now on, we focus on solutions such that $\chi_1=\pm1$. 
The reality conditions on charges and entropy imply that for all $m\in \mathbb{N}$ and 
$n\in\mathbb{Z}$ with~$(m+n) \,\text{mod}\, 3\neq 0$ the following factorisation property holds
\be\label{Constraintp}
\widetilde{p}_0 \= \widetilde{p}_1\widetilde{p}_2\,.
\ee
The factorisation \eqref{Constraintp} implies that the algebraic form of the $\mathcal{S}_{(m,n)}$ is
\be\label{Entrop}
\mathcal{S}_{(m,n)}\=\pi \, \sqrt{3 \,\widetilde{Q}^2 - \frac{4{N^2} \widetilde{J}}{m}}\,. 
\ee
The constraint \eqref{Constraintp} can be used to eliminate one variable in this expression.
An implicit way to do that, is to use~\eqref{E1E2}  in the right-hand side of \eqref{Entrop}. 
In that way one obtains the following parametric expression for the entropy,
\bea\label{EntropTau}
\begin{split}
\mathcal{S}_{(m,n)}(\wt\tau_1)
&\=  \frac{2\pi N^2}{m}\,\frac{\sqrt{{  1\,-\,8\, \wt\tau _1^2  \bigl(8\, \wt\tau _1\, \chi _1\,+\,6\, \wt\tau_1^2+3 \bigr) }}}{48 \sqrt{3} \, \wt\tau _1^2} \\
&\=\ \frac{2\pi N^2}{m}\,\frac{ \sqrt{1-8 \, (m \tau _1+n)^2  \bigl(8\, \chi _1 (m \tau_1+n )+6  (m \tau _1+n )^2\,+\,3 \bigr)}}{48\, \sqrt{3}\,  
(m \tau_1+n )^2}\,,
\end{split} 
\eea   
where the real parameter $\tau_1$ ranges in the domain \eqref{rangetau1}.\footnote{The expression~\eqref{EntropTau} 
is independent of the phase $\varphi$ even if we reinstate it in the preceding discussion.}

\begin{figure}[h]\centering
  \includegraphics[width=8cm]{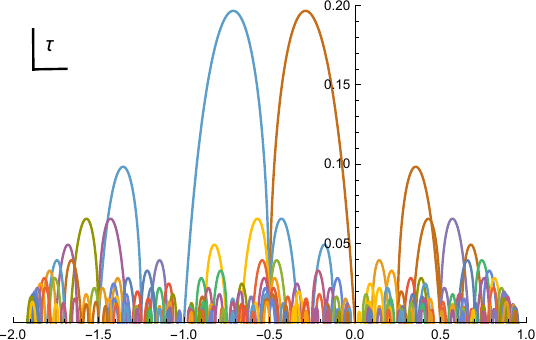}
    \caption{ The extremal curves in the fundamental domain $-2\leq\tau_1<1$, from $m=1,\ldots,\,30$ and $n=-30,\,\ldots,\,30$. }
  \label{ExamplesExtC}
\end{figure}
From Equations \eqref{E1E2} and \eqref{EntropTau} it follows that when the extremal $(m,n)$ curve~\eqref{curves2} 
approaches the rational number~$-\frac{n}{m}$ within the range~\eqref{rangetau1},
the expectation value of charges as well as the entropy (in units of~$N^2$) grow to infinity.  
We refer to such limits as extremal Cardy-like limits (see Figure~\ref{EveryRational}). 
As the set of rationals~$\mathbb{Q}$ is dense, it follows that~$\tau_2=0$ is an accumulation line for such limits. 
Figure \ref{ExamplesExtC} shows a set of extremal curves for $m=1,\ldots, 30$ and $n=-30,\ldots,30$.  
In the~$J$-$Q$ plane, the extremal solution for~$(1,0)$ is the one-dimensional curve~\eqref{Constraint} or, equivalently,~\eqref{JQReal}.
Now we see that there is a dense set of such extremal curves~\eqref{Constraintp} in the~$J$-$Q$ plane labelled by rational numbers.

It would be very interesting to find the gravitational interpretation of the generic~$(m,n)$ saddles. These should be 
asymptotically $AdS_5$ supersymmetric solutions whose Bekenstein-Hawking entropy 
should coincide with our predictions \eqref{Entrop}, \eqref{EntropTau}.\footnote{The hairy black holes 
explored in~\cite{Basu:2010uz,Markeviciute:2016ivy,Markeviciute:2018cqs} could be relevant to this discussion.}
An interesting feature of the~$(m,n)$ saddles is that the labels of the set of saddle-points i.e.~$(m,n)$ with~$\text{gcd}(m,n)=1$, 
can be equivalently thought of as the coset~$\Gamma_\infty\backslash \Gamma$,
i.e.~the set of matrices in~$\Gamma = SL(2,\IZ)$ (acting as usual via fractional linear transformations on~$\t$) modulo the 
subgroup~$\Gamma_\infty$ that fixes the point at infinity. 
This is the same structure as the~AdS$_3$ black hole Farey Tail~\cite{Dijkgraaf:2000fq}, and perhaps points to a similar 
structure of Euclidean saddle-points as in~\cite{Maldacena:1998bw}.

\subsection{Phase structure}

Now we analyze the relative dominance of the saddles for a given value of~$\t$. Firstly, recall that~$\CI(\t)$ is invariant 
under~$\t \to \t+3$, and correspondingly we analyze the region~$-2 \le \Re(\t) < 1$. 
The idea is to compare the real parts of the action at the different saddles at a generic point in the~$\t$-plane. 
The dominant phase (if a unique one exists)
is the saddle with least real action. The phase boundaries where the real parts of the actions of two saddles
are equal are called the anti-Stokes lines. 

As~$\imt \to \infty$, the real part of the action~\eqref{StotRe1} 
of all the phases with~$(m+n) \, \text{mod} \, 3 = 0$ equals~$-\frac{8}{27}\pi N^2 \imt$.
There are an infinite number of such phases and the imaginary parts and sub-dominant effects will be 
important to evaluate the path integral. (Curiously, the entropy of these saddles actually vanish as shown in the 
previous subsection.)
When $(m+n) \, \text{mod} \, 3 \neq 0$, then the third term in~\eqref{StotRe1} 
is always positive, and it dominates over the second term for large enough~$\imt$. Therefore these 
saddles are sub-dominant compared to those with~$(m+n) \, \text{mod} \, 3 = 0$ in that region.
When we start to reduce~$\imt$ we begin to hit regions where one phase dominates. 
For example, the action~$\Seff(1,0;\t)$ blows up at~$\t=0$ and the dominant phase near~$\t =0$ is~$(1,0)$ (the black hole).
The anti-Stokes lines between the black hole and the region at~$\imt \to \infty$ is given by 
\be \label{circle10}
\Re \,\Seff(1,0;\t) \= -\frac{8}{27}\pi N^2 \imt \; \; \Longrightarrow \;   \;   \bigl(\t_1+\frac16 \bigr)^2 + \t_2^2 = \frac{1}{36} \,.
\ee
Similarly, as~$\t$ approaches a rational point~$-\frac{n}{m}$,~$-\Re \, S_\text{eff}(m,n;\t)$ becomes very large and that is the 
corresponding dominant phase near the point.

\begin{figure}[h]\centering
  \includegraphics[width=9cm]{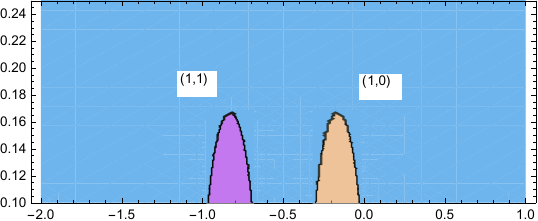}
  \caption{The phase structure for~$\t_2 \ge 0.1$. Apart from the phases~$\imt \to \infty$, there are 
  two dominant phases~$(1,0)$ and~$(1,1)$ in the semi-circular regions. Here we have 
  scanned~$\{(m, n)\mid m=0,\dots 30, n = -30, \; \dots, 30\}$ with~$\text{gcd}(m,n)=1$. 
  }
  \label{Phases}
\end{figure}

\begin{figure}[h]\centering
  \includegraphics[width=8cm]{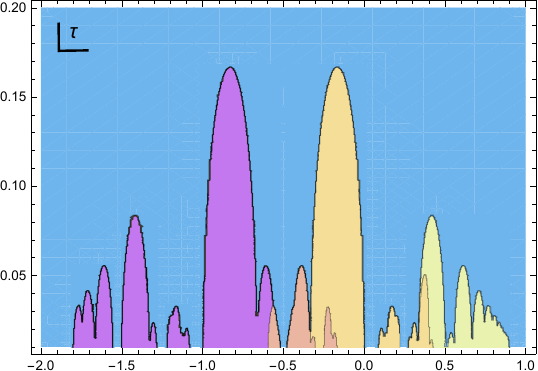}
  \caption{The phase structure for~$\t_2 \ge 0.01$. Here we see that new 
  phases appear close to the real axis, near rational points. Here we have 
  scanned~$\{(m, n)\mid m=0,\dots 10, n = -10, \; \dots, 10\}$ with~$\text{gcd}(m,n)=1$.   }
  \label{Phases2}
\end{figure}

In Figures~\ref{Phases}, \ref{Phases2}, we have plotted the phase diagram 
for~$\{(m, n)\mid m=0,\dots 10, n = -10, \; \dots, 10\}$. 
Figure~\ref{Phases} is plotted for~$\imt > 0.1$ where we only see the dominance of~$(1,0)$ 
(which dominates below the semi-circle~\eqref{circle10})\footnote{This region lies inside the region in red 
in the Figure \ref{NonExtremalBHs} to the right which means that there is a supersymmetric (but not necessarily extremal) 
black-hole-like solution that exists in supergravity).} and~$(1,1)$
(which dominates below a semi-circle isomorphic to~\eqref{circle10} but translated to the left by~$\frac23$)
apart from the ones that dominate as~$\imt \to \infty$ (i.e.~$(m+n) \, \text{mod} \, 3 = 0$).\footnote{A partial 
phase diagram for $\mathcal{N}=4$ SYM was first presented in~\cite{Benini:2018ywd}. 
It focused on the subset of $(1,r)$ configurations, with $r\in\mathbb{Z}$ which is equivalent to our Figure~\ref{Phases}
after mapping conventions (see Footnote \ref{MatchingDict}).}

\subsection{Cardy-like limit(s) revisited \label{sec:Cardy}}

In this subsection we revisit the Cardy-like limit at finite $N$ of the $SU(N)$ $\mathcal{N}=4$ SYM 
superconformal index~\eqref{RelInd0}, 
with $n_0=-1$. As mentioned in the introduction, this limit has been analyzed for a class of $\mathcal{N}=1$ SCFTs 
including $\mathcal{N}=4$ SYM. Here the scope is to address the problem from a different angle, 
by using the results presented so far.

First of all, it is clear from the formula~\eqref{StotRe1} that the dominant contribution to the large-$N$ 
expansion of the index~\eqref{ISaddles}
in the Cardy-like limit,~$\tau\to0$ with $\ret <0$,\footnote{The second condition is important because the sign of the 
second term in~\eqref{StotRe1}, which dominates in absolute value as~$\t \to 0$, is controlled by~$\text{sgn}(\ret)$.} 
comes from the~$(1,0)$ saddle 
which we identified with the black hole in the previous section. 
The large-$N$ action of $(1,0)$ saddle in this limit is
\be\label{Cardy10}
S_{\text{eff}}(1,0;\t)  \; \underset{\t\to 0}{\longrightarrow} \;  \frac{\pi \i N^2}{27} \biggl(\frac{1}{\t^2} + \frac{6}{\t}  \biggr) \,.
\ee

In fact one can do better and go beyond the large-$N$ limit in the Cardy-like limit. 
We start with the formulas~\eqref{KroneckerLimit}, \eqref{Bloch}, \eqref{QKron} and take the~$\t\to 0$ limit.
We obtain
\be 
\begin{split}
\log Q(z;\t) & \; \underset{\t\to 0}{\longrightarrow} \;  
\frac{1}{4\pi^2 \t^2} \, {\underset{m \in \IZ \atop m \neq 0}{\sum}} \; \frac{\e(- m z_1)}{m^3} \,, \\
 & \=  \frac{\pi \i}{3} \frac{1}{\t^2} \, B_3 \bigl(\{z_1\} \bigr) \,.
\end{split}
\ee
Similarly, for the~$P$ function we obtain, from~\eqref{KroneckerLimit}, 
\be 
\log |P(z;\t)|
  \; \underset{\t\to 0}{\longrightarrow} \;  - \frac{\pi \imt}{|\t|^2} \, B_2 \bigl(\{z_1\} \bigr) \,,
\ee
with holomorphic part 
\be 
\log P(z;\t)
  \; \underset{\t\to 0}{\longrightarrow} \;  - \frac{\pi \i}{\t} \, B_2 \bigl(\{z_1\} \bigr) \,.
\ee
From~\eqref{SPQ} we have
\be  \label{CardylikeLimitMatMod}
\begin{split}
S(u)& \; \underset{\t\to 0}{\longrightarrow} \;   \pi \i \sum_{i,j} 
\biggl( \frac{1}{3\t^2} \bigl( B_3 \bigl(\{{u_1}_{ij}\} \bigr) + 3 B_3 \bigl(\{{u_1}_{ij} +\tfrac13 \} \bigr) \bigr)  
+  \frac{1}{\t}\bigl( B_2 \bigl(\{{u_1}_{ij}\} \bigr) - B_2 \bigl(\{{u_1}_{ij} +\tfrac13 \} \bigr) \bigr)  \biggr)\,.
\end{split}
\ee
 The formula \eqref{CardylikeLimitMatMod} is precisely the one obtained in the Cardy-like limit studied in~\cite{Cabo-Bizet:2019osg}. 
We note, however, that our starting point here is different and that we do not need to 
go through the subtleties of the Cardy-like limit that were used in~\cite{Cabo-Bizet:2019osg}.  
The action~\eqref{CardylikeLimitMatMod} can be then used for the finite~$N$ answer in the Cardy-like limit 
as in~\cite{Cabo-Bizet:2019osg}, to obtain~\eqref{Cardy10}.
The advantage of our formalism is that we can calculate further corrections in that limit in a systematic way,
using the Fourier expansions (in~$\t_1$) of the Eisenstein series~\cite{DukeImamoglu}. 

The considerations in Section~\ref{sec:Saddles} allows to clarify one more aspect of the Cardy-like limit. 
In~\cite{Cabo-Bizet:2019osg} we analyzed the different saddles of the theory in the large-$N$ limit along the the real line. 
Now we can repeat that analysis for finite~$N$. 
The right-hand side of \eqref{CardylikeLimitMatMod} only depends on $u_1$, it is differentiable, and periodic with 
period $1$, and thus obeys the conditions of the Lemma in Subsection~\ref{sec:Lemma}. 
For $N$ finite and prime, its Cardy-like saddles are~$(0,0)$ and~$(0,1)$. 
Their actions (after properly dealing with the~$i=j$ modes in~\eqref{CardylikeLimitMatMod}) 
are\footnote{Here we have used the following identity that holds for~$z \in \IR$ 
and any integer $n$ with~$\text{gcd}(N,|n|)=1$ and $k>1$
\be
\sum_{i,j=1}^{N} B_k \Bigl(\bigl\{z+\frac{ n\, (i-j)}{N}\bigr\} \Bigr) \;=\; \frac{1}{N^{k-2}} B_k\bigl( \{N z \}\bigr)\,. 
\ee
}, (as~$\t \to 0$ and exact in~$N$), 
\bea\label{SpCardy}
S^{(0,0)}_{\text{eff}}(\t) &\=& \frac{\pi \i N^2}{27} \biggl(\frac{1}{\t^2} + \frac{6}{\t}  \biggr)-\frac{\pi i  N}{6 \tau }+O(\t^0)\,, \\\label{ActionSaddleMinus}
S^{(0,1)}_{\text{eff}}(\t) &\=&\frac{ \pi\i }{27} \biggl(\frac{\chi _1(N)}{N}\frac{1}{\t^2}+\frac{6}{\t}  \chi _0(N)\biggr)-\frac{\pi i  N}{6 \tau }+O(\t^0)\,.
\eea
These results are consistent with the intuition of a large-$N$ limit followed by Cardy-like limit: 
in the Cardy-like limit the~$(1,0)$ saddle approaches the $(0,0)$ and the other large-$N$ saddles approach the $(0,1)$. 
Indeed the term proportional to~$N^2$ in~\eqref{SpCardy} coincides with~\eqref{Cardy10}, 
and that in~\eqref{ActionSaddleMinus} vanishes  
as does~$S_{\text{eff}}(0,1;\t)$.
In Appendix~\ref{sec:00saddle} we expand on the role of the~$(0,0)$ saddle further.
At finite $N>2$, and in the Cardy-like limit~$\tau\to0$ with $\ret<0$, the saddle \eqref{SpCardy} dominates, 
as shown in \cite{Cabo-Bizet:2019osg} for a generic family of $\mathcal{N}=1$ SCFTs including $\mathcal{N}=4$ SYM. 
For the limit $\tau\to0$ with $\ret>0$, the solution~\eqref{ActionSaddleMinus} dominates. 
A more refined statement about this case involves the use of a higher order expansion in $\tau$ 
with $N$-dependent coefficients. We postpone such an analysis.

More generally, there exists a Cardy-like limit for all $(m,n)$ saddles 
(corresponding to a rational $-\frac{n}{m}$ with $(m+n)\,\text{mod}\,3\neq0$), which is~$\wt\t\to0$. 
From Formula~\eqref{StotRe} it follows that the~$(m,n)$ saddle is the leading configuration in that limit.
As for the~$(1,0)$ case mentioned above, the result is sensitive to the actual limiting procedure. 
As mentioned below Equation~\eqref{domaincurves2} we can define the Cardy-like limit to be~$\wt\t\to0$ 
with~$-\chi_1(m+n) \, \wt \t_1 >0$. In this limit, the real part of the effective action goes to $-\infty$. 
For $\chi_1(m+n)=+1(-1)$, the left(right) extremal curve drawn in Figure~\ref{EveryRational} defines 
a particular example of this limit.

It is interesting to recast the discussion in the above paragraph in terms of the microcanonical ensemble. 
Naively, the Cardy-like limit is one in which all the charges scale to infinity and, correspondingly,~$\t \to 0$.
However, as we saw above, there is actually a family of interesting limits in which
only~$\imt \to 0$ and~$\ret$ approaches 
a rational number. The microcanonical analog of these new limits has to do with the ambiguity in 
taking the ``large-charge limit" when there are multiple charges. In our situation, Equations~\eqref{inddeg}, \eqref{InI0rel} 
imply that~$\t$ couples to~$2J+Q$, which should scale to infinity. However, the charges~$J$ and~$Q$ are 
a priori independent and one needs to specify in what relative manner they scale to infinity. Demanding that 
the entropy is real specifies a certain relation between the charges. As we discussed in Section~\ref{sec:mnent}, there 
is a consistent relation for every~$(m,n)$ which defines a one-dimensional sector of BPS states in the charge space, 
such that the corresponding entropy is~$\mathcal{S}_{m,n}(J,Q)$.
Recalling from~\cite{Cabo-Bizet:2019osg} that the Cardy limit for~$(1,0)$ is 
given by~$Q \sim \Lambda^2$, $J \sim \Lambda^3$, 
it is easy to see that the Cardy-like limit for a given~$(m,n)$ is given simply by scaling the appropriate linear 
combinations in the same manner, i.e.~$\wt Q \sim \Lambda^2$,~$\wt J \sim \Lambda^3$.

\section{Summary and discussion \label{sec:Discussion}}

In this paper we have found a family of saddle points of the matrix model governing the index of~$\CN=4$ 
SYM on~$S^3 \times S^1$. The solution of this problem is governed by the Bloch-Wigner elliptic dilogarithm 
function which is a well-defined real-analytic function on the torus~$\IC/(\IZ\t+\IZ)$
and transforms like a Jacobi form with weight~$(0,1)$ under~$SL(2,\IZ)$ transformations of the torus.
The Bloch formula, which expresses the elliptic dilogarithm as a real-analytic Kronecker-Eisenstein series 
(a particular type of lattice sum), makes the analysis very elegant. The action at any saddle can be calculated by 
reading off the zeroth Fourier coefficient of the series along the direction of the saddle. We also develop a 
second method to calculate the action which uses a series representation of the meromorphic elliptic Gamma functions 
involved in the original definition of the SYM index. We show that deforming the original action to the Bloch-Wigner 
type function away from the real axis in this approach agrees with our first approach. Using our formalism we 
find a family of solutions labelled by~$(m,n)$ with~$\text{gcd}(m,n)=1$, and 
analyze the competition among these phases of the theory at finite~$\t$. 

The analysis of this paper relates in a natural way to the various approaches to the problem 
that have been used so far, and shows that they are all consistent with each other. 
Firstly, the contact with supergravity comes from the fact that the action of the saddle~$(1,0)$ 
is precisely the supergravity action of the black hole. We also calculate the action and entropy 
of the generic~$(m,n)$ saddle. It is an interesting question to identify the corresponding gravitational 
configurations. Secondly, the calculations in the Cardy-like limit get clarified and simplified 
as the limit~$\t \to 0$ of the finite-$\t$ answer. The formalism is powerful and can be used to calculate 
corrections from the Cardy-like limits. 
Thirdly, there seems to be a relation to the Bethe-ansatz type analysis which is intrinsically a meromorphic analysis.
We find that the action of the original mermomorphic elliptic gamma function at the Bethe roots agrees with the 
action at the saddle-points derived from our approach. This makes it clear that the Bethe roots should indeed be identified 
with Euclidean saddle points of the matrix model. 

The relation between the meromorphic approach and our real-analytic approach remains experimental 
and it would be nice to understand it at a deeper level. 
In particular, the global analysis of the contour deformations and of the contributions of the 
saddle-points at different points in moduli space remains to be done.
Towards this end it may be important to understand the meromorphic function~$\log Q$ 
(as opposed to just the real part~$\log |Q|$) better. It seems to us that this should be related to a 
multi-valued function studied in~\cite{Levin,BrownLevin}. The meromorphic function in question 
discussed in~\cite{Levin,BrownLevin} satisfies a certain first-order differential equation in~$\t$ (it is the primitive of 
a holomorphic Kronecker-Eisenstein series), and trying to find such an equation may be one way to completely 
fix the phase ambiguities that we discussed in this paper. It is interesting to note that there is also a version 
of the dilogarithm~\cite{Nahm:1992sx, ZagierGangl, Nahm:2004ch} which seems to be closer to the 
original elliptic gamma function.

The elliptic dilogarithm (in its real-analytic as well as its meromorphic avatars) has been understood in recent years to
play a crucial role in string scattering amplitudes (see e.g.~\cite{DHoker:2015wxz,Broedel:2018izr}). 
It would be remarkable if there is a direct physical relation between
these two appearances. One natural speculation along these lines would be about a \emph{holographic} relation,
with the string amplitudes (perhaps suitably modified) being observables in a bulk theory dual to the SYM  
theory that we discuss in this paper. A less striking, but nevertheless interesting, possibility would be the 
following syllogism: the appearance of the same elliptic dilogarithm in string scattering amplitudes and
the partition function of~AdS$_5$ black holes could be compared with the appearance of the theta function
in string amplitudes and the partition functions of black holes in flat space.

Our analysis also clarifies the role of modular symmetry in this problem. 
The SYM index itself is not a modular form, instead the Bloch-Wigner function is essentially
a period function for the real part of the action of the SYM. More precisely, writing the action as 
a simple linear combination of the functions~$\log P$ and~$\log Q$ (see Equation~\eqref{SPQ}), 
we have that~$\log |P|$ is Jacobi invariant while the Bloch-Wigner dilogarithm is a period function 
of~$\log |Q|$ (see Equation~\eqref{QtlFrel}). 
In particular, this explains the fact that the SYM action on~$S^1 \times S^3$ as~$\t \to 0$ has 
a~$\t^{-2}$ behavior in contrast with the~$\t^{-1}$ behavior of holomorphic modular forms appearing 
in theories on~$S^1 \times S^1$.

There are clearly many interesting things to do. 
For one, our analysis can be enlarged to include 4d~$\CN=1$ SCFTs. In another direction,
these ideas could lead to progress in similar problems in other dimensions. 
It would be nice to find the correct mathematical structures to extend the ideas in this paper 
away from the restriction~$\t=\s$.
At a technical level, it would be nice to find a direct physical interpretation of the elliptic deformation in the complex plane of the 
SYM index, the obvious thing to look for would be a supersymmetry-exact operator. 
Similarly, it would be nice to find a direct relation to the Bethe-ansatz approach, for which
there are clues mentioned in this paper. 

Finally, it is tempting to speculate that deeper number-theoretical aspects of the functions appearing here play
a role in physics. 
The original motivations of Bloch in uncovering his beautiful formulas arose from algebraic K-theory.
Relatedly, the values of the elliptic dilogarithm at particular (algebraic) values of~$\t$ are closely related to values of 
certain L-functions of Hecke characters,
and the so-called Mahler measure~\cite{Villegas}. 
It would be really interesting if the physics of SYM and~AdS$_5$ black holes is directly related to these objects, 
perhaps in a manner that extends the beautiful ideas described in~\cite{Moore:1998pn}.

\section*{Acknowledgements}
It is a pleasure to thank Dionysios Anninos, Francesco Benini, Francis Brown, Davide Cassani, Atish Dabholkar, 
Yan Fyodorov, Stavros Garoufalidis, Mahir Hadzic, Dario Martelli, Oliver Schlotterer, Don Zagier for useful and interesting discussions.
This work was supported by the ERC Consolidator Grant N.~681908, ``Quantum black holes: A macroscopic window into the 
microstructure of gravity'', and by the STFC grant ST/P000258/1.

\appendix

\section{Some special functions and their properties \label{App:theta}}

\ndt {\bf Bernoulli polynomials}
The Bernoulli polynomials $B_{k}(z)$ are defined through the following generating function,
\bea
\frac{t\, \rme^{z \,t}}{ (\rme^{t}-1)}\;=\;\sum_{k=0}^{\infty} B_{k}(z)\, \frac{t^k}{k!}\,.
\eea
 The first three Bernoulli polynomials are
 \bea
 B_1(z) &\=& z-\frac{1}{2}\,,\\
 B_2(z) &\=& z^2-z+\frac{1}{6}\,,\\
 B_3(z) &\=& z^3-\frac{3 \,z^2}{2}+\frac{z}{2}\,.
 \eea
These polynomials obey the following properties for $z\in \mathbb{C}$, 
 \bea
 B_k(z) \= (-1)^{k}B_k(z-1)\,, \\
 B_k(z\,+\,1)-B_k(z)\=\,k \,z^{k\,-\,1} \,. \label{BernPeriod}
 \eea
Their Fourier series decomposition, for~$k\ge 2$ and~$0 \le x<1$ is 
\be
B_k(x) \= - \frac{k!}{(2\pi \i)^k} \, \sum_{j \neq 0} \frac{\e(j x)}{j^k} \,.
\ee
Clearly this also equals~$B_k(\{ x\})$ where $\{x\}\coloneqq x-\lfloor x \rfloor$ is the fractional part of $z$.
In particular we have for~$x \in \mathbb{R}$
\bea
B_2(\{x\}) & \= & \frac{1}{2 \pi^2} \, \sum_{j \neq 0} \frac{\e(jx)}{j^2} \,, \label{B2Four} \\
B_3(\{x\}) & \= & - \frac{3\,\i}{4\pi^3} \, \sum_{j \neq 0} \frac{\e(jx)}{j^3} \,. \label{B3Four} 
\eea

\paragraph{Multiple Bernoulli Polynomials}

The multiple Bernoulli Polynomials $B_{r,n}(z |\, \underline{\omega})$ are defined  
for $z\in\mathbb{C}$, $\underline{\omega}=(\omega_{r-1},\ldots,\omega_0)$ with $\omega_j\in \mathbb{C}-{\{0\}}$ 
through the  following generating function~\cite{10019542457,Narukawa}
(we follow the conventions of~\cite{Narukawa}):
\bea
\frac{t^r e^{z \,t}}{\prod_{j=0}^{r-1} (e^{\omega_j\, t}-1)}\= \sum_{k=0}^{\infty} B_{r,k}(z|\,\underline{\omega}) \, \frac{t^k}{k!}\,.
\eea
The first three cases are
\bea
B_{1,1}(z|\,\omega_0)&\=&\frac{z}{\omega_1}-\frac{1}{2} \,, \\
B_{2,2}(z|\,\omega_1,\,\omega_0)&\=&\frac{z^2}{\omega _0\, \omega _1}-\frac{\left(\omega _0+\omega _1\right) z}{\omega
   _0\, \omega _1}+\frac{\omega _0^2+3\, \omega _0 \, \omega _1+\omega _1^2}{6 \,\omega _0\, \omega
   _1} \,,
\\\nonumber
B_{3,3}(z|\,\omega_2,\,\omega_1,\,\omega_0)&\=&
\frac{\left(z+\frac{1}{2} (-{\omega_0}-{\omega_1}-\omega_2)\right)^3}{{\,\omega_0} {\,\omega_1} {\, \omega_2}}\nonumber\\
&& -\frac{\left({\omega_0}^2+{\omega_1}^2+{\omega_2}^2\right)
   \left(z+\frac{1}{2} (-{\omega_0}-{\omega_1}-{\omega_2})\right)}{4
   {\,\omega_0} {\,\omega_1} {\,\omega_2}} \,.\label{B33def}
\eea
The multiple Bernoulli polynomials are symmetric under permutations of components in~$\underline{\omega}$. 
They obey the following periodicity property
\be\label{propBnn}
B_{k,k}(z\,|\,\underline{\omega}[j]) \= - B_{k,k}(z+\omega_j\,|\,\underline{\omega}) \,,
\ee
where $\omega[j] \coloneqq \left(\omega_{k-1},\ldots, -\omega_{j},\ldots,\omega_0 \right)$.
Moreover, they can be expanded in terms of usual Bernoulli polynomials. 
In particular, we have
\bea\label{Decomp1}
B_{1,1}(z-1\,|\,-1)&\=& -\,B_{1}(z) \,,\\\label{Decomp2}
B_{2,2}(z-1\,| \,\omega_1,-1) &\=& -\frac{1}{\omega_1 }\, B_2(z)\,+\,B_1(z)\,-\,\frac{\omega_1 }{6}\,,\\\label{Decomp3}
B_{3,3}(z-1\,|\,\omega_2,\, \omega_1,-1) &\=&-\,\frac{1}{\,\omega_1\,  \omega_2 }\,B_3(z)\,+\,
\frac{3\,  (\omega_1 +\omega_2 ) }{2 \,\omega_1 \, \omega_2} \, B_2(z) \nonumber \\ 
&&\quad -\,\frac{1}{2} \left(\frac{\omega_1 }{\omega_2 }\,+\,\frac{\omega_2 }{\omega_1}
	+3\right)  B_1(z)\,+\,\frac{\omega_1 +\omega_2 }{4} \,.
\eea
One has similar formulas which express $B_{n,n}(z|\,\underline{\omega})$  with $\omega_0=1$ in terms of usual 
Bernoulli polynomials, to obtain those, one uses \eqref{Decomp1},  \eqref{Decomp2}, \eqref{Decomp3} and applies 
the property~\eqref{propBnn} on the corresponding left-hand sides. The above equation will be convenient to 
analytically-match the results of the action in 
Appendix~\ref{App:AnalyticAction}  and Section~\ref{sec:Action}.

\paragraph{Theta functions}
The odd Jacobi theta function is defined as (with $q=\e(\t)$, $\z = \e(z)$)
\be
\vth_1(z) \= \vth_1(z;\t) \defeq -\i\sum_{j\in \IZ} \, (-1)^j \, q^{\half(j+\half)^2} \z^{(j+\half)}  \,.
\ee
In the main text we use the following function
\be  
\th_0(z;\t) \defeq -\z^\half \, q^{\frac{1}{24}}  \, \frac{\vth_1(z,\t)}{\eta(\t)} \=  (1-\z) \prod_{j=1}^\infty (1-q^j \z) \, (1-q^j \z^{-1}) \,.
\ee
The function~$\th(z)$ is holomorphic and has simple zeros at~$z = j \t + \ell$, $j,\ell \in \IZ$.
The following elliptic transformation properties follows easily from the definition,
\bea
\theta_0(z+1;\,\tau) &\= & \theta_0(z;\,\tau) \,, \label{thellprop1} \\
\theta_0(z+\tau;\,\tau) &\= & -\e (-z) \, \theta_0(z) \= \theta_0(-z;\,\tau) \,. \label{thellprop2}
\eea

\paragraph{Modular properties of~$\th_0$ and~$\Ge$} 
In the main text we also introduced the elliptic gamma function in~\eqref{GammaeDef} which obeys the quasi-periodicity 
property~\eqref{Geperiod1}. 
The function~$\Ge(z,\t,\s)$ is meromorphic, it has simple zeros at~$z=(j+1)\t+(k+1)\s+\ell$ and simple poles at~$z=-j\t-k\s+\ell$,
$j,k\in \mathbb{N}_0$, $\ell \in \IZ$.
Next, we introduce a set of identities that involve modular transformations of parameters~$\tau$,~$\sigma$ and~$z$.
Recall that both~$\th_0$ and~$\Ge$ are invariant under the transformation~$z\mapsto z - k_0$ for~$k_0 \in \mathbb{Z}$.
We will combine this symmetry with the modular transformation properties to obtain new identities. 
For $\theta_0$ we have, for~every~$k_0 \in \IZ$,
\be\label{Modth}
\theta_0(z;\tau) \=\exp\bigl(\pi \i B_{2,2}(z-k_0|\tau,-1)\bigr) \, \theta_0 \biggl(\frac{z-k_0}{\tau };-\frac{1}{\tau } \biggr).
\ee
Using~\eqref{thellprop1}, \eqref{thellprop2}, we can also write this as  
\be\label{Modth2}
\theta_0(z;\tau)\=\exp\bigl(-\pi \i B_{2,2}(z-k_0|\tau,1)\bigr) \, \theta_0 \biggl(-\frac{z-k_0}{\tau };-\frac{1}{\tau } \biggr).
\ee
The analogous identities for elliptic Gamma functions are, for $\text{Im}(\frac{\sigma}{\tau})>0$,~$k_0 \in \IZ$, which are 
minor variations of the ones presented in
\cite{Felder,Narukawa},
\be \label{TripleId}
\Ge(z;\tau,\sigma) \= \exp \biggl(\frac{\pi i}{3} B_{3,3}(z-k_0|\tau,\,\sigma,-1) \biggr) \,  
\frac{\Gamma \Bigl(\frac{z-k_0}{\tau};-\frac{1}{\tau},\frac{\sigma}{\tau} \Bigr)}{ 
\Gamma \Bigl(\frac{z-k_0-\t}{\sigma};-
\frac{1}{\sigma},-\frac{\tau}{\sigma} \Bigr)}\,,
\ee
\be \label{TripleId2}
\Ge(z;\tau,\sigma)\= \exp \biggl(-\frac{\pi i}{3} B_{3,3}(z-k_0|\tau,\sigma,1) \biggr) \,  
\frac{\Gamma \Bigl(-\frac{z-k_0}{\sigma};
	-\frac{1}{\sigma},-\frac{\tau}{\sigma} \Bigr)}{
		\Gamma \Bigl(-\frac{z-k_0-\s}{\tau};-\frac{1}{\tau},\frac{\sigma}{\tau} \Bigr)}\,.
\ee

\subsection*{Factorization properties}

For our purposes it will be important to use the following ``factorization" identity of $\theta_0$. 
For all $m\in \mathbb{N}$ we have \footnote{This identity can be proven with the use of the representation
\eqref{ExpRepTh} and basic trigonometric identities.}
\be\label{facttheta0}
\theta_0(z;\,\tau)\= \prod_{\ell_1=0}^{m-1} \theta_0(z+\ell_1 \tau;\, m \tau) \,,  
\ee
Upon using this factorization identity, the symmetry of~$\theta_0$ under integer shifts of its second argument, 
and the modular identity~\eqref{Modth}, one obtains, for any~$k_0(\ell)\in\IZ$, $\ell = 0,\dots, m-1$,
\be\label{Facttheta2}
\theta_0(z;\,\tau)\=\prod_{\ell=0}^{m-1}\exp \bigl(\pi \i \,B_{2,2}(z-k_0(\ell)+\ell \tau |  m\t+n,-1) \bigr)   \, 
\theta_0 \biggl(\frac{z-k_0(\ell)+\ell\tau}{m\t+n};-\frac{1}{m\t+n } \biggr) \,.
\ee

We also have similar factorization identities of $\Ge$~\cite{Felder} \footnote{This identity can be proven with the use of the representation
\eqref{ExpRepG} and basic trigonometric identities.}
\be\label{FactGamma}
\begin{split}
\Ge(z;\tau,\tau)
&\=\prod_{\ell_1,\ell_2=0}^{m-1} \Ge(z+(\ell_1+\ell_2)\tau\,; \,m\, \tau,\,m\, \tau)\,,  \\
&\=\prod_{\ell=0}^{2 (m-1)} \Ge(z+\ell \tau \,;\, m\, \tau,\,m\, \tau)^{m-|\ell-m+1|}\,.
\end{split}
\ee
Upon using this factorization identity, the symmetry of $\Ge$ under integer shifts of its last two arguments, 
and the appropriate limit of modular identity \eqref{TripleId}, 
one obtains, with $\tau^\varepsilon \=\tau\,+\,\varepsilon$, $\varepsilon \rightarrow 0$,
and for any~$k_0(\ell)\in\IZ$, $\ell = 1,\dots, m$,
\be \label{FactElliptic}
\begin{split}
\Ge(z;\,\tau,\,\tau)
\=& \prod_{\ell=0}^{2(m-1)}\exp \biggl(\frac{\pi \i}{3} \,B_{3,3}(z-k_0(\ell)+\ell\tau|m\t+n,m\t+n,-1) \biggr)  \\
&\qquad\qquad\times\, \frac{\Gamma_e \Bigl(\frac{z-k_0+\ell\t}{m\tau+n};-\frac{1}{m\tau+n},
\frac{m\tau^\varepsilon+n}{m\tau+n} \Bigr)}{ 
 \Gamma_e \Bigl(\frac{z-k_0\,+\,\ell\t\,-\,(m\t+n)}{m\tau^\varepsilon+n};-
\frac{1}{m\tau^\varepsilon+n},-\frac{m\tau+n}{m\tau^\varepsilon+n} \Bigr)}\,.
\end{split}
\ee

\section{Elliptic extension: zeroes and poles \label{App:zeropole}}

In this appendix we comment on the singularities of the elliptically extended action that was introduced in Section~\ref{sec:ellext}. 
The integrand of the SYM index~\eqref{SYMIndex1again} 
can be rewritten in terms of the function~$Q_{a,b}$ which was introduced in~\eqref{Qab} as, for $u\in \mathbb{C}$,
\be\label{DefCalQ}
\mathcal{Q}(u) \defeq \bigl|{Q}_{1,0}(u) \, {Q}_{1,0}(-u) \bigr|^{-1} \,
\bigl|{Q}_{-\frac{1}{3},\frac{1}{3}}(u) \, {Q}_{-\frac{1}{3},\frac{1}{3}}(-u) \bigr|^{-3} \,.
\ee
Here the ${Q}_{1,0}$ factors arise from the vector multiplet, and the ${Q}_{-\frac{1}{3},\frac{1}{3}}$ 
factors arise from the chiral multiplets.

\begin{figure}[h]\centering
  \includegraphics[width=5.8cm]{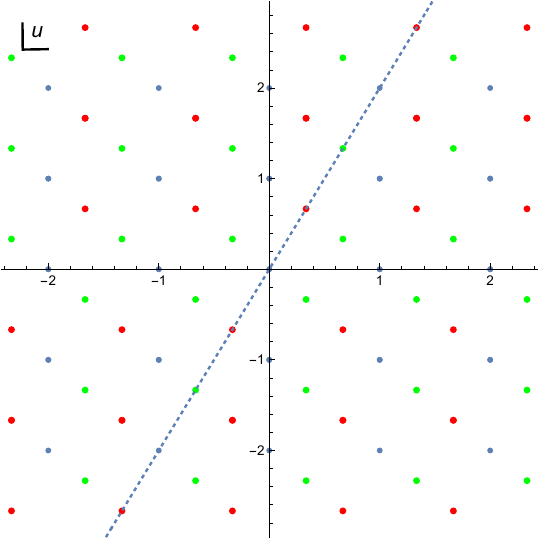}\qquad\qquad
  \includegraphics[width=4.5cm]{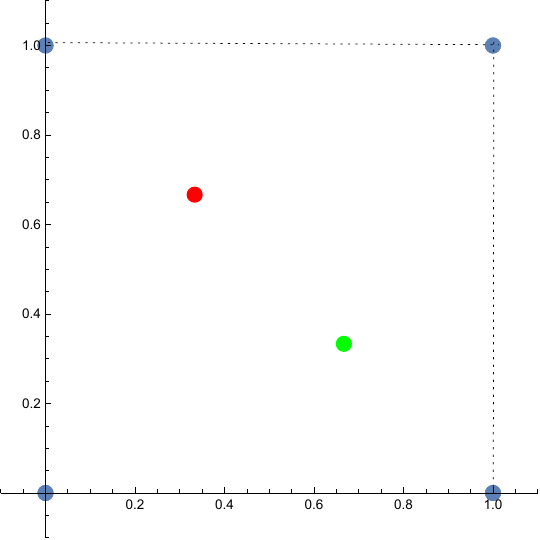}
  \caption{The zeroes and singular points of $\mathcal{Q}$. Blue points are zeroes of the vector multiplet contribution. 
  Red and green points are singularities of the chiral multiplet contributions. The red ones are 
  singularities of the undressed elliptic Gamma functions, while the green ones are singularities that come from 
  the dressing factor $P(z)^{z_2}$ in equation \eqref{defQtl} (see also \eqref{Qab}). 
  Both singularities are of order one and the zeroes are of order two. 
  The horizontal unit stands for~$1$ and the vertical unit stands for~$\tau$.}\label{figZeroesPoles1}
\end{figure}

The action has singularities whenever the function~$\mathcal{Q}$ has singularities or zeros.
In Figure \ref{figZeroesPoles1} we present the positions of zeros (blue) and singularities (red and green) 
of~$\mathcal{Q}$ in the $u$-plane. The zeros come from the vector multiplet, the singularities come from 
the chiral multiplets. 
As $\bigl|{Q}_{-\frac{1}{3},\frac{1}{3}}(u) \, {Q}_{-\frac{1}{3},\frac{1}{3}}(-u) \bigr|^{-1}$ is elliptic, 
the full set of zeros and poles are lattice translates of those in the fundamental domain $0\leq u_1<1$ and $0\leq u_2<1$.
In this domain,  $|\G_e(u+\frac{2}{3}\t+\frac{1}{3})\G_e(-u+\frac{2}{3}\t+\frac{1}{3})|$ has a single pole 
at $\frac{2\t+1}{3}$ and it has no zero, 
and $|P( u -\frac{1}{3}\t+\frac{1}{3})|^{u_2}$ has a zero of order $\frac{1}{3}$ at~$\frac{\t+2}{3}$
and it has no singularity.  
Further, in this domain, $|P( -u -\frac{1}{3}\t+\frac{1}{3})|^{-u_2}$ has a singularity of order $\frac{2}{3}$ at~$\frac{2\t+1}{3}$ and it has no zero. 
It thus follows that  $\bigl|{Q}_{-\frac{1}{3},\frac{1}{3}}(u) \, {Q}_{-\frac{1}{3},\frac{1}{3}}(-u) \bigr|^{-1}$  has two singularities of 
order~$\frac{1}{3}$, located at the green and red points, respectively, in Figure~\ref{figZeroesPoles1}.  
An analogous analysis shows that the only zeroes of \eqref{DefCalQ} in the fundamental domain 
come from the vector multiplet contribution, i.e.~from the first factor in right-hand side of \eqref{DefCalQ} (the blue points in the figure).

For configurations that pass through the zeroes or singularities of $\mathcal{Q}$, 
the effective action~$S(u)$ is $+\infty$ or $-\infty$, respectively. 
Indeed, the original contour of integration defining the SYM index crosses the zeroes of $\mathcal{Q}$. 
In order to deal with this issue, one can turn on a regulator~$\epsilon$, following the prescription introduced in~\cite{Cabo-Bizet:2018ehj},
(and which will be recalled in the following section). 
This regulator shifts the position of zeroes and singularities in the~$u$-plane by an amount proportional to~$\epsilon$ 
in a direction parallel to the dashed line,  
thus making the contour integrals that define the actions of saddles~$(0,1)$ and~$(1,0)$ well-defined\footnote{The double zero at the origin  
splits into two single zeroes located at distances~$\pm \frac{\epsilon}{2}$ from the origin along the dashed line 
in Figure \ref{figZeroesPoles1}.}. 
We note that this regulator does not work if the saddle is along the direction
$2\t+1$ (indicated by the dashed lines in Figure~\ref{figZeroesPoles1}). 
For this saddle a different regularization must be used. For instance, one such possibility is to turn on 
flavour fugacities that shift the position of the  zeroes and singularities off the dashed line, and in the end take the fugacities to zero.

\section{Series representations of $\log \theta_0$ and $\log \Ge$ \label{sec:ExpRep}}

In the main text we calculated the effective action as integrals of elliptic functions constructed out of~$\log \th_0(u)$ and~$\log \Ge(u)$,
along various directions in the complex~$u$-plane, using the double Fourier expansion in~$u$ provided by 
the Bloch formula (see the discussion in Section~\ref{subsec:Action}). 
In the first two subsections we develop an alternative method to integrate ~$\log \th_0(u)$ and~$\log \Ge(u)$, using certain series 
expansions that will be called \emph{$(m,n)$~representations}. 
We will often use the notion of a \emph{$(m,n)$ ray in the $u$-plane}, 
by which we mean a ray emanating from the origin, in direction of the  
vector $m\t+n$ with $m \in \mathbb{N}$ and~$n\in \mathbb{Z}$
In the last subsection we study the definition of the elliptic extensions $Q$ from the perspective of $(m,n)$ representations, 
thus obtaining a series representation for the even (in $u$) product of $Q$'s. 
We end by briefly commenting upon an ambiguous overall phase function that appears in the definition of $Q$. 

\subsection*{Basic series representations}

We start by presenting the representations that are building blocks of our forthcoming analysis. 
For $\theta_0$ that representation is \cite{Felder}
\be\label{ExpRepTh}
\log \theta_0(z;\tau) \= -\i \, \sum_{j=1}^\infty\frac{\cos \bigl( j \pi (2 z -\tau) \bigr)}{j \sin{\pi j \tau}} \,,
\ee
where the series on the right-hand side is absolutely convergent for
\be
 0<\text{Im}(z)<\text{Im}(\tau) \,. 
\ee
The analogous representation for elliptic Gamma functions is \cite{Felder} 
\be\label{ExpRepG}
\log \Ge(z;\tau,\sigma) \=
 -\frac{\i}{2} \, \sum_{j=1}^\infty\frac{\sin \bigl(j \pi (2z -\tau-\sigma)\bigr)}{j \sin{\pi j \tau}\sin{\pi j \sigma}}  \,,
\ee
where the series on the right-hand side is absolutely convergent for 
\be\label{DomainGammaE}
  0<\text{Im}(z)<\text{Im}(\tau)+\text{Im}(\sigma) \,. 
\ee

\subsection*{Some notation}

In the presentation that follows we use the notation
\be \label{zpp}
z \= z_{\perp \t} + z_{\parallel \t} \t 
\ee
for the real components of any complex number~$z$ along~$1$ and~$\t$.
This is very similar to the notation~$z=z_1+z_2 \t$ that we use in the main text, the difference being that here in~\eqref{zpp} 
we consider~$\t$ to be variable, and so in particular we can replace~$\t$ by~$m\t+n$. We can write the components explicitly,
for any~$\t$ and~$z$, as
\be
z_{\parallel \t}  \= \frac{\Im \, z}{\imt} \,, \qquad z_{\perp \t} \= z-z_{\parallel \t} \t \,.
\ee
It is clear from definition \eqref{zpp} that
\be \label{zkperp}
(z+k)_{\perp \t} \= z_{\perp \t} +k \,,  \qquad  (kz)_{\perp \t} \= k z_{\perp \t}  \,, \qquad k \in \IR \,.
\ee

%
%

\vspace{0.4cm}
\noindent {\bf The $(m,n)$ representations of~$\log \th_0$}
\vspace{0.1cm} 

\noindent The series representation~\eqref{ExpRepTh} implies the following formula, 
\be  \label{th0repmod10}
\log  \theta_0 \Bigl(\frac{z-k_0}{\tau};-\frac{1}{\tau} \Bigr) \= \i \, \sum_{j=1}^\infty  
 \frac{ \cos  \bigl( \pi j\, \frac{2(z-k_0) +1}{\tau} \bigr)}{j \sin{\frac{\pi j}{\tau}}}   \,,
\ee
which converges in the region 
\be
 0<\text{Im}\Bigl(\frac{z-k_0}{\tau} \Bigr)<\text{Im} \Bigl(-\frac{1}{\tau} \Bigr) \,. 
\ee
Expanding~$z$ as in~\eqref{zpp}, we see that this convergence condition is  equivalent to 
\be  \label{k0reg}
0< (-z+k_0)_{\perp \tau}<1 \quad \Longrightarrow \quad 0< -z_{\perp \tau} +k_0 <1\,,
\ee
where we have used~\eqref{zkperp} to reach the second set of inequalities.  For $k_0 \in \mathbb{Z}$
the inequalities~\eqref{k0reg} are solvable if and only if~$z_{\perp \tau}\notin\mathbb{Z}$.
For fixed~$k_0$ the region in the~$z$-plane given by~\eqref{k0reg} is an infinite ribbon 
parallel to, and enclosing, the ray~$\t$. The unique value 
of~$k_0$ that satisfies this condition is, 
\be \label{k0zperp}
k_0 \= \lfloor z_{\perp \tau} \rfloor +1= -\lfloor -z_{\perp \tau} \rfloor \,.
\ee
The second equality in \eqref{k0zperp} follows from the fact that $z_{\perp \tau} \notin \mathbb{Z}$.

Now, combining the formula~\eqref{th0repmod10} with the modular transformation~\eqref{Modth} we obtain,
in the region~\eqref{k0reg},
\be \label{seriesth10}
\log \theta_0(z;\tau) \=  
\pi\, \i \, B_{2,2}(z-k_0|\tau,-1) +  \i \, \sum_{j=1}^\infty  
 \frac{ \cos  \bigl( \pi j\, \frac{2(z-k_0) +1}{\tau} \bigr)}{j \sin{\frac{\pi j}{\tau}}} \,.
\ee
In exactly the same manner
we can combine the basic series representations with the factorization formulas~\eqref{Facttheta2}
to obtain, for any given~$(m,n)$, 
\be \label{logthcdrep}
\begin{split}
\log \theta_0(z;\tau) & \=  \pi \,\i \, \sum_{\ell=0}^{m-1} \,B_{2,2}(z-k_0(\ell,z)+\ell \tau |  m\t+n,-1) \, + \cr
& \qquad\qquad +\,  \i \, \sum_{j=1}^\infty  
 \frac{1}{j \sin{\frac{\pi j}{ m\t+n}}} \,  \sum_{\ell=0}^{m-1}   \cos  \Bigl( \pi j\, \frac{2(z-k_0(\ell,z)+\ell \tau) +1}{ m\t+n} \Bigr) \,,
\end{split}
\ee
which converges when the following condition is satisfied for $\ell = 0, \dots, m-1$,
\be  \label{k0regcd}
0< \bigl(-z+k_0(\ell,z) \bigr)_{\perp m\t+n}<1 \, 
\quad \Longleftrightarrow \quad 
k_0(\ell,z) \=- \lfloor (-z-\ell\t)_{\perp m\t+n} \rfloor \,. 
\ee
Representations \eqref{logthcdrep} will be called $(m,n)$ representations of $\log \theta_0(z)$.

In what follows we will encounter the combination 
\be\label{twoterms}
 u_2 \log \theta_0(z_I(u); \tau) - u_2  \log \theta_0(z_I(-u); \tau) \,,
\ee
where the index~$I$ stands for either chiral $(C)$ or vector $(V)$ multiplets, and 
\be \label{zIrI}
z_{I}(u)\= u +\frac{r_I}{2}  (2\tau+1) \,,
\ee
with~$r_I$ the R-charge of multiplet~$I$. 

Next, we will use representations \eqref{logthcdrep} in both terms in \eqref{twoterms}, for which the choice of $k_0$ is
\be\label{k0upm}
k_0(\ell,z_I(\pm u)) 
\=- \Big\lfloor\,\mp u_{\perp m\tau+n}\,+\,\frac{d}{c}\left(r_I+\ell\right)-\frac{r_I}{2}\,\Big\rfloor\,,
\ee
when the argument of the function $\lfloor\,\,\rfloor$ is  not an integer. 
We note that in the domain  
\bea \label{udomain}
|u_{\perp m\tau+n}| < \kappa(r_I,\,\ell,\,m,\,n) \,,
\eea 
with
\bea
\kappa(r_I,\,\ell,\,m,\,n)=\text{min}\Big(\Big\{\frac{n}{m}\left(r_I+\ell\right)-\frac{r_I}{2}\Big\},\, 
1-\Big\{\frac{n}{m}\left(r_I+\ell\right)-\frac{r_I}{2}\Big\}\Big) \,,
\eea
and for
\bea
\frac{n}{m}\left(r_I+\ell\right)-\frac{r_I}{2} \,\notin\, \mathbb{Z} \, ,
\eea
the functions $k_0(\ell,z_I(u))$ and $k_0(\ell,z_I(-\,u))$ 
are both equal to
\be\label{k0Value}
k_0(\ell)\=- \Big\lfloor\,\frac{n}{m}\left(r_I+\ell\right)-\frac{r_I}{2}\,\Big\rfloor\,,
\ee
and, in particular, they are independent of~$u$.

Each value of R-charge~$r_I$ defines the width of the ribbon-like domain of absolute convergence \eqref{udomain}. 
The middle axis of this ribbon passes through the origin of $u$-plane, and is parallel to the vector $m\tau+n$. 
For a theory with various building-block multiplets, we linearly combine the $(m,n)$ representations of \eqref{twoterms}. 
The domain of absolute convergence of the $(m,n)$ representation of such linear combinations is typically determined 
by the R-charge of one of the building-block multiplets.  
For instance, we can ask for the domain of absolute convergence of the $(c,0)$ representation of a linear combination 
of terms like \eqref{twoterms}, associated to $\mathcal{N}=4$ SYM, for which the spectrum of R-charges is $r_V\=2$ 
and $r_C=\frac{2}{3}$.  For that case, from equality $\kappa(r_V,\ell,m,0)=0$ and \eqref{udomain}, 
it follows that the domain of absolute convergence of the latter linear combination of $(m,0)$ representations is \emph{empty}. 
To deal with this issue we use the infinitesimal regulator $\epsilon=0^+$ that was introduced in~\cite{Cabo-Bizet:2019osg} 
and mentioned in Appendix \ref{App:zeropole}. 
 For the vector and chiral multiplets, such regulator is introduced as the following infinitesimal deformation of the 
 R-charge spectrum,  $r_V=2-\epsilon$ and $r_C=\frac{2+\epsilon}{3}$. In that case, the $(c,0)$-representation 
 of the corresponding linear combination of \eqref{twoterms} is absolutely convergent in 
 $|u_{\perp c \tau}|<\frac{\epsilon}{2}=0^+$, i.e., in a ribbon of infinitesimal width $\epsilon$ directed along the~$(1,0)$-ray.

To summarize, after using the formula~\eqref{logthcdrep} we obtain the following absolutely convergent
series representation in the domain \eqref{udomain} and for the choice of $k_0(\ell)$ given in~\eqref{k0Value},
\be \label{ulogthseries}
\begin{split}
& u_2 \, \log \theta_0(z_I(u); \tau) - u_2 \, \log \theta_0(z_I(-u); \tau) \\
& \qquad\qquad\qquad \qquad  \=  \,2\, \mathcal{F^{\,'}}^{(m,n)}(u)
+  \sum_{j=1}^\infty\, \frac{2}{j} \, c^{\prime}_{2}(j;m,n) \, 
u_2 \sin{\frac{2 \pi\, j\, u}{m\t+n}}   \,.
\end{split}
\ee
Here the pre-factor $\mathcal{F^{\,'}}^{(m,n)}(u)$ is
\be \label{FcdTheta}
2\,\mathcal{F^{\,'}}^{(m,n)}(u) \= \pi \,\i\,  \sum_{\ell=0}^{m\,-\,1}\, 
\Bigl(u_2\,B_{2,2}(z_I(u)-k_0(\ell)+\ell \tau|\, m\t+n,-1)+ \left(u \rightarrow -u\right)\Bigr) \,,
\ee
and
\be \label{cnTheta1}
c^{\prime}_{2}(j;m,n) \= -\,\i\,  \sum_{\ell=0}^{m-1}
\frac{\sin{\frac{\pi\, j\, ( {r_I} (2\tau+1)+1-2k_0(\ell)+2\ell \tau )}{m\tau+n}}}{\sin \frac{\pi j}{m\tau+n}} \,.
\ee

From an explicit expansion of \eqref{FcdTheta} for several values of $m$, $n$, and R-charge $r_I$, 
we have checked that for values of $u$ in an $(m,n)$ ray i.e~for $u=\frac{u_2}{m} (m\t+n )$ the quantity
\be\label{auxeq1}
2\,\mathcal{F^{\,'}}^{(m,n)}(u)\,+\, 2 \pi  \i \,( 2 r_I-1)\,u_2^2 \, \tau 
\ee
is purely imaginary and is independent of~$\tau$. 
After summing \eqref{auxeq1} over the matter content of $\mathcal{N}=4$ SYM, 
and taking the regulator $\epsilon$ to zero, one obtains
\be\label{EqScd1}
\sum_{I}\,2\,\mathcal{F^{\,'}}^{(m,n)}(u)\,+\, 8 \pi\i \,u_2^2 \, \tau \=  2\pi\i\, u_2^2 \, \eta' (m,n)\,,
\ee
where~$\eta'$ is a real function.
The index $I$ runs over the labels of vector and three chiral multiplets of $\mathcal{N}=4$ SYM, 
each of them with their corresponding charge asignment $r_I$. We have checked the following 
three properties in numerous examples (not proven):  
(a) that the coefficient~$\eta'(m,n)$ is rational, (b) that it depends only on the ratio~$n/m$, 
and (c) that $m^2\,\eta'(m,n)\in \mathbb{Z}$ (see Table~\ref{tablaKappacd}).  
\begin{table}[h]
\begin{center} 
\begin{tabular}{||cccccccccccc||}\hline $m \, \backslash\, n$  &$0$&$1$&$2$&$3$&$4$&$5$&$6$&$7$&$8$&$9$&$10$ \\ \hline\hline
$1$&$ 0 $&$ -4 $&$ -10 $&$ -10 $&$ -16 $&$ -22 $&$ -22 $&$ -28 $&$ -34 $&$ -34 $&$ -40 $\\$
2$&$0 $&$ 
\quad\square $&$ -4 $&$ -6 $&$ -10 $&$ -9 $&$ -10 $&$ -15 $&$ -16 $&$ -18 $&$ -22 $\\$
3$&$ 0 $&$ -\frac{4}{3} $&$ -\frac{8}{3} $&$ -4 $&$ -\frac{14}{3} $&$ -\frac{20}{3} $&$ -10 $&$
   -\frac{26}{3} $&$ -\frac{32}{3} $&$ -10 $&$ -\frac{38}{3} $\\$
4$&$0 $&$ -\frac{3}{2} $&$ 
\quad\square $&$ -\frac{5}{2} $&$ -4 $&$ -\frac{11}{2} $&$ -6 $&$ -7 $&$ -10 $&$
   -\frac{17}{2} $&$ -9 $\\$
5$&$ 0 $&$ -\frac{2}{5} $&$ -\frac{8}{5} $&$ -\frac{12}{5} $&$ -\frac{18}{5} $&$ -4 $&$ -\frac{24}{5} $&$
   -6 $&$ -6 $&$ -\frac{36}{5} $&$ -10 $\\$
6$&$ 0 $&$ -\frac{2}{3} $&$ -\frac{4}{3} $&$ 
\quad\square $&$ -\frac{8}{3} $&$ -\frac{10}{3} $&$ -4 $&$
   -\frac{13}{3} $&$ -\frac{14}{3} $&$ -6 $&$ -\frac{20}{3} $\\$
7$&$0 $&$ -\frac{6}{7} $&$ -\frac{6}{7} $&$ -\frac{12}{7} $&$ -\frac{16}{7} $&$ -\frac{22}{7} $&$
   -\frac{22}{7} $&$ -4 $&$ -\frac{34}{7} $&$ -\frac{34}{7} $&$ -\frac{40}{7} $\\$
8$&$ 0 $&$ -\frac{1}{4} $&$ -\frac{3}{2} $&$ -\frac{7}{4} $&$ 
\quad\square $&$ -\frac{9}{4} $&$ -\frac{5}{2} $&$
   -\frac{15}{4} $&$ -4 $&$ -\frac{9}{2} $&$ -\frac{11}{2} $\\$
9$&$ 0 $&$ -\frac{4}{9} $&$ -\frac{10}{9} $&$ -\frac{4}{3} $&$ -\frac{16}{9} $&$ -\frac{20}{9} $&$
   -\frac{8}{3} $&$ -\frac{26}{9} $&$ -\frac{32}{9} $&$ -4 $&$ -\frac{38}{9} $\\$
10$&$ 0 $&$ -\frac{3}{5} $&$ -\frac{2}{5} $&$ -\frac{6}{5} $&$ -\frac{8}{5} $&$
\quad\square$&$
   -\frac{12}{5} $&$ -\frac{14}{5} $&$ -\frac{18}{5} $&$ -\frac{17}{5} $&$ -4 $\\   \hline
\end{tabular}
\end{center}
\caption{Table of values of $\eta'(m,n)$. 
Notice that $m^2\,\eta'$ is always an integer and that $\eta^\prime$ only depends on the combination $n/m$. The squared boxes mark the slots corresponding to pairs $(m,n)$ that are proportional to the pair $(2,1)$ for which the regulator $\epsilon$ fails (see the last paragraph of Appendix \ref{App:zeropole}).}
\label{tablaKappacd}
\end{table}


\vspace{0.4cm}
\noindent {\bf The $(m,n)$ representations of $\log \Ge(z;\tau,\tau)$}
\vspace{0.1cm} 

\begin{figure}[h]\centering
  \includegraphics[width=8cm]{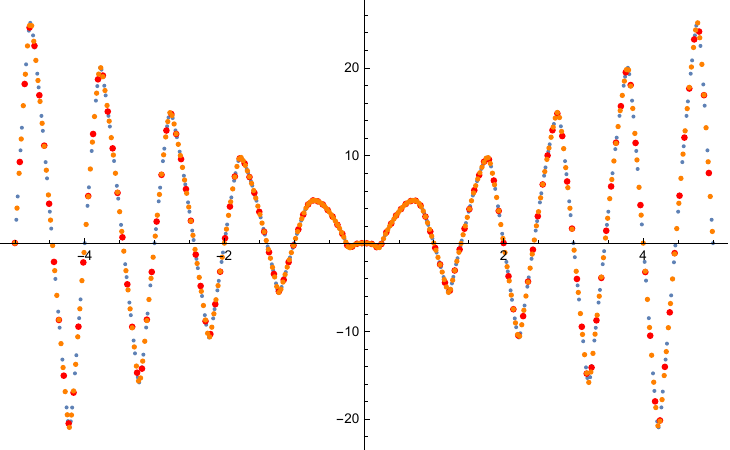}
  \caption{The numerical plots of $\log \Big|\prod_I \Ge(z_I( u(\upsilon));\,\tau,\,\tau)\Ge(z_I(-u(\upsilon));\,\tau,\,\tau)\Big|$ 
  as a function of~$\upsilon\in \mathbb{R}$, for $\tau =1+2 \,\i$ and~$u(\upsilon)\,=\,\upsilon \,\tau$. 
 The blue points are obtained by using the product representation~\eqref{GammaeDef} with~$\epsilon=0.2$. 
 We have truncated this representation at level~$j+k=30$, where the indices $j$ and $k$ are the mute ones in the 
 product~\eqref{GammaeDef}. 
  The orange points are obtained by using~\eqref{SeriesReppm} for $m=2$, $n=0$, $\epsilon=0.2$. 
  We truncate this series at level~$j=30$, where the index $j$ is the mute one in the $(2,0)$ 
  representation~\eqref{SeriesReppm}. The red points are obtained by using the representation~\eqref{SeriesReppm} 
  with~$m=1$,  $n=0$, and $\epsilon=0.2$. We truncate this  series at level~$j=30$, where the index $j$ 
  is the mute one in the~$(1,0)$ representation \eqref{SeriesReppm}.
}
  \label{GellFnPlot}
\end{figure}

\noindent We now develop similar series representations for the function~$\log \Ge(z;\t,\t)$ using 
the basic series representation~\eqref{ExpRepG} for~$\log \Ge(z;\t,\s)$ and the 
modular factorization identities~\eqref{FactElliptic}.
We start with an identity analogous to~\eqref{seriesth10} for the~$\Ge$-function
by combining~\eqref{ExpRepG} with the basic modular identity~\eqref{TripleId}.
The locus~$\sigma=\tau$, however, is not in the range of validity of~\eqref{TripleId}, 
so we need to take a limit $\sigma\rightarrow \tau$. 
After substituting the two elliptic gamma functions in the right-hand side of \eqref{TripleId} by their 
respective series representations~\eqref{ExpRepG} and taking the limit
$\varepsilon \to 0$ with~$\sigma = \tau+\varepsilon$
in the resulting expressions, one obtains the following representation, for any~$k_0 \in \IZ$,
\bea 
\log \Ge(z;\, \tau,\,\tau) & \=  & \frac{\pi i}{3} B_{3,3}(z-k_0|\tau,\t,-1) \nn \\
&& \; +\,{\i}\,\Bigl(\,\sum_{j=1}^{\infty}{  \frac{-\,2\, \tau +2 (z-k_0)+1 }{2 \,j\, \tau\, \sin \left(\frac{\pi  j}{\tau }\right) }\,
	 \cos  \frac{\pi j (2(\,z\,-\,k_0)+1)}{\tau }}  \quad  \label{ExpGammaNew} 
\\\nonumber
&&\qquad\quad - \, \frac{\pi  j \cot \left(\frac{\pi  j}{\tau }\right)+\tau }{2 \pi\,  j^2\, \tau \, \sin \left(\frac{\pi  j}{\tau }\right)} \,
	\sin{ \frac{\pi  j (2(z\,-\,k_0) +1)}{\tau}}\,\Bigr) \,. \nn
\eea
From domain of convergence of the series \eqref{ExpRepG}, given in \eqref{DomainGammaE}, it follows that the series on the right-hand side of \eqref{ExpGammaNew} are absolutely convergent in  
$ 0< -z_{\perp \tau} +k_0 <1$. 
We~have numerically checked the veracity of~\eqref{ExpGammaNew} in this ribbon,    
using the defining product representation~\eqref{GammaeDef} of the~$\Ge$-function. 
From representations \eqref{ExpGammaNew} and the factorization property \eqref{FactGamma}, one obtains the following representations
\be \label{repmnG}
\begin{split}
\log \Ge(z;\, \tau,\,\tau) & \= \sum_{\ell=0}^{2 (m-1)}   \frac{\pi i}{3} B_{3,3}(z+\ell \t-k_0|m\tau+n,m \t+n,-1)  \\
& \; \,+\,  \sum_{\ell=0}^{2 (m-1)} {\i}\,\Bigl(\,\sum_{j=1}^{\infty}{  \frac{-\,2\, (m\tau+n) +2 (z+\ell \t-k_0)+1 }{2 \,j\, (m\tau+n)\, \sin \left(\frac{\pi  j}{m\tau+n }\right) }\,
	 \cos  \frac{\pi j (2(\,z\,+\,\ell\t\,-\,k_0)+1)}{m\tau+n }} \\
& - \, \frac{\pi  j \cot \left(\frac{\pi  j}{m\tau+n }\right)+(m\tau+n) }{2 \pi\,  j^2\, (m\tau+n) \, \sin \left(\frac{\pi  j}{m\tau+n }\right)} \,
	\sin{ \frac{\pi  j (2(z\,+\,\ell \t\,-\,k_0) +1)}{m\tau+n}}\,\Bigr) \,.
\end{split}
\ee
We call them $(m,n)$~representations~of~$\log \Ge(z;\t,\t)$.
Since our eventual goal is to use these formulas to represent the SYM~index, wherein only even combinations of~$\log \Ge(z_I(u);\t,\t)$
appear, we move directly to a representation of such functions 
which will be very useful for us in the following. For every~$(m,n)$, 
\be\label{SeriesReppm}
\begin{split}
& \log \Ge(z_I(u);\t,\t) + \log \Ge(z_I(-u);\t,\t) \= \\
& \qquad -\,2\, \mathcal{F}^{(m,n)}(u) \,+ \,\sum_{j=1}^{\infty} \frac{2}{j} 
\Bigl(c_{1} (j;m,n) \, g(j u; m\tau+n)+c_{2} (j;c,d) \,\widehat{g}(j; u,  m\tau+n ) \Bigr)\,,
\end{split}
\ee
which follows from the identity \eqref{repmnG} and the definition \eqref{zIrI}. 
Analogously to the $(m,n)$ representations \eqref{ulogthseries}, the series in the right-hand side 
of \eqref{repmnG} and \eqref{SeriesReppm} are absolutely convergent in domain \eqref{udomain} for the $k_0(\ell)$ in \eqref{k0Value}. 
In Figure~\ref{GellFnPlot} we have plotted the real part of a couple of series 
representations~\eqref{SeriesReppm} and compared them with the representation of the left-hand side 
of \eqref{SeriesReppm} that one obtains after using the product definition of $\log\Ge(z;\t,\t)$.

The pre-factor~$\CF^{(m,n)}(u)$ is defined as
\be \label{Fcd}
- \,2\,\mathcal{F}^{(m,n)}(u) \=\frac{\pi\,\i }{3}  \, \sum_{\ell=0}^{2(m-1)} \, (m-|\ell -m+1|) \,
\Big( B_{3,3} \bigl(z_I(u)-k_0(\ell)+ \ell \tau| m \tau+{n},m \tau+{n},-1 \bigr) + \left(u \rightarrow -u\right)\Big)\,.
\ee
The function $g(u;\tau)$ (which is periodic under~$u \mapsto u+\t$) and the function $\widehat{g}(u;\tau)$ (which is not periodic under this shift) are 
\be \label{Defgghat}
g(u;\tau)\=  \cos{ \frac{2\pi u}{\tau}}\,, \qquad 
\widehat{g}(j;u,\tau) \= \frac{u}{\tau} \, \sin{\frac{2 \pi  j u}{\tau}}\,,
\ee
and the series coefficients in \eqref{SeriesReppm} are 
\bea
&& c_{1} (j;m,n) \=  \nn \\ 
&&  -\,\frac{\i}{2} \, \sum_{\ell=0}^{2(m-1)} (m-| \ell-m+1|) \times \,  \label{c1ncd} \\
&&  \times \Biggl(\, \frac{ { \cos \frac{\pi  j}{m\tau+n }
+\frac{m\tau+n}{\pi j} \sin \frac{\pi j}{m\tau+n}}}{ (m\tau+n)\sin^2 \frac{\pi  j}{m\tau+n }} \;  
\sin \Bigl(\frac{\pi  j (r_I(2  \tau +1)-2 k_0(\ell)+2 \,\ell\, \tau+1)}{m\tau+n }\Bigr) \qquad \qquad \qquad \qquad \nn 
\eea
\bea
&&   - \,\frac{ \bigl(-2 (m\tau+n)+r_I( 2  \tau+1)-2 k_0(\ell)+2 \ell \tau +1 \bigr)}{  (m\tau+n) \sin \frac{\pi  j}{m\tau+n } } \; 
\cos \Bigl(\frac{\pi  j (r_I(2  \tau
   +1)-2 k_0(\ell)+2 \ell \tau+1)}{m\tau+n } \Bigr) \,\Biggr), \nn
\eea
and 
\be
\begin{split}
\label{c2nGamma}
c_{2} (j;m,n)
&\=-\, 
\i\, \sum_{\ell=0}^{2(m-1)}\,(m-| \ell-m+1|)\, \frac{\sin{\frac{\pi j\left({r_I}\left(2 \tau+1\right)-2 k_0(\ell)
+2 \ell \tau+1\right)}{m\tau+n}}}{\sin{\frac{\pi j}{m\tau+n}}}  \\
&\= -\,
 \i \, m\,\sum_{\ell=0}^{m-1}\frac{\sin{\frac{\pi j\left({r_I}\left(2 \tau+1\right)-2 k_0(\ell)
 +2 \ell \tau+1\right)}{m\tau+n}}}{\sin{\frac{\pi j}{m\tau+n}}} \,.
\end{split}
\ee
In going from the first to the second line in \eqref{c2nGamma} we have used the identity $k_0(\ell+m)=k_0(\ell)-n$ (which follows immediately from \eqref{k0Value}), and the identity
\bea
(m-| \ell-m+1|)+(m-|\ell+1|)=m \,, 
\eea
which is valid for $0\leq \ell < m$.

For $u$ along an $(m,n)$ ray i.e.~$u \= \frac{u_2}{m} (m\tau+n )$, and several values of $m$, $n$ and R-charges~$r_I$, 
we have checked that the quantity
\be\label{auxeq2}
- 2\, \,\mathcal{F}^{(m,n)} (u) \,+\,2\,\mathcal{F}^{(m,n)} (0)\,+\,   2\pi \i\, \tau \, u_2^2 \, (r_I-1)\,,
\ee
is purely imaginary. After summing \eqref{auxeq2} over the matter content of $\mathcal{N}=4$ SYM, 
and taking the regulator $\epsilon$ to zero, one obtains
\be\label{u2term}
 - 2\,\sum_I \,\mathcal{F}^{(m,n)} (u) \,+\,2\, \sum_I\,\mathcal{F}^{(m,n)} (0) \= 2\pi \i \, \eta (m,n) \, (u_2)^2\,.
\ee
Before moving on, let us comment briefly about the constant $\eta(m,n)$. As for~$\eta'$
we have checked the following three properties of $\eta(m,n)$ in numerous examples: (a) it is rational, (b) it depends only 
on the ratio~$n/m$, and (c) $m^2\,\eta(m,n)\in \mathbb{Z}$ (See Table~\ref{tablaEtaGamma}).  
The last property will be useful in the next appendix.

\begin{table}[h]
\begin{center} 
\begin{tabular}{||cccccccccccc||}\hline $m \, \backslash\, n$  & $0$ &$1$&$2$&$3$&$4$&$5$&$6$&$7$&$8$&$9$&$10$ \\ \hline\hline
$1$&$0 $&$ 0 $&$ -1 $&$ 1 $&$ 0 $&$ -1 $&$ 1 $&$ 0 $&$ -1 $&$ 1 $&$ 0 $\\$
2$&$ 0 $&$ 
\,\square $&$ 0 $&$ 0 $&$ -1 $&$ \frac{1}{2} $&$ 1 $&$ -\frac{1}{2} $&$ 0 $&$ 0 $&$ -1 $\\$
3$&$ 0 $&$ 0 $&$ 0 $&$ 0 $&$ \frac{1}{3} $&$ 0 $&$ -1 $&$ \frac{1}{3} $&$ 0 $&$ 1 $&$ \frac{1}{3} $\\$
4 $&$0 $&$ -\frac{1}{4} $&$ 
\,\square $&$ \frac{1}{4} $&$ 0 $&$ -\frac{1}{4} $&$ 0 $&$ 0 $&$ -1 $&$ \frac{1}{4} $&$
   \frac{1}{2} $\\$
5$&$ 0 $&$ \frac{1}{5} $&$ 0 $&$ 0 $&$ -\frac{1}{5} $&$ 0 $&$ 0 $&$ -\frac{1}{5} $&$ \frac{1}{5} $&$ 0 $&$ -1 $\\$
6$&$ 0 $&$ 0 $&$ 0 $&$
\,\square$&$ 0 $&$ 0 $&$ 0 $&$ \frac{1}{6} $&$ \frac{1}{3} $&$ 0 $&$ 0 $\\$
7$&$ 0 $&$ -\frac{1}{7} $&$ \frac{1}{7} $&$ 0 $&$ 0 $&$ -\frac{1}{7} $&$ \frac{1}{7} $&$ 0 $&$ -\frac{1}{7}
   $&$ \frac{1}{7} $&$ 0 $\\$
8$&$ 0 $&$ \frac{1}{8} $&$ -\frac{1}{4} $&$ -\frac{1}{8} $&$ 
\,\square $&$ \frac{1}{8} $&$
   \frac{1}{4} $&$ -\frac{1}{8} $&$ 0 $&$ 0 $&$ -\frac{1}{4} $\\$
9$&$ 0 $&$ 0 $&$ -\frac{1}{9} $&$ 0 $&$ 0 $&$ 0 $&$ 0 $&$ \frac{1}{9} $&$ 0 $&$ 0 $&$ \frac{1}{9} $\\$
10$&$ 0 $&$ -\frac{1}{10} $&$ \frac{1}{5} $&$ 0 $&$ 0 $&$ 
\,\square$&$ 0 $&$ 0 $&$ -\frac{1}{5} $&$
   \frac{1}{10} $&$ 0 $\\   \hline
\end{tabular}
\end{center}
\caption{Table of values of $\eta(m,n)$. Note that $m^2\,\eta(m,n)$ is always integer and that $\eta(m,n)$ is only a function 
of the quotient $\frac{n}{m}$. The squared boxes mark the slots corresponding to pairs $(m,n)$ that are proportional to the 
pair $(2,1)$ for which the regulator $\epsilon$ fails (see the last paragraph of Appendix \ref{App:zeropole}).}
\label{tablaEtaGamma}
\end{table}


\vspace{0.4cm}
\noindent {\bf The $(m,n)$ representations of the elliptic extension}
\vspace{0.1cm} 

\noindent In Section~\ref{sec:Extension} we showed that although $\Ge(u)$ is not elliptic invariant one can dress 
it with~$\theta_0(u)^{u_2}$ to produce the elliptic function~$Q^{-1}(u)$ that coincides with~$\Ge(u)$ at~$u_2=0$, 
and is holomorphic in $\tau$ for finite values of $u_2$. The function $Q$ was represented by the double Fourier 
series expansion \eqref{QKron} which includes the~$\t$-independent purely imaginary 
term~$\i \Psi$ which was not fixed. 
In this subsection, we find an alternate representation of the even-in-$u$ product (from now even product) 
of $Q$'s that follows from the $(m,n)$ representations of $\theta_0$ and $\Ge$ that 
we have developed in the previous appendices. As before, we work with the even 
combination~$\Ge(z_I(u);\t,\t)\,\Ge(z_I(-u);\t,\t)$ 
appearing in the index formula. We show that, along an~$(m,n)$~ray,  the even product of~$Q$'s, 
coincides with the~$(m,n)$~representation of the even product of~$\Ge$'s, once the non-periodic 
part~$\widehat{g}$ of the $\Ge$'s is projected out\footnote{Recall that the series~\eqref{SeriesReppm} 
has three parts the pre-factor~$\CF$, the periodic part (governed by $g(n\,u; \t)$), and the non-periodic 
part ($\widehat{g}(n \,u; \t)$).}. 
We end this section by using this result to deduce an identity that will be used in Section \ref{App:AnalyticAction} 
to relate the elliptic and meromorphic actions of $(m,n)$ configurations.

We start by proving a useful identity. Upon comparing the coefficient of the third term in~\eqref{SeriesReppm} 
with the coefficient of the second term in~\eqref{ulogthseries}, 
we conclude that along an~$(m,n)$ ray i.e. for $u= \frac{u_2}{m} (m\tau+n) $, 
\be
\begin{split}
\sum_{j=1}^{\infty} \frac{2}{j} \, c_{2} (j;m,n) \; \widehat{g}(j; u,  m\tau+n ) 
\=    \sum_{j=1}^\infty \frac{2}{j} \, c^{\prime}_{2}(j;m,n) \, u_2\,
 \sin{\frac{2 \pi\, j\, u}{m\t+n}} \\
\= u_2 \, \log \theta_0(z_I(u); \tau) - u_2 \, \log \theta_0(z_I(-u); \tau) \,-\, 2\, \mathcal{F^{\,'}}^{(m,n)}(u)\,.
\end{split}
\ee 
We are ready to reach an $(m,n)$ representation for the logarithm of the even product of the 
function~$Q$ defined in~\eqref{Qab}. More specifically, we focus on the logarithm of the object in the left-hand side of
\bea\label{RepQTilde}
\begin{split}
\frac{1}{{Q}_{\left(r_I-1,\, \frac{r_I}{2}\right)}(u){Q}_{\left(r_I-1,\, \frac{r_I}{2}\right)}(-u)}& 
\= \e\bigl(\wt \Psi(u)\bigr)\, \e\bigl(-r_I \,\tau \, u_2^2\bigl)
\\ &\qquad \times\frac{\Ge(z_I(u);\tau,\tau)\Ge(z_I(-u);\tau,\tau)}{\theta_0(z_I(u);\tau)^{u_2}\theta_0(z_I(-u);\tau)^{-u_2}}.
\end{split}
\eea
The $\tau$-independent function $\wt\Psi(u)$ is a real-analytic, perhaps multi-valued, function 
obeying the following two conditions. Firstly, it needs to obey the initial condition
\be\label{PhasePsi}
\wt \Psi(u_2=0)\=0
\ee
so that the left-hand side of  \eqref{RepQTilde} reduces to $\Ge(z_I(u_1))\Ge(z_I(-u_1))$ for $u_2=0$. 
Secondly, its ``lack of ellipticity" needs to be such that \eqref{RepQTilde} is an elliptic function in $u$-plane.

Finally, we move on to write down the $(m,n)$ representation of the logarithm of~\eqref{RepQTilde}\footnote{Before entering into definition of 
logarithms, it worths noticing that the $\theta_0(z_I(\pm u))^{\pm u_2}$ in the right-hand side of \eqref{RepQTilde} must be defined with care, 
as the power function is multivalued when acting over the field of complex numbers (except for integer powers).}. 
We will do so in three steps. First, we define such logarithm, using representations~\eqref{ulogthseries} and~\eqref{SeriesReppm}. 
For $u\,=\, \frac{u_2}{m}(m\tau+n)$ one obtains
\bea \label{eqQplethystic}
&& -\log  Q_{\bigl(r_I-1,\, \frac{r_I}{2}\bigr)}(u) - \log  Q_{\bigl(r_I-1,\, \frac{r_I}{2}\bigr)}(-u) \qquad\qquad\qquad\qquad\qquad\\
&&  \qquad \= 2\pi\i\,\bigl( \,\wt\Psi(u)\,-\,\tau \, r_I\, u_2^2 \,\bigr)\, - \,2\,\mathcal{F}^{(m,n)}(u) \,- \, 2\, \mathcal{F^{\,'}}^{(m,n)}(u) 
+ \sum_{j=1}^{\infty} \frac{2}{j} c_{1} (j;m,n) \, g(j u; m\tau+n)  \,. \nn
\eea
Second, before reaching the final equation, we must note an interesting property.  Representations \eqref{eqQplethystic} are labelled by $m\in\mathbb{N}$ and $n\in \mathbb{Z}$.
From observations made on expressions \eqref{auxeq1} and \eqref{auxeq2}, it follows that, along an $(m,n)$ ray i.e.~for $u\,=\, \frac{u_2}{m}(m\tau+n)$
\be\label{C42}
\begin{split}
- 2\pi \i\, \tau \, r_I\, u_2^2 \, - \,2\,\mathcal{F}^{(m,n)}(u) \,- \, 2\, \mathcal{F^{\,'}}^{(m,n)}(u) \qquad\qquad\qquad\\ \= - \,2\,\mathcal{F}^{(m,n)}(0) \,+\, 2 \pi\i \, \bigl(\eta(m,n) -\eta^\prime(m,n)\bigr)\, u_2^2.
\end{split}
\ee
The second term in the second line is a $\tau$-independent imaginary term proportional to~$u_2^2$.  
Third, and final, we plug \eqref{C42} in \eqref{eqQplethystic} and 
redefine $\wt\Psi(u) \to \wt\Psi(u) - (\eta(m,n) -\eta^\prime(m,n))\, u_2^2$, 
so that, for all $m\in\mathbb{N}$ and $n\in\mathbb{Z}$, and 
for $u\,=\, \frac{u_2}{m}(m\tau+n)$
\be \label{eqQplethystic2}
\begin{split}
&-\log  Q_{\bigl(r_I-1,\, \frac{r_I}{2}\bigr)}(u) - \log  Q_{\bigl(r_I-1,\, \frac{r_I}{2}\bigr)}(-u) \qquad\qquad\qquad\qquad\qquad\\
& \qquad\qquad\= \,2\pi \i\, \wt\Psi(u)\, - \,2\,\mathcal{F}^{(m,n)}(0)  \, +\,\sum_{j=1}^{\infty} \frac{2}{j} \, 
c_{1} (j;m,n) \, g(j u; m\tau+n)  \,. 
\end{split}
\ee
This is the $(m,n)$ representation we were looking for. It shows that along an $(m,n)$~ray, 
the even sum of elliptic functions $-\log{Q}$ in the left-hand side, can be seen, up to a redefinition 
of $\wt\Psi(u)$\footnote{The new~$\wt \Psi$ is doubly periodic, respects the constraint~\eqref{PhasePsi},
 and is real-valued.}, as the result of
projecting out the non periodic part $\wt g$ in $(m,n)$~representation \eqref{SeriesReppm} of the 
corresponding even sum of $\log{\Ge}$. 
Notice that $\wt\Psi(u)$ in \eqref{eqQplethystic2} is still ambiguous, but it is elliptic, 
satisfies~\eqref{PhasePsi}, and it does not depend on $\tau$.

Equation \eqref{eqQplethystic2} and the definition of the elliptic extension of the action~$S(\underline{u})$ given in~\eqref{SPQ}, 
implies the following identity
\bea\label{IdentityCompActions}
S_{\text{eff}}(m,n;\t)\=\, N^2\,\sum_I\,\mathcal{F}^{(m,n)} (0) \,+\,\pi\i N^2 \vt(m,n) \,,
\eea
where $\frac{\vt(m,n)}{4}\in\mathbb{R}$ are the Fourier coefficients of $\wt\Psi(u)$ along the $(m,n)$ ray. 
They are arbitrary constants that could be matched to $\v(m,n)$.
We have checked that the right-hand side of~\eqref{IdentityCompActions} coincides with the 
right-hand side of~\eqref{actionEll} for several values of~$m\in \mathbb{N}$,~$n\in \mathbb{Z}$ and gcd$(m,n)=1$. 
This will be used in Section \ref{App:AnalyticAction} to match the elliptic and meromorphic actions 
(up to a $\tau$-independent imaginary quantity).

\section{The role of the saddle~$(0,0)$ \label{sec:00saddle}}
\begin{figure}[h]\centering
  \includegraphics[width=8cm]{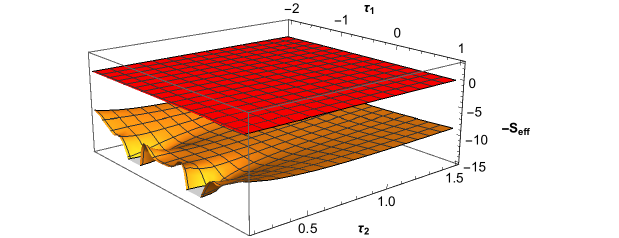}
  \includegraphics[width=6cm]{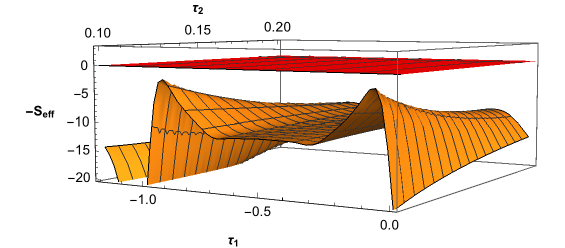}
  \caption{$-\text{Re} \, S(0,0)$ (Orange) vs $-\text{Re} \, S(0,1) =0$ (Red). The unit of the vertical axis is $\frac{N^2}{2}$.
   We have used the representation~\eqref{SeriesReppm} with $m=1$ and $n=0$, with a truncation at $j=2000$. 
 The plot on the left covers the domain $\tau_1 \in [-2,\,1]$ and $\tau_2 \in [0.2,\, 1.5]$, and  
  the plot on the right covers the domain $\tau_1 \in [-1.2,\,0]$ and $\tau_2 \in  [0.1,\, 0.2]$. Both plots have~$\epsilon\,=\,0.001$
  which is small enough for the ranges of~$\t$ considered here, so that the $(0,0)$ solution is suppressed with respect to~$(0,1)$.
  }
  \label{fig00Comparison}
\end{figure}

We now clarify the role (and the lack of role) of the saddle~$(0,0)$.
First let's turn off the regulator~$\epsilon$ so that~$u_i=0$ strictly. 
In this case the vector multiplet has~$z_V(0)=2 \t +1$ (see Equation~\eqref{zIrI}).
From the product representation~\eqref{GammaeDef} we see that this value of~$z_V$ 
is a zero of the elliptic gamma function and therefore the index actually vanishes. 
Now let's turn on the regulator~$\epsilon$. From the representation~\eqref{SeriesReppm} 
we see that, for small enough $\epsilon$ for fixed~$\tau \in \mathbb{H}$, the quantity $-\text{Re} \, S(0,0)$
is negative and thus the contribution of saddle~$(0,0)$ is suppressed. 
These observations are summarized in Figure \ref{fig00Comparison}.
For the~$(0,0)$ saddle, all values of~$(m,n)$ in the representation~\eqref{SeriesReppm} clearly give the same answer.  
We have used~$m=1$, $n=0$ to generate the figure. 

In contrast, for fixed~$\epsilon$ and in the Cardy-like limit ($\tau \to 0$ with $\ret<0$) the quantity~$-\text{Re} \, S(0,0)$
becomes positive. 
Furthermore, as shown in~\cite{Cabo-Bizet:2019osg}, the full action asymptotes to the action of the black hole.  
 In Section~\ref{sec:Cardy} we have revisited and confirmed the results of \cite{Cabo-Bizet:2019osg}, 
 as to how any asymptotic limit~$\tau \to 0$ of the action of the 
black hole saddle $(1,0)$ coincides, in that very same limit, with the $\epsilon$-regulated action of the~$(0,0)$ saddle.

\section{The $(m,n)$ meromorphic action \label{App:AnalyticAction}}

In this appendix we evaluate the analytic continuation of the integrand of~\eqref{SYMIndex} 
on the $(m,n)$~configurations $u_i=(m\t+n) \frac{i}{N}$ and compare it 
with the valuation of the elliptic extension of the action calculated in Section~\ref{sec:Action} on the same configurations.
To ease the language, we will call the negative of the logarithm of these two quantities, 
the \emph{meromorphic action} and the \emph{elliptic action}, respectively. 
We find that, remarkably, although the $(m,n)$ configurations are not saddles of the meromorphic action, 
the \emph{valuations} of the meromorphic and elliptic actions on these configurations are essentially the 
same\footnote{The difference of the two actions is a purely imaginary $\tau$-independent term, that 
depends on $m$ and $n$.}!

The meromorphic action  is defined as the logarithm of the integrand of~\eqref{SYMIndex} at large~$N$, i.e.~(the 
relevant notation has been summarized near~\eqref{zIrI}), 
\be \label{SmeroD}
S_\text{mero}(\underline{u})\defeq - \sum_I \sum_{i,j} \, \log \Ge \bigl(u_{ij} + \frac{r_I}{2}(2\t+1); \t,\t \bigr)   \,.
\ee
In particular, using the regulator discussed in the previous appendix, we should take
$r_V=2-\epsilon$ and $r_C=\frac{2+\epsilon}{3}$, and~$\epsilon \to 0^+$ at the end.

In order to calculate the meromorphic action we simply evaluate the series representation
of the elliptic gamma functions developed in the previous section on the $(m,\,n)$~configuration, 
\be\label{cdansatz}
u_{ij} \= (m\t+n) \, \frac{i-j}{N} \,.
\ee
We recall that the $(m,n)$ series representation~\eqref{SeriesReppm} has three parts, the pre-factor~$\CF$, 
the periodic part (governed by $g(j\,u; \t)$), and the non-periodic part ($\widehat{g}(j \,u; \t)$). We proceed to evaluate them 
on the configuration~\eqref{cdansatz}.

From the identity, 
\bea\label{D11}
\sum_{i,\,j=1}^{N}\; g\,\Bigl(k_1 (m\tau\,+\,n)\frac{i-j}{N}; \, m\, \tau+\, n\Bigr) \;=\;  \sum_{k_2 \in \mathbb{Z}}\, N^2 \,\delta_{k_1, \,k_2\, N} \,,
\eea
it follows that the only non-vanishing contributions in the series
\be \label{seriesg}
 \sum_{k=1}^\infty \, \frac{1}{k} \,c_{1}(k;m,n)\, \sum_{i,\,j=1}^{N}\;  g\Bigl(k (\,m\t+n\,)\frac{(i-j)}{N}; \,m\, \tau\,+\,n\Bigr)
\ee
are when~$k$ equals integer multiples of~$N$. Noting that the coefficients~$c_{1}(k;m,n)$~\eqref{c1ncd} are 
exponentially suppressed as a function of~$k$ for large~$k$, we have that the double sum~\eqref{seriesg} 
behaves like~$e^{-N}$ and therefore vanishes in the large-$N$ limit.

The non-periodic function $\widehat g(u;m\t+n)$ vanishes as well when evaluated on \eqref{cdansatz}. 
This follows from the identity, $k\in \mathbb{Z}$,
\be\label{Identity1}
\sum_{i,\, j=1}^{N} \Bigl(\frac{i-j}{N}\Bigr) \sin \Bigl( \frac{2 \pi k (i-j)}{N} \Bigr)\=0\,,
\ee
which is proved as follows
\be \label{Identity1}
\begin{split}
 \sum_{i,j=1}^N \, (i-j) \, \e \bigl(\tfrac{i-j}{N} k \bigr)  
& \= \sum_{i=1}^N \, i \, \e \bigl(\tfrac{i}{N} k \bigr)\sum_{j=1}^N  \, \e \bigl(\tfrac{-j}{N} k \bigr) \; - \;  
 \sum_{j=1}^N \, j \, \e \bigl(\tfrac{-j}{N} k \bigr)\sum_{i=1}^N  \, \e \bigl(\tfrac{i}{N} k \bigr)  \\
&  \= \sum_{i=1}^N \, i \, \e \bigl(\tfrac{i}{N} k \bigr) \, \delta_{k,0} \; - \;  
 \sum_{j=1}^N \, j \, \e \bigl(\tfrac{-j}{N} k \bigr) \, \delta_{k,0} \\
& \=  \delta_{k,0} \Bigl( \, \sum_{i=1}^N \, i \; - \; \sum_{j=1}^N \, j  \Bigr) \= 0 \,.
\end{split}
\ee

The only term contributing to the action~\eqref{finalFormula} is the pre-factor~$\CF$.
Using Formula~\eqref{u2term}, summing over all the~$\Ge$-functions that contribute to the index, 
and using the fact that on the~$(m,n)$ ansatz  
\be \label{Phasecd}
\sum_{i,j=1}^{N} ({u_ {ij}}_2)^2 \;=\; m^2 \,\frac{\sum_{i,j=1}^{N} (i-j)^2}{N^2}
\= \frac{m^2}{6} (N^2-1)\=\frac{m^2}{6} \,N^2\,+\, O \Bigl(\frac{1}{N} \Bigr)\,,
\ee
we obtain that the value of the meromorphic action $S_{\text{mero}}(\underline{u})$ on the~$(m,n)$ configuration is
\be \label{finalFormula}
S_\text{eff}^\text{mero}(m,n;\t) \= N^2 \,\sum_I\,\mathcal{F}^{(m,n)} (0) \,
\,-\, \frac{ \pi\, \i}{6}\, N^2 \, m^2 \,\eta(m,n)  \,,
\ee
where $\mathcal{F}^{(m,n)} (0) $ is defined by \eqref{Fcd}, and the index $I$ runs over 
the matter content of $\mathcal{N}=4$ SYM. 
The real function $\eta(m,n)$ is the pure imaginary term discussed in \eqref{u2term}.

Upon comparing the formulas~\eqref{finalFormula} and~\eqref{IdentityCompActions}, it follows that 
the elliptic and meromorphic $(m,n)$ effective actions, for relatively prime~$(m,n)$, coincide up to a $\tau$-independent imaginary quantity. 
Thus, for every coprime~$(m,n)$, the meromorphic action~\eqref{finalFormula} can be written in the 
form\footnote{For $N$ not a prime, recall that there are other solutions 
along $(m,n)$ rays denoted by $(K|m,n)$ with $K|N$, and for $m$,~$n$~co-prime. 
We can also compute their large-$N$ action by similar methods to be 
\be
S_{\text{eff}}^{\text{mero}}(K|m,n)\;=\;S_{\text{eff}}^{\text{mero}}(m,n)\,-\, \frac{N^2}{K}\,\sum_{j=1}^{\infty} \frac{c_{1}(jK;m,n)}{j}\, 
+\,\text{$\tau$-independent phase}\,.
\ee
The gap between the actions of $(N|m,n)\equiv (m,n)$ and $(K|m,n)$ vanishes exponentially 
fast for large values of $K\sim N$ because the coefficient $c_1$ is exponentially suppressed for 
large values of its first argument (the~$\epsilon$-regulator is important to do such calculations). 
It would be interesting to understand the meaning of the large-$N$ limits of these solutions.}   
\be\label{Actionmero}
S_\text{eff}^\text{mero}(m,n;\tau)  
\= \frac{N^2 \,\pi\, \i }{27\,m} \,\frac{  \bigl(2 (m\t+n)  +  \chi_1(m+n) \bigr)^3}{(m\t+n)^2} + N^2 \,\pi\, \i  \, \v_\text{mero}(m,n)  \,,
\ee
where~$\v_\text{mero}(m,n)\in \mathbb{R}$. In particular we find that~$\v_\text{mero}(1,0)=0$ as mentioned 
below Equation~\eqref{JQReal}.

We end this section with a comment about a subtle issue regarding the comparison between the elliptic and meromorphic approaches. 
Recall from the discussion in Section~\ref{sec:Saddles} that, for $N$ prime, the saddles of the elliptic action are labelled 
by pairs~$(m,n)$ such that~$\text{gcd}(m,n)=1$. We could have chosen to reparameterize the saddle~$(m,n)$ as~$(km,kn)$, where~$k$
is any integer---even depending on~$m,n$---and this would not have made any difference to the elliptic action because 
elliptic symmetry forces the action to be a function of the quotient $n/m$, and not of $m$ and $n$ independently. 
In the meromorphic approach, however, we find that the action is not a function of~$n/m$ only. In particular, 
we have checked in several examples that the first summand in the right-hand side of~\eqref{finalFormula} 
only depends on the ratio~$n/m$ but the second summand in the right-hand side of~\eqref{finalFormula} 
is not just a function of $n/m$. 
If we think of this reparameterization as a  ``local rescaling" of the space of saddles with a function $k(m,n)$, 
we are drawn to conclude that the meromorphic action is anomalous under such a local scale-reparametrization. 
The resolution of this puzzle is tied to a better understanding of the imaginary term in the 
action (see Footnote~\ref{footPhidef}).

\section{Relation to the $\mathcal{N}=4$ SYM Bethe Ansatz formula \label{sec:BA}}

In this subsection we make contact with the approach of~\cite{Benini:2018ywd,Benini:2018mlo}.
The Bethe Ansatz formula in these papers, introduced in a related context in~\cite{Closset:2017bse,Closset:2018ghr}, 
has the following form for the~$\CN=4$ SYM index,
\be\label{BA}
\mathcal{I}_{BA}\=\kappa_N \sum_{\underline{u}^\star \in \text{BAEs}} \,
\mathcal{Z}(\underline{u}^\star) \, \mathcal{H}^{-1}(\underline{u}^\star) \,.
\ee
Here the function~$\mathcal{Z}(\underline{u})$ is the analytic continuation in~$\underline{u}$ of the integrand of the 
index~\eqref{SYMIndexst}---which is precisely what we discussed in Appendix~\ref{App:AnalyticAction}, i.e.~in our 
language~$\mathcal{Z}(\underline{u})=\exp\bigl(-S^\text{mero}_\text{eff}(\underline{u})\bigr)$ defined in~\eqref{SmeroD}.
The factor $\mathcal{H}^{-1}$ is a Jacobian factor which is subleading in~$\frac{1}{N}$ compared to the effective action.
The factor $\kappa_N$ is precisely the pre-factor in the right-hand side of our equation \eqref{SYMIndex}, which is also
sub-leading in~$\frac{1}{N}$.
The numbers~$\underline{u}^\star=\{u^\star_i\}$, $i \= 1,\ldots N$ entering the formula, called Bethe roots, 
are solutions to a set of algebraic equations called the Bethe Ansatz Equations (BAEs), and obey~$\sum_i u_i^\star=0$.

The BAEs are equations for the positions of poles of the integrand in a certain integral representation of the SYM index (which does not look 
a priori the same as~\eqref{SYMIndex}). These equations are of the form
\be \label{BAEs}
\prod_{j=1}^N \, f(u_{ij}^\star) \=1\,, \qquad i \= 1, \dots, N \,. 
\ee
where~$f$ is a certain combination of~$\theta_0$-functions. 
Solutions to these equations were found in~\cite{Hosseini:2016cyf}, \cite{Hong:2018viz}, \cite{Benini:2018mlo}
and, in particular, it was shown that the~$(m,n)$ configuration i.e.~$u^\star_{ij} = \frac{i-j}{N}(m\t+n)$ 
with~$\text{gcd}(m,n)=1$ is a solution to the BAEs. 
It was also shown that the~$(1,0)$ configuration\footnote{\label{MatchingDict}The conventions of \cite{Benini:2018ywd} can be 
matched to the ones used in this paper  under the identifications~$ \tau_{\text{here}} \leftrightarrow  \tau_{\text{there}}\,-\,1$, 
$\Delta_{1,2,3}= (2\tau_{\text{there}}-1)/3$, and 
$\CI_{\text{here}}(\t_{\text{here}}, \t_{\text{here}};n_0=-1)\, 
\leftrightarrow \,\CI_{\text{there}}(\Delta_1, \Delta_2, \Delta_3, \t_{\text{there}},\t_{\text{there}})$. 
This leads to the following map between saddles:~$(m,n)_{\text{there}}\,\leftrightarrow (m,m+n)_{\text{here}}$.
} 
has the same entropy as the AdS$_5$ black hole.

From our point of view, we notice that the logarithm of the BAEs~\eqref{BAEs} is precisely of the form~\eqref{Vpr} 
for some~$V(u)$ that is even in~$u$ and double-periodic on the lattice generated by~$1$ and~$\t$. 
Thus, in virtue of our discussion in Section \ref{sec:Saddles}, it follows that all the~$(m,n)$ 
configurations discussed in this paper are Bethe roots. 
Further, from the discussion in Appendix~\ref{App:AnalyticAction}, it follows that the action of these roots is given by the 
formula~\eqref{Actionmero}. As mentioned below that equation, it agrees with the elliptic action introduced in this paper 
evaluated at the same saddle, up to a~$\t$-independent imaginary term.

%

\providecommand{\href}[2]{#2}\begingroup\raggedright\endgroup

\end{document}